\documentclass[aps,prx,twocolumn,showpacs,psfig,superscriptaddress,longbibliography]{revtex4-2}
\usepackage{textcomp}
\usepackage{times}
\usepackage{graphicx}
\usepackage{float}
\usepackage{latexsym,amsmath,amssymb,bm,euscript}
\usepackage{color}
\usepackage{subfigure}
\usepackage{epstopdf}
\usepackage[colorlinks=true,linkcolor=blue,citecolor=blue,urlcolor=blue]{hyperref}
\usepackage{hyperref}
\usepackage{soul}
\usepackage[normalem]{ulem}
\usepackage{mathrsfs}
\usepackage{amsmath,amssymb}
\usepackage{lettrine}
\usepackage{xfrac}
\usepackage{xspace}
\usepackage{textcomp}
\usepackage{wasysym}
\usepackage{xr}
\usepackage{lipsum}
\usepackage{mhchem}
\usepackage{xcolor}
\usepackage{booktabs}
\usepackage{float}
\usepackage{bibunits}

\graphicspath{{./Figs/}}
\AtBeginDocument{\newwrite\bibnotes
\def\bibnotesext{Notes.bib}
\immediate\openout\bibnotes=\jobname\bibnotesext
\immediate\write\bibnotes{@CONTROL{REVTEX41Control}}
\immediate\write\bibnotes{@CONTROL{apsrev41Control,author="08",editor="1",pages="1",title="0",year="1"}}
\if@filesw
\immediate\write\@auxout{\string\citation{apsrev41Control}}\fi
}

\begin{document}

\title{Pseudo-Goldstone Modes at Finite Temperature}
\author{Xiyue Lin}
\author{Tao Shi}
\email{tshi@itp.ac.cn}

\affiliation{CAS Key Laboratory of Theoretical Physics,
Institute of Theoretical Physics,
Chinese Academy of Sciences, Beijing 100190, China} 
\affiliation{School of Physical Sciences, University of Chinese
Academy of Sciences, Beijng 100049, China}
\begin{abstract}
Goldstone’s theorem and its extension to pseudo-Goldstone (PG) modes have profound implications across diverse areas of physics, from quantum chromodynamics to quantum magnetism. PG modes emerge from accidental degeneracies lifted by quantum and thermal fluctuations, leading to a finite gap—a phenomenon known as "order by disorder." In this paper, we derive a general curvature formula for the PG gap at finite temperature, applicable to both collinear (e.g., ferromagnets and anti-ferromagnets) and noncollinear magnetic orders (e.g., coplanar orders in frustrated magnetic systems). After validating our formula against known models, we apply it to the XXZ model on the triangular lattice, which hosts coplanar magnetic orders in equilibrium and is relevant to materials such as \ce{Na2BaCo(PO4)2} and \ce{K2Co(SeO3)2}, known for their supersolid phases and giant magnetocaloric effects. Our results reveal a distinct scaling behavior: a linear decrease of the PG gap with temperature, driven by entropy effects from magnon scattering across multiple bands. This stands in stark contrast to the high-temperature scaling recently proposed for systems with a single magnon band. This work establishes a
general framework for investigating PG modes at finite temperatures and opens an avenue to explore rich quantum phases and dynamics in frustrated systems with noncollinear magnetic orders.

\end{abstract}

\date{\today }
\maketitle

\section{Introduction} 
Goldstone’s theorem establishes a fundamental connection between the spontaneous breaking of continuous symmetries and the emergence of gapless Nambu-Goldstone excitations~\cite{Nambu1960su,Goldstone1961,Goldstone1962}. When accidental degeneracies, not enforced by symmetries, arise at the classical level, an intriguing class of excitations known as pseudo-Goldstone (PG) modes can emerge~\cite{Burgess2000}. Unlike true Goldstone modes, PG modes acquire a mass gap~\cite{Weinberg1972} because quantum and thermal fluctuations lift the accidental degeneracy through a mechanism commonly referred to as ``order by disorder"~\cite{Villain1980,Shender1982,Henley1989}. This phenomenon has been widely studied across various fields, including high-energy physics~\cite{Nambu1960,Nambu1961}, atomic physics~\cite{Deng2020,Xiaoqi2012,Li2021}, high-temperature superconductors~\cite{Demler2004,Fernandes2017}, and quantum magnetism~\cite{Villain1980,Shender1982,Henley1989,Murthy1997o,Belorizky1980,Yildirim1999o}. PG modes are crucial for understanding exotic dynamics and spectroscopy related to low-lying excitations, which are experimentally measured
via neutron and time-of-flight inelastic scattering techniques in various compounds, e.g.,
\ce{Na2BaCo(PO4)2}~\cite{Gao2024,Xiang2024}, \ce{K2Co(SeO3)2}~\cite{zhu2024,chen2024}, \ce{Er2Ti2O7}~\cite{Ross2014,Savary2012}, \ce{LaTiO2}~\cite{Keimer2000}, \ce{CoTiO3}~\cite{Elliot2021}, \ce{Sr2Cu3O4Cl2}~\cite{Kim1999}, \ce{Fe2Ca3(GeO4)3}~\cite{Brueckel1988} and \ce{Yb2Ge2O7}~\cite{Sarkis2020}.

\begin{figure}[!t]
\includegraphics[angle=0,width=1\linewidth]{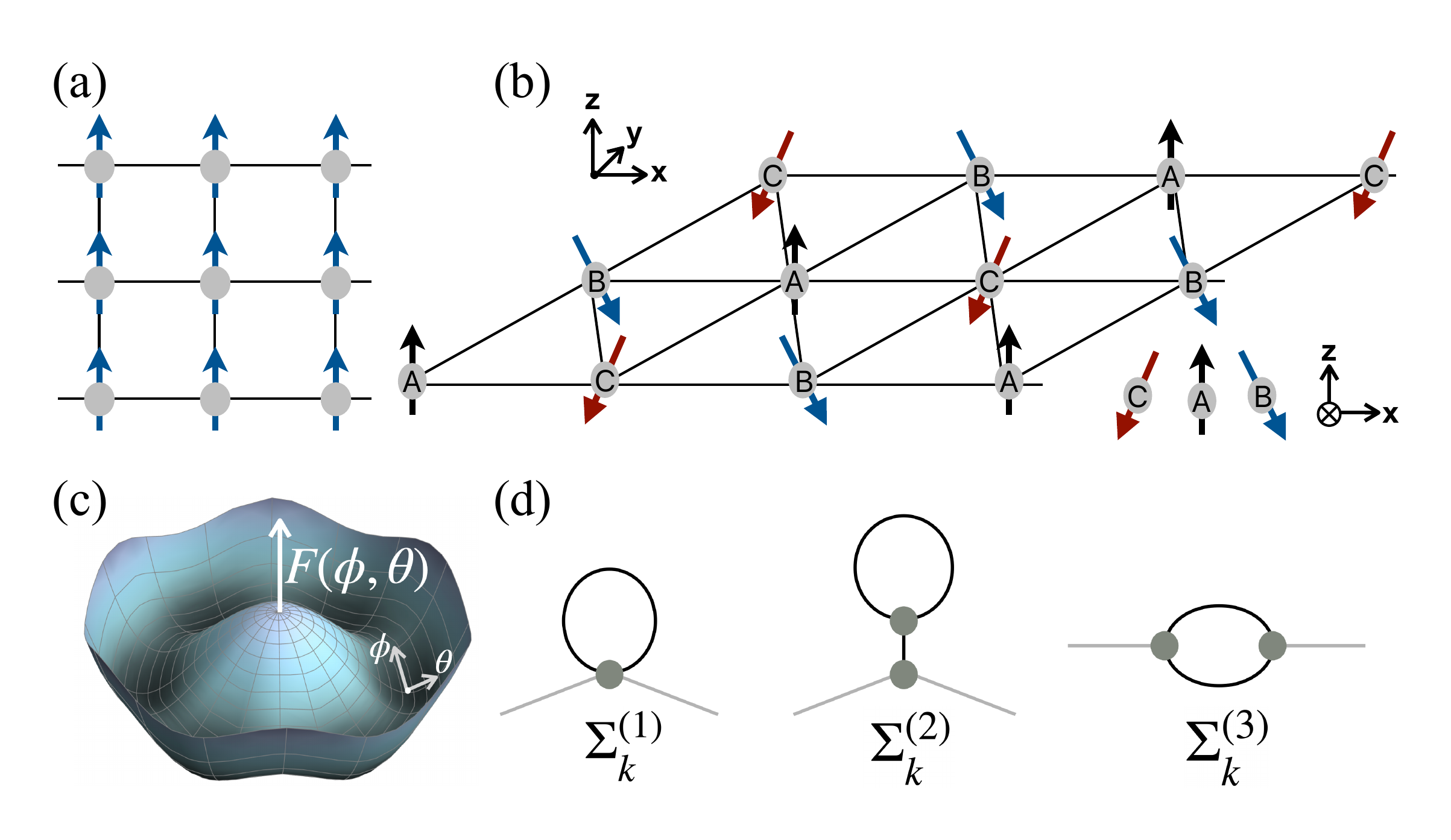}
\caption{(a) Ferromagnet of Heisenberg-compass model on square lattice. (b)
Three sublattice structure of the triangular lattice XXZ antiferromagnet.
(c) Schematic illustration of free energy $F(\protect\phi,\protect\theta)$,
where the curvatures of the free energy are related to the self-energy and
pseudo-Golstone gap. (d) The three classes of diagrams contribution to the $O(S^{(0)})$ magnon self-energy. }
\label{Fig:Model}
\end{figure}

The PG gap at zero temperature is closely linked to the curvature of the ground-state energy landscape. However, the curvature formula~\cite{Rau2018} has only been justified for systems with collinear magnetic order, e.g., (anti-)ferromagnets. A scaling relation for the PG gap as a function of temperature has been proposed at high temperatures and further validated for the Heisenberg-compass model [cf. Fig.~\ref{Fig:Model}(a)] with collinear order~\cite{Khatua2023}. However, extrapolating this scaling relation to intermediate and low temperatures remains challenging. More importantly, the universality of this scaling relation across a wide range of magnetic systems at low temperatures is questionable, particularly for those with noncollinear orders. A notable example is the Heisenberg XXZ model on the triangular lattice [cf. Fig.~\ref{Fig:Model}(b)], which is of particular interest due to its relevance to materials such as \ce{Na2BaCo(PO4)2} and \ce{K2Co(SeO3)2}, known for exhibiting a supersolid phase and giant magnetocaloric effects~\cite{Gao2024,Xiang2024,zhu2024,chen2024}.

In this paper, we present a general curvature formula for the PG gap at finite temperature, applicable to both collinear and noncollinear orders. After validating this formula with known models~\cite{Rau2018,Khatua2023}, we apply it to the XXZ model on the triangular lattice, which exhibits coplanar orders in equilibrium. Our curvature formula reveals that, in such systems, the PG gap decreases linearly with temperature, in contrast to predictions~\cite{Khatua2023} for systems with collinear orders. We demonstrate that this behavior is driven by an entropy effect arising from magnon scattering across multiple bands, distinct from systems with a single magnon band, e.g., the Heisenberg-compass model. Our results provide a more general framework for understanding the PG gap in frustrated systems with noncollinear orders at low and intermediate temperatures.

\section{Model and equilibrium state configurations}

We consider an interacting spin system in $d$ dimensions, consisting of $N = N_c N_s$ sites, which exhibits translational symmetry. Here, $N_c$ is the number of unit cells, and $N_s$ is the number of sublattices per unit cell. The system Hamiltonian reads 
\begin{equation}
H=\frac{1}{2}\sum_{\mathbf{r}\mathbf{r}^{\prime }}\sum_{\mu\mu^{\prime }}J_{%
\mathbf{r}-\mathbf{r}^{\prime }}^{\mu\mu^{\prime }} \hat{S}_{\mathbf{r}%
}^{\mu}\hat{S}_{\mathbf{r}^{\prime }}^{\mu^{\prime }},  \label{eq:H}
\end{equation}
where $\hat{S}_{\mathbf{r}}^{\mu} \equiv \hat{S}_{j,\alpha}^{\mu}$ is the
spin operator along the $\mu=x,y,z$ direction at site $\mathbf{r}=%
\mathbf{r}_{j,\alpha}$ of the sublattice $\alpha$ within unit cell $j$, and $%
J_{\mathbf{r}-\mathbf{r}^{\prime }}^{\mu\mu^{\prime }}$ is the interaction
strength.

In magnetically ordered phases, the ground state can be approximated as a spin coherent state $|\{\phi_{\mathbf{r}}^{c},\theta_{%
\mathbf{r}}^{c}\}\rangle$~\cite{Radcliffe1971,Perelomov1972,Arecchi1972}
parametrized by the polar and azimuthal angles $(\phi_{\mathbf{r}%
}^{c},\theta_{\mathbf{r}}^{c})\equiv(\phi_{j,\alpha}^{c},\theta_{j,%
\alpha}^{c})$, where the average values
\begin{align}
    \langle\hat{S}_{\mathbf{r}%
}^{x}\rangle &=S\sin(\theta_{\mathbf{r}}^{c})\cos(\phi_{\mathbf{r}}^{c}) \nonumber\\
   \langle\hat{S}_{\mathbf{r}}^{y}\rangle &=S\sin(\theta_{\mathbf{r}%
}^{c})\sin(\phi_{\mathbf{r}}^{c}) \nonumber\\
\langle\hat{S}_{\mathbf{r}%
}^{z}\rangle &=S\cos(\theta_{\mathbf{r}}^{c}).
\end{align}
The classical configuration $%
(\phi_{\mathbf{r}}^{c},\theta_{\mathbf{r}}^{c})$ is determined by minimizing
the classical energy $E_c=\sum_{\mathbf{r}\mathbf{r}^{\prime }}J_{\mathbf{r}-%
\mathbf{r}^{\prime }}^{\mu\mu^{\prime }} \langle\hat{S}_{\mathbf{r}%
}^{\mu}\rangle \langle\hat{S}_{\mathbf{r}^{\prime }}^{\mu^{\prime }}\rangle/2
$. The state $|\{\phi_{\mathbf{r}}^{c},\theta_{\mathbf{r}}^{c}\}\rangle=\hat{%
U}_{R}|S\rangle$ is related to the highest-weight eigenstate $|S\rangle$ of $%
\hat{S}_{\mathbf{r}}^z$ via a unitary rotation 
\begin{equation}
    \hat{U}_{R}(\phi_{\mathbf{r}}^{c},\theta_{\mathbf{r}}^{c})=e^{-i\sum_{\mathbf{r}}\phi_{\mathbf{r}}^{c}\hat{S}^{z}_{\mathbf{r}}}e^{-i\sum_{\mathbf{r}}\theta_{\mathbf{r}}^{c}\hat{S}^{y}_{\mathbf{r}}}.
\end{equation}
For convenience, we transform
the Hamiltonian into this rotated frame as $\hat{U}_{R}^{\dagger} H \hat{U}%
_{R}=H(\phi^{c}_{j,\alpha},\theta^{c}_{j,\alpha})$. In this frame, all spins in the classical state are aligned along the $z$-axis,
allowing us to employ the Holstein-Primakoff (HP) representation: $\hat{S}%
_{j,\alpha}^{z}=S-\hat{b}_{j,\alpha}^{\dagger}\hat{b}_{j,\alpha}$ and $\hat{S%
}_{j,\alpha}^{+}=(\hat{S}_{j,\alpha}^{-})^{\dagger}=\hat{S}_{j,\alpha}^{x}+i%
\hat{S}_{j,\alpha}^{y}=\sqrt{2S-\hat{b}_{j,\alpha}^{\dagger}\hat{b}%
_{j,\alpha}}\hat{b}_{j,\alpha}$, where $%
\hat{b}_{j,\alpha}^{\dagger}(\hat{b}_{j,\alpha})$ is the creation (annihilation) operator for magnon excitations. Following the spirit of the large-$S$ expansion, we expand the Hamiltonian as 
\begin{equation}
H(\phi^{c}_{j,\alpha},\theta^{c}_{j,\alpha}) =\sum_{n=0}^{\infty}S^{2-n/2}%
\mathcal{H}_{n}.  \label{eq:HS}
\end{equation}
Here, we emphasize that $\mathcal{H}_{n}$ and $\hat{b}_{j,\alpha}$ depend on
the choice of the configuration $(\phi^{c}_{j,\alpha},%
\theta^{c}_{j,\alpha})$. For a classical state $|\{\phi_{\mathbf{r}}^{c},\theta_{\mathbf{r}%
}^{c}\}\rangle$ with translational symmetry, $(\phi^{c}_{j,\alpha},%
\theta^{c}_{j,\alpha})=(\phi^{c}_{\alpha},\theta^{c}_{\alpha})$ are
$j$-independent. Due to the translational invariance of $%
H(\phi^{c}_{\alpha},\theta^{c}_{\alpha})$, we can express $\mathcal{H}_{n}$
by $\hat{b}_{\mathbf{k},\alpha}\equiv\sum_{j}e^{-i\mathbf{%
k}\cdot\mathbf{r}_{j,\alpha}}\hat{b}_{j,\alpha}/\sqrt{N_{c}}$ in the momentum space, as shown in App.~\ref{AppHam}.
Since the classical energy $E_c=S^{2}\mathcal{H}_{0}$ is achieved at the minimum in the
variational manifold, the linear term vanishes, i.e., $\mathcal{H}_{1}=0$.
The most relevant contributions arise from the terms $\mathcal{H}_{2\sim4}$.

(i) At first order in $S$, 
\begin{equation}
    \mathcal{H}_{2}=\mathcal{H}_{0}+\frac{1}{2}\sum_{%
\mathbf{k}}B_{\mathbf{k}}^{\dagger }M_{\mathbf{k}}B_{\mathbf{k}}
\end{equation}
can be
expressed in the Nambu basis $B_{\mathbf{k}}=(\hat{b}_{\mathbf{k},\alpha },%
\hat{b}_{-\mathbf{k},\alpha }^{\dagger })^{T}$, where the dispersion matrix $%
M_{\mathbf{k}}$ is obtained using linear spin wave theory, as shown in App.~\ref{AppHam} . If $M_{\mathbf{k}}$ is positive-definite, it can be diagonalized as $U_{\mathbf{k}}^{\dagger }M_{\mathbf{k}}U_{\mathbf{k}%
}=I_{2}\otimes D_{\mathbf{k}}$ via a Bogoliubov transformation $U_{%
\mathbf{k}}$ satisfying $U_{\mathbf{k}}^{\dagger }(\sigma ^{z}\otimes
I_{N_{s}})U_{\mathbf{k}}=\sigma ^{z}\otimes I_{N_{s}}$, where $I_{N_{s}}$ ($I_{2}$) is the identity matrix in the sublattice
(Nambu) basis, $\sigma^z$ is the Pauli matrix in Numbu space, and the element $d_{\mathbf{k},m}$ of the diagonal matrix $D_{%
\mathbf{k}}$ denotes the magnon dispersion in the band $m$. If $M_{\mathbf{k}_{0}}$ is semi-positive definite at some momentum $\mathbf{k}_{0}$%
, positive magnon frequencies can be obtained analogously to the positive-definite case.
Thus, we focus on $M^{(0)}_{%
\mathbf{k}_{0}}=U_{\mathbf{k}_{0}}^{{(0)}\dagger }M_{\mathbf{k}_{0}}U_{%
\mathbf{k}_{0}}^{(0)}$ projected in the zero-mode subspace spanned by $%
U^{(0)}_{\mathbf{k}_{0}}=(u_{\mathbf{k}_{0}},v_{\mathbf{k}_{0}})$, where $u_{%
\mathbf{k}_{0}}$ and $v_{\mathbf{k}_{0}}=\sigma ^{x}u_{-\mathbf{k}%
_{0}}^{\ast }$ describe the modes for particle and hole excitations. For $m_0$ type-I zero modes, $M^{(0)}_{\mathbf{k}_{0}}=\left( 
\begin{array}{cc}
1 & 1 \\ 
1 & 1%
\end{array}%
\right) \otimes D_{\mathbf{k_{0}}}^{(0)}/2$ is determined by a diagonal
matrix $D_{\mathbf{k_{0}}}^{(0)}$ with elements $\widetilde{d}_{\mathbf{k}%
_{0},m=1,...,m_{0}}$. This structure reveals that the zero modes in $\mathcal{H}_{2}$ have vanishing canonical momentum operators, implying that $M^{(0)}_{%
\mathbf{k}_{0}}$ is not diagonalizable via a Bogoliubov transformation. For
type-II zero modes, $M^{(0)}_{\mathbf{k}_{0}}=0$.

(ii) The interactions involving three and four magnons are described by the cubic and quartic terms 
\begin{eqnarray}
\mathcal{H}_{3} &=&\sum_{\mathbf{p},\mathbf{q},\{l_{i}\}}\frac{%
V_{l_{1}l_{2}l_{3}}^{(3)}(\mathbf{p},\mathbf{q})}{3!\sqrt{N_{c}}}B_{\mathbf{p%
},l_{1}}^{\dagger }B_{\mathbf{q},l_{2}}B_{\mathbf{p}-\mathbf{q},l_{3}}, 
\\
\mathcal{H}_{4} &=&\sum_{\mathbf{k},\mathbf{p},\mathbf{q},\{l_{i}\}}\frac{%
V_{l_{1},l_{2},l_{3}l_{4}}^{(4)}(\mathbf{k},\mathbf{p},\mathbf{q})}{4!N_{c}}%
B_{\mathbf{k},l_{1}}^{\dagger }B_{\mathbf{p},l_{2}}^{\dagger }B_{\mathbf{k}+%
\mathbf{q},l_{3}}B_{\mathbf{p}-\mathbf{q},l_{4}}, \notag 
\end{eqnarray}
where $l_i$ labels the component of $B_{\mathbf{k}}$, and the explicit expressions for the interaction tensors $V^{(3,4)}$ are provided in App.~\ref{AppHam}.

In frustrated spin systems \cite%
{Chubukov1992o,Sachdev1992o,Murthy1997o,Yildirim1999o,Champion2003o,Savary2012o,Zhitomirsky2012o,Consoli2020o,Rau2018,Khatua2023}, certain symmetries are preserved at the classical level but broken by quantum corrections. More precisely, two states $\hat{U}_{R,i=0,1}|S\rangle$ generated by the rotations $\hat{U}%
_{R,i=0,1}=\hat{U}_{R}(\phi^{c}_{\mathbf{r},i}, \theta^{c}_{\mathbf{r},i})$ yield the same classical energy $E_c$. However, the Hamiltonian~\eqref{eq:H} is not invariant under the rotation $\delta U_R=\hat{U}_{R,1}\hat{U}_{R,0}^{\dagger}$, i.e., $\delta U_R
^{\dagger}H\delta U_R \neq H$. Since $\delta U_R$ is not a true symmetry of the system, the accidental degeneracy present at the classical level is lifted when quantum fluctuations are included. For instance, to first order in $S$, the ground state energy $E_g=E_c+S\mathcal{H}_0+E_{\mathrm{zp}}$ includes
a contribution from the zero-point energy $E_{\mathrm{zp}}=S\sum_{\mathbf{k}}%
\text{Tr}D_{\mathbf{k}}/2$ of $\mathcal{H}_{2}$, where zero-modes are excluded. In most cases, minimizing $E_g$ selects the
ground state configuration $(\phi^{0}_{\alpha},\theta^{0}_{\alpha})$ from
the classically degenerate configurations $(\phi^{c}_{\alpha},\theta^{c}_{\alpha})$.

In some special cases, classical degeneracy cannot be lifted solely by quantum
fluctuations. For instance, although the ferromagnetic Heisenberg models $H_{\rm{DM}}$~\cite{hickey2024} with
Dzyaloshinskii–Moriya (DM) interactions do not preserve spin $SU(2)$ symmetry, all ferromagnetic states breaking $SU(2)$ symmetry not only minimize $E_c$ but also remain exact degenerate ground states of $H_{\rm{DM}}$. This anomaly arises from the vanishing quantum corrections in $\mathcal{H}_{n>0}$ for ferromagnetic states. However, quantum corrections do affect excited states due to the explicit symmetry 
breaking in $H_{\rm{DM}}$. Therefore, the accidental degeneracy is lifted by thermal fluctuations at finite temperatures. The equilibrium state at temperature $T$ is $\rho_G=\exp{(-H/T)}/Z$ with $Z=%
\text{Tr}\exp{(-H/T)}$ being the partition function.
To first order in $S$ in Eq.~(\ref{eq:HS}), the free energy $F=-T\ln Z$ is 
\begin{equation}
F=E_g+T\sum_{\mathbf{k},m}\ln(1-e^{-\frac{S}{T}d_{\mathbf{k},m}}),
\label{eq:F}
\end{equation}
where $d_{\mathbf{k},m}\neq0$. Minimizing $F$ including both quantum and thermal
fluctuations, we obtain the optimal equilibrium configuration $(\phi^{0}_{\alpha},\theta^{0}_{\alpha})$ that lifts the accidental degeneracy~\cite{fn1,Bernier2008,Chernyshev2014,Danu2016,Schick2020}. 

\section{Pseudo-Goldstone gap} 

Similar to the emergence of Goldstone zero modes~\cite%
{Goldstone1962}, the spontaneous breaking of approximate symmetries, i.e., accidental degeneracy at the classical level, gives rise to zero modes of $M_{\mathbf{k}_0}$, known as PG modes. However, quantum corrections and
thermal fluctuations lift this accidental degeneracy, generating a mass gap for PG modes. Here, we focus on the PG mode at zero momentum $\mathbf{k}_0=0$, with the general cases
analyzed in App.~\ref{Appproof} following the same procedure. For convenience, we denote the elements of \( \widetilde{d}_{\mathbf{0},m} \) corresponding to type-I PG modes as \( (\widetilde{d}_{\mathbf{0}})_{\mathrm{ps}}\) and define the columns of \( u_{\mathbf{0}} \) and \( v_{\mathbf{0}} \) associated with PG modes as \( (u_{\mathbf{0}})_{\mathrm{ps}} \) and \( (v_{\mathbf{0}})_{\mathrm{ps}} \), respectively.

To compute the gap of the PG mode, we consider the equilibrium state $\hat{U}%
_{R,0}|S\rangle$ with $\hat{U}_{R,0}=\hat{U}_{R}(\phi
_{\alpha}^{0},\theta_{\alpha }^{0})$, which minimizes the free energy~(\ref{eq:F}) incorporating
quantum and thermal fluctuations. In the
transformed frame by $\hat{U}_{R,0}$, the Hamiltonian $H(\phi _{\alpha
}^{0},\theta _{\alpha }^{0})=\sum_{n=0}^{\infty }S^{2-n/2}\mathcal{H}_{n}$, and the rotation
\begin{equation}
\hat{R}(\phi,\theta)=\exp [-i\sum_{j,\alpha }\omega _{\alpha }\mathbf{n}_{\alpha }(\phi
,\theta )\cdot \hat{\mathbf{S}}_{j,\alpha }],  \label{eq:Rex}
\end{equation}%
connecting states with identical classical
energy $E_c$, is generated by the superposition of $\hat{S}^{x,y}_{j,\alpha}$.
Without loss of generality, we can parameterize 
\begin{equation}
\omega_{1}\mathbf{n}_{1}\cdot\hat{\mathbf{S}}_{j,1}=\sqrt{\frac{2}{N_c}}(\phi|\chi_{\phi,1}| \widetilde{S}_{j,1}^{x}+\theta|\chi_{\theta,1}|\widetilde{S}_{j,1}^{y})
\end{equation}
for the sublattice $\alpha =1$ by $\phi$ and $\theta$, where $\widetilde{S}_{j,1}^{x(y)}=\{{\text{Re}[\chi_{\phi(\theta),1}]}\hat{S}_{j,1}^{x}+{\text{Im}[\chi_{\phi(\theta),1}]}\hat{S}_{j,1}^{y}\}/|\chi_{\phi(\theta),1}|$. $\chi _{\phi,1}$ and $\chi _{\theta ,1}$ represents the
first elements of 
\begin{align}
     \chi _{\phi }&=\frac{1}{\sqrt{2}}[(u_{\mathbf{0}})_{\mathrm{ps}}-(v_{\mathbf{0%
}})_{\mathrm{ps}}] \nonumber\\
\chi _{\theta }&=\frac{i}{\sqrt{2}}[(u_{\mathbf{0}})_{\mathrm{ps%
}}+(v_{\mathbf{0}})_{\mathrm{ps}}].
\end{align}
We remark that in general $%
\omega _{\alpha }\mathbf{n}_{\alpha }(\phi ,\theta )$ is a nonlinear
function of $\phi $ and $\theta $ for $\alpha \neq 1$.
For the type-I PG mode, only rotations along the direction $\text{Re}(\chi_{\phi,1})\hat{\mathbf{x}} + \text{Im}(\chi_{\phi,1})\hat{\mathbf{y}}$ in sublattice $\alpha=1$ preserve $E_c$, leading to $\theta=0$. For the type-II PG mode, $E_c$ remains unchanged under $\hat{R}(\phi,\theta)$ with arbitrary $\phi$ and $\theta$.

\textit{Theorem: At finite temperature $T$, the PG gap is 
\begin{equation}
\Delta=\frac{1}{S}\sqrt{\left(\frac{\partial^{2}F}{\partial\theta^{2}}%
\right)_{0}\left(\frac{\partial^{2}F}{\partial\phi^{2}}\right)_{0}-\left(%
\frac{\partial^{2}F}{\partial\theta\partial\phi}\right)_{0}^{2}},
\label{eq:cur}
\end{equation}
determined by the curvature of the free energy $F(\phi,\theta)$.} Here, the subindex `0'
indicates derivatives evaluated at $\theta=\phi=0$, and $F(\phi ,\theta )=-T\ln Z(\phi,\theta)$ is obtained from the partition function $Z(\phi,\theta)=\text{Tr}\exp [-H_{R}(\phi ,\theta )/T]$. The quadratic Hamiltonian $H_R(\phi ,\theta)$ is obtained by truncating%
\begin{equation}
\hat{R}^{\dagger }(\phi ,\theta )H(\phi _{\alpha }^{0},\theta _{\alpha }^{0})%
\hat{R}(\phi ,\theta )=\sum_{n=0}^{\infty }S^{2-n/2}\mathcal{H}_{n}(\phi
,\theta)
\end{equation}%
at $n=2$, where $\mathcal{H}_{n}(0,0)=\mathcal{H}_{n}$. We note that for the type-I PG
mode, although only $\hat{R}(\phi ,0)$ preserves the classical
energy, a finite $\theta $ in $\hat{R}(\phi ,\theta )$ is still required to
compute the gap. The free-energy landscape is schematically illustrated in Fig.~\ref{Fig:Model}(c).
The detailed proof of Eq.~\eqref{eq:cur} is provided in App.~\ref{Appproof}, while here we outline the three key steps. 

\begin{table*}[t]
\caption{Comparison between the pseudo-Goldstone gaps obtained from the curvature formula developed in this work and those reported in Refs.~\cite{Rau2018,Khatua2023}. For each model, the table provides the lattice geometry, exchange interaction regime, type of pseudo-Goldstone mode, and the corresponding PG gap values in the low- and high-temperature limits. Additional computational and analytical details are presented in Appendices~\ref{AppXXZ} and~\ref{AppObD} ($S=1$).}
\label{tab:PG gap}\centering
\begin{ruledtabular}
\begin{tabular}{cccccccccc}
\textrm{Model} & \textrm{Parameters} & \textrm{Type}  & \textrm{Linear} & \textrm{$\Delta$(low-T)} & \textrm{$\Delta$}(high-T) & \textrm{$\Delta$(T=0\cite{Rau2018})} & \textrm{$\Delta$(high-T\cite{Khatua2023})}\\
\colrule
Heisenberg-compass & $\kappa=-0.5$ & I & yes & $0.17+0.074T^3$  & $0.17\sqrt{T}$ & $0.17$ & $\sim\sqrt{T}$ \\
(Square, Ferromagnet) & $\kappa=-5$ & I & yes & $3.75+0.044T^3$ & $2.46\sqrt{T}$ & $3.75$ & $2.46\sqrt{T}$ \\
\colrule
Pyrochlore Heisenberg ferromagnet\\
(Dzyaloshinskii-Moriya interaction) & $\kappa=1$ & II & yes & $0.016T^3-0.0039T^2+0.0003T$   & $0.024T$ & 0 & $\sim T$ \\
\colrule
Triangular XXZ & $\kappa=1.2$ & I & no & $0.52-0.58T$ &  & $0.49$  & $\sim\sqrt{T}$ \\

\end{tabular}
\end{ruledtabular}
\end{table*}

(i) We first expand the generator of the rotation (\ref{eq:Rex}) in powers of $\phi$, $\theta$ and $S$:
\begin{equation}
\hat{R}(\phi ,\theta )=\exp \left( -i\sum_{\zeta ,\eta ,\nu =0}^{\infty
}O_{\zeta ,\eta }^{(\nu )}\phi ^{\zeta }\theta ^{\eta }S^{1/2-\nu }\right),
\label{eq:R}
\end{equation}%
where the Hermitian operators $O_{\zeta,\eta }^{(\nu)}$ are determined order by order through imposing the invariance of $E_c$ under $\hat{R}(\phi,\theta )$: $\mathcal{H}_{0}(\phi ,0)=\mathcal{H}_{0}$ ($\mathcal{H}_{0}(\phi ,\theta)=\mathcal{H}_{0}$) for type-I (II) PG modes. Since all $\phi$ minimize the
classical energy, the saddle point condition, i.e., the vanishing
single-magnon term $\mathcal{H}_{1}(\phi,0)=0$ ($\mathcal{H}_{1}(\phi,\theta)=0$) for type-I (II) PG modes, determines $O_{\zeta,\eta }^{(\nu)}$ via the
Baker-Campbell-Hausdorff (BCH) formula. Below, we list several key terms. The canonical position and momentum operators, i.e., $O_{1,0}^{(0)}=B_{\mathbf{0}}^{\dagger}\sigma^{z}\chi _{\phi}$ and $O_{0,1}^{(0)}=B_{\mathbf{0}}^{\dagger
}\sigma^{z}\chi_{\theta }$, satisfy the commutation relation $[O_{1,0}^{(0)},O_{0,1}^{(0)}]=i$. The bilinear operator $O_{2,0}^{(0)}=\chi_{\phi}^{\dagger}\Omega\chi_{\phi}/2$ generates the nonlinear dependence on $\phi$ in $%
\omega _{\alpha }\mathbf{n}_{\alpha }$, where 
\begin{equation}
    \Omega _{l_{1}l_{2}}=\frac{i}{\sqrt{N_{c}}}\sum_{l_{3}}V_{l_{1}l_{2}l_{3}}^{(3)}(\mathbf{0},%
\mathbf{0})\left(M_{\mathbf{0}}^{-1}\sigma ^{z}B_{\mathbf{0}}\right)
_{l_{3}}
\end{equation}
with $M_{\mathbf{0}}^{-1}$ being a pseudo-inverse. The bilinear operators $O_{0,2}^{(0)}=\chi_{\theta }^{\dagger }\Omega \chi_{\theta }/2$
and $O_{1,1}^{(0)}=\chi _{\theta }^{\dagger}\Omega \chi_{\phi }$ are relevant only for type-II PG modes.

(ii) In the second step, we derive the Hamiltonian $H_R(\phi,\theta)=E_c+S\mathcal{H}_{2}(\phi,\theta)$ using the BCH formula. Here, 
\begin{equation}
    \mathcal{H}_{2}(\phi,\theta)=\sum_{\zeta+\eta =0}^{2}h_{\zeta,\eta }\phi
^{\zeta}\theta ^{\eta}
\end{equation}
is expanded up to second order in $\phi$ and $\theta$. Two terms $h_{0,0}=\mathcal{H}_{2}$ and $h_{1,0}=-i[\mathcal{H}_{3},O_{1,0}^{(0)}]$ are useful in the subsequent discussion. The partition function $Z(\phi,\theta)$ can be
expressed in the standard path integral form shown in App.~\ref{Appproof}, allowing an efficient computation of the second derivative $\mathcal{F}_{\phi
}\equiv (\partial ^{2}F/\partial \phi ^{2})_{0}=\sum_{L=1}^{3}\mathcal{F}%
_{\phi ,L}$. Here,
\begin{align}
\mathcal{F}_{\phi ,1}&=-S\langle \lbrack \lbrack \mathcal{H}%
_{4},O_{1,0}^{(0)}],O_{1,0}^{(0)}]\rangle _{0}\nonumber\\
\mathcal{F}_{\phi
,2}&=-2iS\langle \lbrack \mathcal{H}_{3},O_{2,0}^{(0)}]\rangle _{0}\\
\frac{\mathcal{F}_{\phi,3}}{TS^{2}}&=-\int_{0}^{\frac{1}{T}}d\tau d\tau ^{\prime
}\langle \mathcal{T}_{\tau }h_{1,0}(\tau )h_{1,0}(\tau ^{\prime })\rangle
_{0}+\frac{\langle h_{1,0}\rangle _{0}^{2}}{T^{2}}\nonumber
\end{align}
are (imaginary-time ordered $\mathcal{T}_{\tau}$) expectation values in the thermal state $\rho _{0}=e^{-S%
\mathcal{H}_{2}/T}/Z_{0}$ with $Z_{0}=\mathrm{Tr}e^{-S\mathcal{H}_{2}/T}$, and the operator evolution $h_{1,0}(\tau )=e^{S%
\mathcal{H}_{2}\tau }h_{1,0}e^{-S\mathcal{H}_{2}\tau}$. A straightforward calculation yields $\mathcal{F}_{\phi}=S\chi_{\phi }^{\dagger}\bar{\Sigma}_{\mathbf{0}%
}\chi _{\phi }$, where $\bar{\Sigma}_{\mathbf{0}}=\sum_{L=1}^{3}\bar{\Sigma}%
_{\mathbf{0}}^{(L)}$, with explicit form of $\bar{\Sigma}_{\mathbf{0}}^{(L)}$ provided in App.~\ref{Appproof}. For type-I PG modes, at leading order in the large-$S$ expansion, only the generator $O_{0,1}^{(0)}$ contributes, resulting in 
\begin{align}
    \left( \frac{\partial ^{2}F}{\partial \theta ^{2}}\right)_{0}=S^{2}(%
\widetilde{d}_{\mathbf{0}})_{\mathrm{ps}},
( \frac{\partial ^{2}F}{\partial \theta \partial \phi} )_{0}=0.
\end{align}
For type-II PG modes, 
\begin{equation}
    (\frac{\partial
^{2}F}{\partial \theta ^{2}})_{0}=S\chi _{\theta }^{\dagger }\bar{\Sigma}_{%
\mathbf{0}}\chi _{\theta }, (\frac{\partial ^{2}F}{\partial \theta \partial
\phi} )_{0}=S\chi _{\theta }^{\dagger }\bar{\Sigma}_{\mathbf{0}}\chi _{\phi }
\end{equation}
are governed by the same matrix $\bar{\Sigma}_{\mathbf{0}}$.  It follows from Eq.~\eqref{eq:cur} that%
\begin{equation}
\Delta =\left\{ 
\begin{array}{c}
S^{1/2}\sqrt{(\widetilde{d}_{\mathbf{0}})_{\mathrm{ps}}\chi _{\phi
}^{\dagger }\bar{\Sigma}_{\mathbf{0}}\chi _{\phi }} \text{ \ \ \ \ \ \ \ \ \ \ \ \ \
\ \ \ \ \ \ \ type-I} \\ 
\sqrt{\chi _{\theta }^{\dagger }\bar{\Sigma}_{\mathbf{0}}\chi _{\theta }\chi
_{\phi }^{\dagger }\bar{\Sigma}_{\mathbf{0}}\chi _{\phi }-(\chi _{\theta
}^{\dagger }\bar{\Sigma}_{\mathbf{0}}\chi _{\phi })^{2}} \text{ \ \ type-II}%
\end{array}%
\right. .  \label{eq:gap}
\end{equation}%

(iii) Finally, we apply a field-theoretical approach to derive Eq.~\eqref{eq:gap}, thereby proving the curvature formula Eq.\eqref{eq:cur}. We expand $H(\phi_{\alpha }^{0},\theta
_{\alpha }^{0})\sim \sum_{n=0}^{4}S^{2-n/2}\mathcal{H}_{n}$ around the equilibrium configuration $(\phi_{\alpha }^{0},\theta
_{\alpha }^{0})$. By computing the Fourier
transform $G_{\mathbf{k}}(i\omega _{n})=\int_{0}^{1/T}d\tau G_{\mathbf{k}%
}(\tau )e^{i\omega _{n}\tau }$ of the Green function
\begin{equation}
  G_{\mathbf{k}}(\tau
)=-\left\langle \mathcal{T}_{\tau }B_{\mathbf{k}}(\tau )B_{\mathbf{k}%
}^{\dagger }(0)\right\rangle  
\end{equation}
in the thermal
state $\rho =e^{-H(\phi _{\alpha }^{0},\theta _{\alpha }^{0})/T}/Z$, we can determine the magnon spectrum from the
poles of $G_{\mathbf{k}}(i\omega_{n})$. Here, $\omega _{n}=2\pi nT$ ($n\in \mathbf{Z}$) is the Matsubara frequency, $Z=$Tr$\exp [-H(\phi _{\alpha }^{0},\theta _{\alpha }^{0})/T]$ is the partition function, and $B_{\mathbf{k}}(\tau )=e^{H(\phi _{\alpha }^{0},\theta _{\alpha }^{0})\tau }B_{\mathbf{k}}e^{-H(\phi _{\alpha }^{0},\theta _{\alpha }^{0})\tau }$. Using perturbation theory, we obtain
\begin{equation}
    G_{\mathbf{k}}^{-1}(i\omega
_{n})=i\omega_{n}\sigma ^{z}\otimes I_{N_{s}}-SM_{\mathbf{k}}-\Sigma _{%
\mathbf{k}}(i\omega _{n}).
\end{equation}
Here, the self-energy $\Sigma _{%
\mathbf{k}}(i\omega _{n})=\sum_{L=1}^{3}\Sigma _{\mathbf{k}}^{(L)}(i\omega
_{n})$ consists of three terms corresponding to Feynman diagrams shown in Fig.~%
\ref{Fig:Model}(d), which are computed in App.~\ref{Appproof}. The frequency-independent Hartree-Fock term $\Sigma _{%
\mathbf{k}}^{(1)}$ originates from the quartic-magnon interaction $%
\mathcal{H}_{4}$. The frequency-independent tadpole term $\Sigma _{%
\mathbf{k}}^{(2)}$ may contribute significantly when the three-body interaction $\mathcal{H}_{3}\neq 0$. The frequency-dependent term $\Sigma _{\mathbf{k}}^{(3)}(i\omega _{n})$ is generated by the second-order perturbation of $\mathcal{H}_{3}$. At leading order $O(S^{0})$, the self-energy $\Sigma _{\mathbf{k}}(i\omega_{n})$ can be approximated by $\Sigma _{\mathbf{k}}(0)=\sum_{L=1}^{3}\Sigma
_{\mathbf{k}}^{(L)}(0)$. For PG\ modes with $\mathbf{k}=0$, we can identify $\Sigma^{(L)}_{\mathbf{0}}(0)=\bar{%
\Sigma}^{(L)}_{\mathbf{0}}$ by direct comparison. Eventually, by solving det$[G_{\mathbf{0}%
}^{-1}(\omega )]=0$, we reproduce the PG gap, as given by Eq.~\eqref{eq:gap}. Here, the details are left in App.~\ref{Appproof}.

It is important to compare our curvature formula with previous results~\cite{Rau2018,Khatua2023}. At $T=0$, our formula appears to reduce to that in Ref.~\cite{Rau2018}. However, the nonlinear dependence on $\phi$ and $\theta$ in Eq.~(\ref{eq:Rex}), arising from $O_{2,0}^{(0)}$, $O_{0,2}^{(0)}$ and $O_{1,1}^{(0)}$, is entirely neglected in Ref.~\cite{Rau2018} [see Eq.~(7) and $U(\phi,\theta)$ therein]. Consequently, the tadpole term $\Sigma_{\mathbf{0}}^{(2)}$ is omitted~\cite{fn2}. Fortunately, for all models in Ref.~\cite{Rau2018} [see Table I therein], either $\mathcal{H}_{3}=0$ or $[\mathcal{H}_{3},O_{2,0}^{(0)}]=0$ (as in the Heisenberg-Kitaev-$\Gamma$ models and the material Er$_2$Ti$_2$O$_7$), ensuring that $\Sigma _{\mathbf{0}}^{(2)}=0$. Therefore, the
proof in Ref. \cite{Rau2018} remains valid for these models. As a paradigmatic example, we benchmark Eq.~\eqref{eq:cur} against the Heisenberg-compass model in
the first row of Table I, where our result is consistent with that of Ref.~\cite{Rau2018}. The second row of Table I presents results for the Pyrochlore Heisenberg
ferromagnet with DM interactions, where $\mathcal{H}_{3}\neq 0$ even for the equilibrium state with collinear order. However, $O_{2,0}^{(0)}=0$ still ensures $\Sigma _{\mathbf{0}}^{(2)}=0$. More generally, since $\Sigma _{\mathbf{0}}^{(2)}$ is determined by $[\mathcal{H}%
_{3},O_{2,0}^{(0)}]$, the
approach in Ref.~\cite{Rau2018} remains justified, provide that $[\mathcal{H}_{3},O_{2,0}^{(0)}]=0$ (which is the case for most known materials with collinear order). At finite $T$, the PG gap $\Delta $ of the Heisenberg-compass model was analyzed
in the high-temperature limit~\cite{Khatua2023}, i.e., $T\gg d_{\mathbf{k}%
,m} $, yielding results consistent with Eq.~\eqref{eq:cur}, as shown
in the second row of Table I. However, the scaling relation $\Delta \sim T^{1/2}$ ($T$) for type-I (II) PG modes, as identified in Ref.~\cite{Khatua2023}, is specific to this model and may not hold in general, particularly for systems with noncollinear order at low temperatures.

\section{XXZ triangular-lattice model}
We consider the antiferromagnetic
XXZ model%
\begin{equation}
H_{XXZ}=\sum_{\langle \mathbf{r},\mathbf{r}^{\prime }\rangle }(%
\hat{S}_{\mathbf{r}}^{x}\hat{S}_{\mathbf{r}^{\prime }}^{x}+\hat{S}_{\mathbf{r%
}}^{y}\hat{S}_{\mathbf{r}^{\prime }}^{y}+\kappa \hat{S}_{\mathbf{r}}^{z}%
\hat{S}_{\mathbf{r}^{\prime }}^{z})
\end{equation}%
on the triangular lattice with $\kappa >1$, where $\langle \mathbf{r},%
\mathbf{r}^{\prime }\rangle $ denotes nearest-neighbor sites.
The Hamiltonian $H_{XXZ}$ is invariant only under rotations along the $z$-axis, i.e., $[S^{z},H_{XXZ}]=0$, where $S^{z}=\sum_{\mathbf{r}}\hat{S}_{%
\mathbf{r}}^{z}$.

By minimizing the free energy Eq.~\eqref{eq:F},
we obtain $12$ degenerate equilibrium-state configurations in the $x$-$z$ plane with
the classical energy
\begin{equation}
    E_c=-NS^2\frac{\kappa ^{2}+\kappa +1}{1+\kappa }.
\end{equation}
Without loss of generality, we break the $12$-fold symmetry and choose $\theta _{A}^{0}=0$, $\theta
_{B}^{0}=\arccos(-\kappa /(1+\kappa ))$, $\theta _{C}^{0}=2\pi -\theta
_{B}^{0}$, and $\phi_{\alpha }^{0}=0$, representing a coplanar order as shown in Fig.\ref{Fig:Model}(b). The remaining $11$
configurations can be obtained by permuting the three sublattices and applying reflection symmetry with respect to the $x$-$y$ plane. The configuration $(\phi _{\alpha }^{0},\theta _{\alpha }^{0})$ spontaneously
breaks the rotational symmetry along the $z$-axis. Since the rotation $e^{iS^{z}\varphi }$ conserves $H_{XXZ}$, it generates a true Goldstone mode when acting on the equilibrium state. Moreover, the rotation $\hat{R}(\phi
,0)=\exp [-i\sum_{j,\alpha }\omega _{\alpha }(\phi )\hat{S}_{j,\alpha }^{y}]$
along the $y$-axis preserves the classical
energy $E_c$, indicating the degeneracy at the classical level. However, since it is not a symmetry of $H_{XXZ}$, a PG mode emerges, associated with $\hat{R}(\phi,0)$. Importantly, although%
\begin{equation}
\omega _{A}(\phi )=\sqrt{\frac{2(1+\kappa )(1+2\kappa )}{N_c [2\kappa
^{2}(1+\kappa )-1]}}\phi
\end{equation}%
is proportional to $\phi $, the rotation angles $\omega _{B,C}(\phi )$ are
nonlinear functions of $\phi$~\cite{Murthy1997o}, as shown in App.~\ref{AppXXZ}. To second order in $\phi$, 
\begin{equation}
    \omega _{B,C}(\phi )=\kappa \omega _{A}\pm \Lambda _{\kappa
}\omega _{A}^{2},
\end{equation}
where $\Lambda _{\kappa }\equiv \kappa (1-\kappa ^{2})/(2%
\sqrt{1+2\kappa })$. Further analysis around the classical configuration $(\phi _{\alpha }^{0},\theta _{\alpha }^{0})$ shows that the projection of $M_{\mathbf{0}}$ onto the zero-mode subspace spanned by
vectors $U^{(0)}_{\mathbf{0}}=((u_{\mathbf{0}})_{\mathrm{ps}},(v_{\mathbf{0}})_{%
\mathrm{ps}})$ is 
\begin{equation}
U_{\mathbf{0}}^{(0)\dagger }M_{%
\mathbf{0}}U^{(0)}_{\mathbf{0}}=\frac{1}{2}(\widetilde{d}_{\mathbf{0}})_{\mathrm{ps}}\left( 
\begin{array}{cc}
1 & 1 \\ 
1 & 1%
\end{array}%
\right)
\end{equation}
with $(\widetilde{d}_{\mathbf{0}})_{\mathrm{ps}}=3/(2+2\kappa)$, demonstrating a type-I PG mode. 

\begin{figure}[t]
\includegraphics[angle=0,width=1\linewidth]{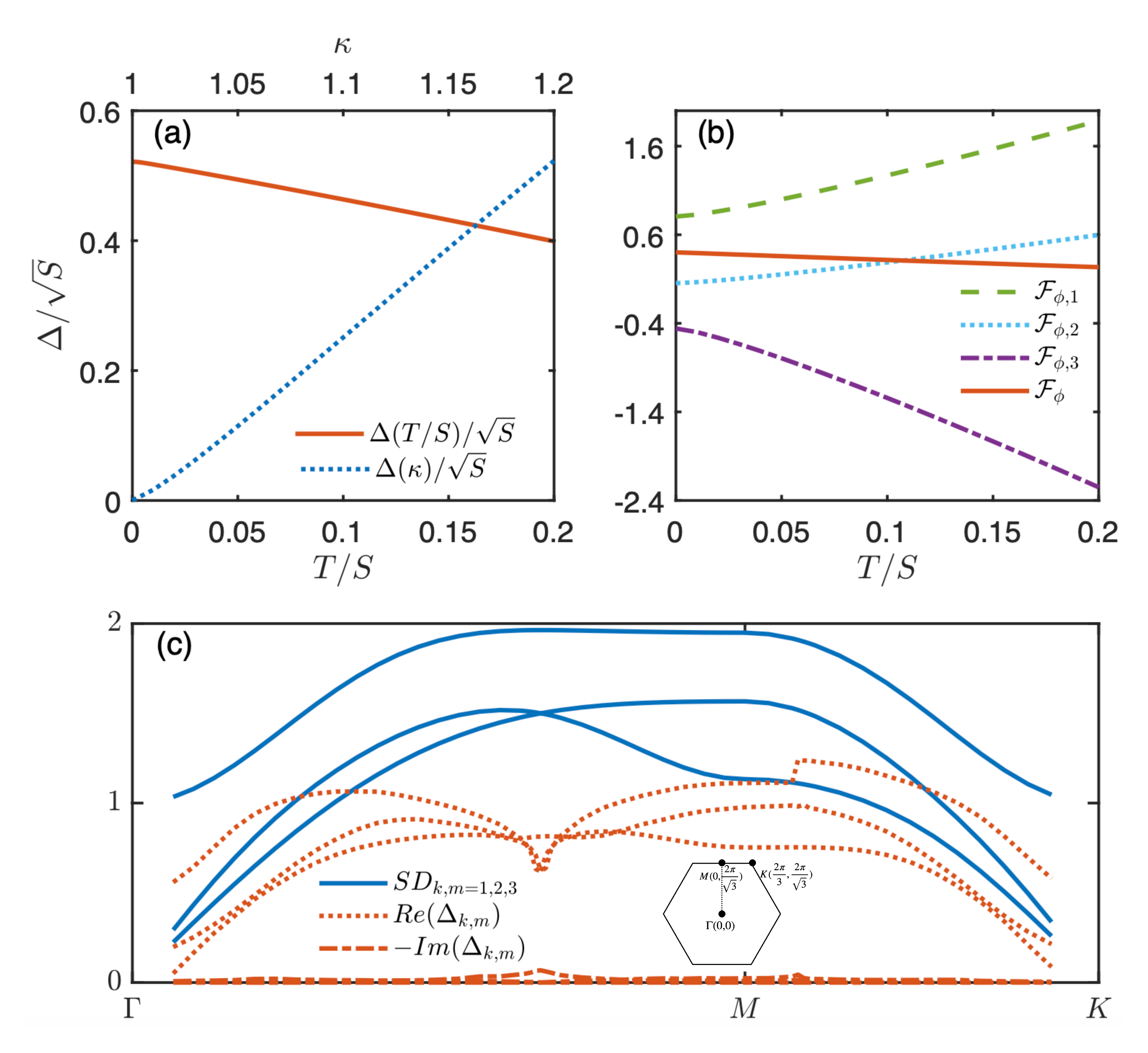}
\caption{
(a) Pseudo-Goldstone gap $\Delta/\sqrt{S}$ increases with $\kappa$ (blue dots). For $\kappa = 1.2$, the temperature dependence of $\Delta/\sqrt{S}$ obtained from the curvature formula (red line) is well fitted by $\Delta/\sqrt{S} = -0.5803T/S + 0.523$ in the low-temperature limit $T \to 0$.
(b) Temperature dependence of $\mathcal{F}_{\phi}$ and its three contributions $\mathcal{F}_{\phi, L=1,2,3}$.
(c) Spin-wave spectrum $S D_{\mathbf{k},m=1,2,3}$ and the $1/S$-corrected spectrum $\Delta_{\mathbf{k},m} = S D_{\mathbf{k},m} + u_{\mathbf{k},m}^{\dagger} \Sigma_{\mathbf{k}}(S D_{\mathbf{k},m}) u_{\mathbf{k},m}$. The renormalized dispersion exhibits a roton minimum at $\mathbf{M} = (0, 2\pi/\sqrt{3})$ for $\kappa = 1.2$, $S = 0.5$.
}
\label{Fig:ThermoQuant}
\end{figure}

For the coplanar order, $[\mathcal{H}_{3},O_{2,0}^{(0)}] \neq 0$ results in a finite $\Sigma _{\mathbf{0}}^{(2)}$. In the third row of Table I, we present $\Delta$ obtained from Eq.~\eqref{eq:cur}. If the rotation angles linearly depend on $\phi$, i.e., $O_{2,0}^{(0)}=0$, $\Delta=-0.18i\sqrt{S}$ would be imaginary at $T=0.12$. In Fig.~\ref{Fig:ThermoQuant}(a), we depict $\Delta$ as a function
of $\kappa$ at $T=0$. For $\kappa =1$, the full $SU(2)$ symmetry
of $H_{XXZ}$ is restored, thereby the PG mode becomes a true Goldstone
mode with $\Delta =0$. For $\kappa >1$, quantum corrections induce a
gapped PG mode, where $\Delta$ increases monotonically with $\kappa$.

At finite $T$, the function $\Delta(T)$ is shown in Fig.~\ref{Fig:ThermoQuant}(a). At low temperatures, we find a distinct
relation $\Delta (T)\sim \Delta _{0}-cT$ with $c>0$. To illustrate the
mechanism behind this relation, we plot $\mathcal{F}_{\phi ,L}$ in Fig.~\ref{Fig:ThermoQuant}(b), where $\mathcal{F}_{\phi,3}$
decreases significantly faster than the increase of $\mathcal{F}_{\phi,1}+\mathcal{F}_{\phi,2}$. By rewriting $F(\phi,0)=E_{g}(\phi)-T f(\phi,T)$, we find that $E_{g}(\phi)$ remains $T$-independent for the present model. However, the term $h_{\zeta,\eta}$, generated by the three-magnon interaction $\mathcal{H}_{3}$, flattens the bands, thereby increasing the entropy and leading to $\partial ^{2}f(\phi,T>0)/\partial\phi ^{2}>0$. Since $f(\phi,T=0)=0$ for all $\phi$, $\partial ^{3}f(\phi,T)/\partial \phi ^{2}\partial T |_{T=0} >0$ leads to $\partial ^{3}F/\partial \phi ^{2}\partial T|_{T=0}<0$. Consequently, it follows from Eq.~\eqref{eq:cur} that $\Delta$
decreases with $T$ at low temperatures. The temperature dependence reflects an entropy effect induced by magnon scattering across multiple bands. Intuitively, $\mathcal{H}_{3}$ describes the decay of a single magnon into two, leading to an increase in entropy. The function $\Delta (T)$ differs from the $T^{1/2}$ dependence of the type-I PG gap in collinear systems~\cite{Khatua2023}. In such systems, which feature a single magnon band, quantum and thermal fluctuations at high temperatures open the PG gap and lift the single band, thereby reducing entropy. As a
result, the curvature of the free energy, i.e., the PG gap, increases with $T$. 

The Green function approach also allows us to compute the magnon spectrum
at finite momentum. In Fig.~\ref{Fig:ThermoQuant}(c), we present the modified magnon dispersion, incorporating the $1/S$-corrections \cite{Starykh2006s}, in comparison to the results from spin-wave theory. Notably, a roton minimum emerges at the $M$ point of the Brillouin zone, which may make substantial contributions to the entropy and affect the thermal properties at low temperatures~\cite{He4,roton2006}. The discontinuities in the modified magnon dispersion, arising from topological transitions of the magnon decay surface, may be removed by accounting for the renormalized $\Sigma_{\mathbf{k}}^{(3)}$ obtained via ladder diagram resummation~\cite{Chernyshev2009}.



\section{Conclusion.}
In this work, we derived a general curvature formula for the PG gap at finite temperature, extending to noncollinear orders. Applying this framework to the XXZ model on the triangular lattice, we demonstrated that the PG gap decreases linearly with temperature, in contrast to the scaling behavior observed in the Heisenberg-compass model~\cite{Khatua2023}. This distinct thermal dependence stems from entropy effects driven by magnon scattering across multiple bands, highlighting a fundamentally different mechanism compared to systems with a single magnon band. Our findings provide a more comprehensive approach to analyzing PG modes at finite temperature in systems with both collinear and noncollinear orders. More broadly, this work paves the way for exploring novel thermal effects, quantum phases, and the rich dynamics in frustrated systems.

\begin{acknowledgments}
We thank Wei Li, Junsen Wang, and Yuan Gao for the stimulating discussion. This work was supported by the NSFC (Grants No. 12135018, No.12047503, and No. 12274331), by National Key Research and Development Program of China (Grant No. 2021YFA0718304), and by CAS Project for Young Scientists in Basic Research (Grant No. YSBR-057).
\end{acknowledgments}

\bibliographystyle{apsrev4-1}
\bibliography{PGMRef}

\begin{thebibliography}{51}
\bibitem [{fn1()}]{fn1}%
  \BibitemOpen
  \href@noop {} {}\bibinfo {note} {It may happen that the equilibrium con-figuration also changes with temperature, e.g., in the Heisenberg anti-ferromagnet on a face-centered cubic lattice.}\BibitemShut {Stop}%
\end{thebibliography}
\clearpage

\widetext

\appendix
\section{The Hamiltonian under Holstein-Primakoff transformation}\label{AppHam}
In this Appendix, we determine the classical configuration of spins in magnetic systems under the mean-field approximation. Using the Holstein-Primakoff (HP) transformation for the spin fluctuations around the classical configuration, we express the spin Hamiltonian in terms of the magnon creation and annihilation operators. 

\subsection{Classical ground-state configurations}
We model the spin system comprising $N$ sites using a translationally invariant Hamiltonian
\begin{equation}
H=\frac{1}{2}\sum_{\mathbf{r}\mathbf{r}'}\sum_{\mu\mu'}\hat{S}_{\mathbf{r}}^{\mu}J_{\mathbf{r}-\mathbf{r}'}^{\mu\mu'}\hat{S}_{\mathbf{r}'}^{\mu'},\label{eqs:H}
\end{equation}
where $\hat{S}_{\mathbf{r}}^{\mu}$ is a spin-$S$ operator along the $\mu=x,y,z$ axis 
at site $\mathbf{r}$ in the laboratory frame, and $J_{\mathbf{r}-\mathbf{r}'}^{\mu\mu'}=J_{\mathbf{r}'-\mathbf{r}}^{\mu'\mu}$
is the strength of exchange interaction. For magnetically ordered phases, at the
mean-field level, the ground state is approximated by a spin coherent state $|\{\phi_{\mathbf{r}}^{c},\theta_{%
\mathbf{r}}^{c}\}\rangle$, where $\hat{S}_{\mathbf{r}}^{\mu}$ acquires average values $\langle\hat{S}_{\mathbf{r}}^{x}\rangle=S\sin(\theta_{\mathbf{r}}^{c})\cos(\phi_{\mathbf{r}}^{c})$,
$\langle\hat{S}_{\mathbf{r}}^{y}\rangle=S\sin(\theta_{\mathbf{r}}^{c})\sin(\phi_{\mathbf{r}}^{c})$
and $\langle\hat{S}_{\mathbf{r}}^{z}\rangle=S\cos(\theta_{\mathbf{r}}^{c})$. The classical ground-state configuration $(\phi_{\mathbf{r}}^{c},\theta_{\mathbf{r}}^{c})$ is determined by minimizing the ground-state energy $E_c=\sum_{\mathbf{r}\mathbf{r}'}\langle\hat{S}_{\mathbf{r}}^{\mu}\rangle J_{\mathbf{r}-\mathbf{r}'}^{\mu\mu'}\langle\hat{S}_{\mathbf{r}'}^{\mu'}\rangle/2$, which satisfies
\begin{align}
\frac{\partial E_c}{\partial\theta_{\mathbf{r}}^{c}} & =\sum_{\mathbf{r}'}\sum_{\mu\mu'}S\left(\begin{array}{ccc}
\cos\left(\theta_{\mathbf{r}}^{c}\right)\cos\left(\phi_{\mathbf{r}}^{c}\right) & \cos\left(\theta_{\mathbf{r}}^{c}\right)\sin\left(\phi_{\mathbf{r}}^{c}\right) & -\sin\left(\theta_{\mathbf{r}}^{c}\right)\end{array}\right)_{\mu}J_{\mathbf{r}-\mathbf{r}'}^{\mu\mu'}\langle \hat S_{\mathbf{r}'}^{\mu'}\rangle=0,\label{eq:Ethe}\\
\frac{\partial E_c}{\partial\phi_{\mathbf{r}}^{c}} & =\sum_{\mathbf{r}'}\sum_{\mu\mu'}S\sin\left(\theta_{\mathbf{r}}^{c}\right)\left(-\begin{array}{ccc}
\sin\left(\phi_{\mathbf{r}}^{c}\right) & \cos\left(\phi_{\mathbf{r}}^{c}\right) & 0\end{array}\right)_{\mu}J_{\mathbf{r}-\mathbf{r}'}^{\mu\mu'}\langle \hat S_{\mathbf{r}'}^{\mu'} \rangle=0.\label{eq:Ephi}
\end{align}

The state $|\{\phi_{\mathbf{r}}^{c},\theta_{\mathbf{r}}^{c}\}\rangle=\hat{%
U}_{R}(\phi_{\mathbf{r}}^{c},\theta_{\mathbf{r}}^{c})|S\rangle$ is related to the highest-weight eigenstate $|S\rangle$ of $%
\hat{S}_{\mathbf{r}}^z$ through a unitary rotation $\hat{U}_{R}(\phi_{\mathbf{r}}^{c},\theta_{\mathbf{r}}^{c})=\exp{(-i\sum_{\mathbf{r}}\phi_{\mathbf{r}}^{c}\hat{S}^{z}_{\mathbf{r}})}\exp{(-i\sum_{\mathbf{r}}\theta_{\mathbf{r}}^{c}\hat{S}^{y}_{\mathbf{r}})}$. For convenience, we transform
the Hamiltonian into this rotated frame as 
\begin{equation}
H(\phi_{\mathbf{r}}^{c},\theta_{\mathbf{r}}^{c})=\hat{U}_{R}^{\dagger} H \hat{U}_{R}=\frac{1}{2}\sum_{\mathbf{r}\mathbf{r}'}\left(\begin{array}{ccc}
\hat{S}_{\mathbf{r}}^{x} & \hat{S}_{\mathbf{r}}^{y} & \hat{S}_{\mathbf{r}}^{z}\end{array}\right)R_{0}^{T}(\phi_{\mathbf{r}}^{c},\theta_{\mathbf{r}}^{c})J_{\mathbf{r}-\mathbf{r}'}R_{0}(\phi_{\mathbf{r}'}^{c},\theta_{\mathbf{r}'}^{c})\left(\begin{array}{c}
\hat{S}_{\mathbf{r}'}^{x}\\
\hat{S}_{\mathbf{r}'}^{y}\\
\hat{S}_{\mathbf{r}'}^{z}
\end{array}\right),
\end{equation}
where the rotation matrix
\begin{equation}
R_{0}(\phi_{\mathbf{r}}^{c},\theta_{\mathbf{r}}^{c})=\left(\begin{array}{ccc}
\cos(\theta_{\mathbf{r}}^{c})\cos(\phi_{\mathbf{r}}^{c}) & -\sin(\phi_{\mathbf{r}}^{c}) & \sin(\theta_{\mathbf{r}}^{c})\cos(\phi_{\mathbf{r}}^{c})\\
\cos(\theta_{\mathbf{r}}^{c})\sin(\phi_{\mathbf{r}}^{c}) & \cos(\phi_{\mathbf{r}}^{c}) & \sin(\theta_{\mathbf{r}}^{c})\sin(\phi_{\mathbf{r}}^{c})\\
-\sin(\theta_{\mathbf{r}}^{c}) & 0 & \cos(\theta_{\mathbf{r}}^{c})
\end{array}\right).
\end{equation}

In the new frame, all spins in the classical state are aligned along the $z$-axis,
allowing us to employ the HP representation for $\hat{S}_{\mathbf{r}}^{z}$
and the spin ladder operators $\hat{S}_{\mathbf{r}}^{\pm}=\hat{S}_{\mathbf{r}}^{x}\pm i\hat{S}_{\mathbf{r}}^{y}$ as
\begin{align}
\hat{S}_{\mathbf{r}}^{z} & =S-b_{\mathbf{r}}^{\dagger}b_{\mathbf{r}},\nonumber \\
\hat{S}_{\mathbf{r}}^{+} & =\sqrt{2S-b_{\mathbf{r}}^{\dagger}b_{\mathbf{r}}}b_{\mathbf{r}},\\
\hat{S}_{\mathbf{r}}^{-} & =b_{\mathbf{r}}^{\dagger}\sqrt{2S-b_{\mathbf{r}}^{\dagger}b_{\mathbf{r}}},\nonumber 
\end{align}
where the bosonic creation (annihilation) operators $b_{\mathbf{r}}^{\dagger}(b_{\mathbf{r}})$
describe deviations from the spin coherent state and are directly associated with magnons as the elementary excitations. Expanding
the spin ladder operators in powers of $b_{\mathbf{r}}^{\dagger}b_{\mathbf{r}}/(2S)$,
the Hamiltonian can be rewritten as 
\begin{equation}
    H(\phi_{\mathbf{r}}^{c},\theta_{\mathbf{r}}^{c}) =\sum_{n=0}^{\infty}S^{2-n/2}\mathcal{H}_{n}.\label{eq:HSa}
\end{equation}
Here, all $\mathcal{H}_{n}$ including the operator $\hat{b}_{j,\alpha}$ depend on the definition of the classical configuration $(\phi_{\mathbf{r}}^{c},\theta_{\mathbf{r}}^{c})$.

The leading term $S^{2}\mathcal{H}_{0}=E_c$
corresponds to the classical ground-state energy. The linear term  is given by
\begin{align}
\mathcal{H}_{1} & =\frac{S^{3/2}}{\sqrt{2}}\sum_{\mathbf{r}\mathbf{r}'}\left(\begin{array}{ccc}
1 & i & 0\end{array}\right)R_{0}^{T}(\phi_{\mathbf{r}}^{c},\theta_{\mathbf{r}}^{c})J_{\mathbf{r}-\mathbf{r}'}R_{0}(\phi_{\mathbf{r}'}^{c},\theta_{\mathbf{r}'}^{c})\left(\begin{array}{c}
0\\
0\\
1
\end{array}\right)b_{\mathbf{r}}^{\dagger}+\rm{H.c.},\label{eq:S1H1}
\end{align}

A straightforward calculation leads to
\begin{equation}
\mathcal{H}_{1}=\sqrt{\frac{S}{2}}\sum_{\mathbf{r}}b_{\mathbf{r}}^{\dagger}\left(\frac{\partial E_c}{\partial\theta_{\mathbf{r}}^{c}}+\frac{i}{\sin\left(\theta_{\mathbf{r}}^{c}\right)}\frac{\partial E_c}{\partial\phi_{\mathbf{r}}^{c}}\right)+\rm{H.c.}.   \label{H1}
\end{equation}
Then the saddle point condition Eqs.~(\ref{eq:Ethe}) and~\eqref{eq:Ephi} ensures the vanishing linear term $\mathcal{H}_{1}=0$.

\subsection{Hamiltonian in momentum space}

We focus on a system of size $N=N_c N_s$, where $N_c$ is the number of unit cells and $N_s$ is the number of sublattices. We introduce $\mathbf{r}_{j,\alpha}$ to label the position of the sublattice site $\alpha$ in the unit cell $j$, and define $\hat{S}_{\mathbf{r}}^{\mu} \equiv \hat{S}_{j,\alpha}^{\mu}$. For the ground state with the translational symmetry, the classical configuration
$(\phi_{\mathbf{r}}^{c},\theta_{\mathbf{r}}^{c})\equiv(\phi_{j,\alpha}^{c},\theta_{j,\alpha}^{c})=(\phi_{\alpha}^{c},\theta_{\alpha}^{c})$ is independent of $j$. The transformed Hamiltonian reads
\begin{equation}
H(\phi_{\mathbf{r}}^{c},\theta_{\mathbf{r}}^{c})=H(\phi_{\alpha}^{c},\theta_{\alpha}^{c})=\frac{1}{2}\sum_{j,j'}\sum_{\alpha,\alpha'}\left(\begin{array}{ccc}
\hat{S}_{j,\alpha}^{+} & \hat{S}_{j,\alpha}^{-} & \hat{S}_{j,\alpha}^{z}\end{array}\right)\widetilde{J}_{j\alpha j'\alpha'}\left(\begin{array}{c}
\hat{S}_{j',\alpha'}^{+}\\
\hat{S}_{j',\alpha'}^{-}\\
\hat{S}_{j',\alpha'}^{z}
\end{array}\right),
\end{equation}
where the exchange interaction matrix in the new frame is
\begin{equation}
\widetilde{J}_{j\alpha j'\alpha'}=\frac{1}{4}\left(\begin{array}{ccc}
1 & -i & 0\\
1 & i & 0\\
0 & 0 & 2
\end{array}\right)R_0^{T}(\phi_{\alpha}^c,\theta_{\alpha}^c)J_{\mathbf{r}_{j,\alpha}-\mathbf{r}_{j',\alpha'}}R_0(\phi_{\alpha'}^c,\theta_{\alpha'}^c)\left(\begin{array}{ccc}
1 & 1 & 0\\
-i & i & 0\\
0 & 0 & 2
\end{array}\right).
\end{equation}

After applying the HP transformation, we express the Hamiltonian in the momentum space, where ${b}_{\mathbf{k},\alpha}\equiv\sum_{j}e^{-i\mathbf{k}\cdot\mathbf{r}_{j,\alpha}}{b}_{j,\alpha}/\sqrt{N_{c}}$ and the Fourier transform $\widetilde{J}_{\mathbf{k},\alpha\alpha'}=\sum_{\mathbf{r}_{j,\alpha}-\mathbf{r}_{j',\alpha'}}\widetilde{J}_{j\alpha j'\alpha'}e^{-i\mathbf{k}(\mathbf{r}_{j,\alpha}-\mathbf{r}_{j',\alpha'})}$: 
\begin{equation}
\widetilde{J}_{\mathbf{k},\alpha\alpha'}\equiv\left(\begin{array}{ccc}
\widetilde{J}_{\mathbf{k},\alpha\alpha'}^{++} & \widetilde{J}_{\mathbf{k},\alpha\alpha'}^{+-} & \widetilde{J}_{\mathbf{k},\alpha\alpha'}^{+z}\\
\widetilde{J}_{\mathbf{k},\alpha\alpha'}^{-+} & \widetilde{J}_{\mathbf{k},\alpha\alpha'}^{--} & \widetilde{J}_{\mathbf{k},\alpha\alpha'}^{-z}\\
\widetilde{J}_{\mathbf{k},\alpha\alpha'}^{z+} & \widetilde{J}_{\mathbf{k},\alpha\alpha'}^{z-} & \widetilde{J}_{\mathbf{k},\alpha\alpha'}^{zz}
\end{array}\right)
\end{equation}
is introduced for convenience. We consider the expansion Eq.~\eqref{eq:HSa} up to the fourth order $n=4$.

For $n=0$, the leading term $E_c=S^{2}\mathcal{H}_{0}=S^{2}\sum_{j,j'}\sum_{\alpha,\alpha'}\widetilde{J}_{j\alpha j'\alpha'}^{zz}/2$
represents the classical ground-state energy. The minimum of $E_c$ results in the absence of the linear term $\mathcal{H}_{1}$ [cf.~\eqref{H1}]. For $n=2$, The quadratic term
\begin{equation}
S\mathcal{H}_{2}=S\mathcal{H}_{0}+\frac{S}{2}\sum_{\mathbf{k}}B_{\mathbf{k}}^{\dagger}M_{\mathbf{k}}B_{\mathbf{k}}
\end{equation}
is expressed in the Nambu representation $B_{\mathbf{k}}=(
{b}_{\mathbf{k},\alpha},{b}_{-\mathbf{k},\alpha}^{\dagger})^{T}$. The matrix $M_{\mathbf{k}}$ is expressed as $M_{\mathbf{k}}=\left(\begin{array}{cc}
\varepsilon_{\mathbf{k}} & \Delta_{\mathbf{k}}\\
\Delta_{\mathbf{k}}^{\dagger} & \varepsilon_{-\mathbf{k}}^{\mathrm{T}}
 \end{array}\right)$, where the matrix elements of $\varepsilon_{\mathbf{k}}$ and $\Delta_{\mathbf{k}}$ are
\begin{align}
\varepsilon_{\mathbf{k}}^{\alpha\alpha'} & =2\widetilde{J}_{\mathbf{k},\alpha\alpha'}^{-+}-\delta_{\alpha\alpha'}\sum_{\alpha_1}\widetilde{J}_{\mathbf{k}=\mathbf{0},\alpha\alpha_1}^{zz},\\
\Delta_{\mathbf{k}}^{\alpha\alpha'} & =2\widetilde{J}_{\mathbf{k},\alpha\alpha'}^{--}.
\end{align}

For positive-definite $M_{\mathbf{k}}$, the magnon spectrum can be obtained by the symplectic diagonalization 
\begin{equation}
U_{\mathbf{k}}^{\dagger}M_{\mathbf{k}}U_{\mathbf{k}}=\mathrm{diag}(D_{\mathbf{k}},D_{-\mathbf{k}}), \label{B}   
\end{equation}
where the Bogoliubov transformation $U_{\mathbf{k}}$
satisfies $U_{\mathbf{k}}^{\dagger}(\sigma^{z}\otimes I_{N_s})U_{\mathbf{k}}=\sigma^{z}\otimes I_{N_s}$, $I_2$ and $I_{N_s}$ are identity matrices in the Numbu space and the sublattice basis, and the element $d_{\mathbf{k},m}$ of the diagonal matrix $D_{\mathbf{k}}$ denotes the energy of the magnon excitation with momentum $\mathbf{k}$ in the band $m$.

For a semi-positive definite $M_{\mathbf{k}_0}$ at some momentum $\mathbf{k}_0$, the most general approach to obtaining the linear spin-wave spectrum and corresponding quasiparticles is to analyze the system in a quadrature basis $R_{\mathbf{k}_0}=(\hat{x}_{\mathbf{k}_0,\alpha},\hat{p}_{\mathbf{k}_0,\alpha})^T$, 
 where   $\hat{x}_{\mathbf{k}_0,\alpha}=(\hat{b}_{\mathbf{k}_0,\alpha}+\hat{b}_{\mathbf{k}_0,\alpha}^{\dagger})/\sqrt{2}$ and $\hat{p}_{\mathbf{k}_0,\alpha}=-i(\hat{b}_{\mathbf{k}_0,\alpha}-\hat{b}_{\mathbf{k}_0,\alpha}^{\dagger})/\sqrt{2}$. The quadratic Hamiltonian at momentum $\mathbf{k}_0$
in the quadrature basis becomes
\begin{equation}
    \frac{S}{2}\sum_{\mathbf{k}_0} B_{\mathbf{k}_0}^{\dagger} M_{\mathbf{k}_0}  B_{\mathbf{k}_0}=\frac{S}{2}\sum_{\mathbf{k}_0} R_{\mathbf{k}_0}^{T} W M_{\mathbf{k}_0} W^{\dagger} R_{\mathbf{k}_0},\label{eq:H2k0}
\end{equation}
where $W=w\otimes I_{N_s}$ and $w=\left(\begin{array}{cc}
1 & 1\\
-i & i
\end{array}\right)/\sqrt{2}$. The matrix $WM_{\mathbf{k}_0}W^{\dagger}$ is diagonalized via the symplectic transformation $U_{\text{R},\mathbf{k}_0}^{T}WM_{\mathbf{k}_0}W^{\dagger}U_{\text{R},\mathbf{k}_0}=D_{\text{R},\mathbf{k}_0}$, where $U_{\text{R},\mathbf{k}_0}\equiv(u_{\text{R},\mathbf{k}_0},v_{\text{R},\mathbf{k}_0})$
satisfies $U_{\text{R},\mathbf{k}_0}(\sigma^{y}\otimes I_{N_s}) U_{\text{R},\mathbf{k}_0}^{T}=(\sigma^{y}\otimes I_{N_s})$, and $D_{\text{R},\mathbf{k}_0}=\text{diag}(D_{\mathbf{k}_0}^{(0)},D_{\mathbf{k}_0})$ is a diagonal matrix. The matrices $D_{\mathbf{k}_0}^{(0)}$ and $D_{\mathbf{k}_0}$ are diagonal with elements $\widetilde{d}_{\mathbf{k}_0,m=1,...,N_s}$ and $d_{\mathbf{k}_0,m}$ respectively.
The eigenvalue pairs $\widetilde{d}_{\mathbf{k}_0,m}$
and $d_{\mathbf{k}_0,m}$ fall into two categories. (i) The first type consists of pairs of equal eigenvalues $\widetilde{d}_{\mathbf{k}_0,m}=d_{\mathbf{k}_0,m}\geq0$, similar to the diagonalization of a positive-definite $M_{\mathbf{k}}$. (ii) The second type involves one eigenvalue being zero, which indicates the loss of degrees of freedom for the conjugate momentum or coordinate. Without loss of generality, we consider the case where $\widetilde{d}_{\mathbf{k}_0,m'}>0$
and $d_{\mathbf{k},m'}=0$, in which $(\sigma_z\otimes I_{N_s}) M_{\mathbf{k}_0}$ is non-diagonalizable.

In the new basis $\widetilde{B}_{\mathbf{k}_0}=U_{\mathbf{k}_0}^{-1}B_{\mathbf{k}_0}$ with the Bogoliubov transformation $U_{\mathbf{k}_0}=W^{\dagger}U_{\text{R},\mathbf{k}_0}W$ satisfying $U_{\mathbf{k}_0}^{\dagger}(\sigma^{z}\otimes I_{N_s})U_{\mathbf{k}_0}=\sigma^{z}\otimes I_{N_s}$, the quadratic term becomes
\begin{equation}
      \frac{S}{2}\sum_{\mathbf{k}_0} B_{\mathbf{k}_0}^{\dagger} M_{\mathbf{k}_0}  B_{\mathbf{k}_0}=\frac{S}{2}\sum_{\mathbf{k}_0} \widetilde{B}_{\mathbf{k}_0}^{\dagger}W^{\dagger}D_{\text{R},\mathbf{k}_0}W\widetilde{B}_{\mathbf{k}_0}.
\end{equation}
Here, the columns $u_{\mathbf{k}_0}$ and $v_{\mathbf{k}_0}$ in the matrix $U_{\mathbf{k}_0}=(u_{\mathbf{k}_0},v_{\mathbf{k}_0})$ describe the particle and hole excitations, which satisfy $v_{\mathbf{k}_0}=\sigma^{x}u_{-\mathbf{k}_0}^{*}$. For eigenvalue pairs $\widetilde{d}_{\mathbf{k}_0,m}=d_{\mathbf{k}_0,m}\geq0$, 
the dispersion matrix in the subspace spanned by $(u_{\mathbf{k}_0,m},v_{\mathbf{k}_0,m})$ is diagonalized: 
\begin{equation}
(u_{\mathbf{k}_0,m},v_{\mathbf{k}_0,m})^{\dagger}M_{\mathbf{k}_0}(u_{\mathbf{k}_0,m},v_{\mathbf{k}_0,m})=w^{\dagger}\text{diag}(\widetilde{d}_{\mathbf{k}_0,m},d_{\mathbf{k}_0,m})w=d_{\mathbf{k}_0,m}I_2. 
\label{D}
\end{equation}
For $\widetilde{d}_{\mathbf{k}_0,m'}>0$ and $d_{\mathbf{k}_0,m'}=0$, the dispersion matrix in the subspace spanned by 
$(u_{\mathbf{k}_0,m'},v_{\mathbf{k}_0,m'})$ is not diagonalizable: 
\begin{equation}
(u_{\mathbf{k}_0,m'},v_{\mathbf{k}_0,m'})^{\dagger} M_{\mathbf{k}_0} (u_{\mathbf{k}_0,m'},v_{\mathbf{k}_0,m'})=w^{\dagger}\text{diag}(\widetilde{d}_{\mathbf{k}_0,m'},d_{\mathbf{k}_0,m'})w=\left(\begin{array}{cc}
1 & 1\\
1 & 1
\end{array}\right)\widetilde{d}_{\mathbf{k}_0,m'}/2.
\label{ND}
\end{equation}
There are two types of zero modes:  
(i) For the type-I zero mode, the eigenvalue $\widetilde{d}_{\mathbf{k}_0,m'}>0$ and the dispersion matrix in the subspace spanned by  $(u_{\mathbf{k}_0,m'},v_{\mathbf{k}_0,m'})$ is not diagonalizable, as shown in Eq.~\eqref{ND}.
(ii) For the type-II zero mode, the eigenvalue pair $\widetilde{d}_{\mathbf{k}_0,m}=d_{\mathbf{k}_0,m}=0$ in Eq.~\eqref{D}.

For $n=3$, the interaction involving three magnons is described by the cubic term
\begin{equation}
\mathcal{H}_{3}=\frac{-\sqrt{2}}{\sqrt{N_{c}}}\sum_{\alpha,\alpha'}\sum_{\mathbf{p},\mathbf{q}}\widetilde{J}_{\mathbf{p},\alpha\alpha'}^{-z}b_{\mathbf{p},\alpha}^{\dagger}b_{\mathbf{q},\alpha'}^{\dagger}b_{\mathbf{p}+\mathbf{q},\alpha'}+\rm{H.c.},
\end{equation}
which can be written in compact forms
\begin{align}
\mathcal{H}_{3}=&\sum_{\mathbf{p},\mathbf{q},l_1,l_2,l_3}\frac{V_{l_1 l_2 l_3}^{(3)}(\mathbf{p},\mathbf{q})}{3!\sqrt{N_{c}}}B_{\mathbf{p},l_1}^{\dagger}B_{\mathbf{q},l_2}B_{\mathbf{p}-\mathbf{q},l_3}=\sum_{\mathbf{p},\mathbf{q},l_1,l_2,l_3}\frac{\overline{V}_{l_1 l_2 l_3}^{(3)}(\mathbf{p},\mathbf{q})}{3!\sqrt{N_{c}}}B_{\mathbf{p},l_1}^{\dagger}B_{\mathbf{q},l_2}B_{\mathbf{q}-\mathbf{p},l_3}^{\dagger},
\end{align}
with interaction tensors $V_{l_1 l_2 l_3}^{(3)}(\mathbf{p},\mathbf{q})$
and $\overline{V}_{l_1 l_2 l_3}^{(3)}(\mathbf{p},\mathbf{q})=\sum_{l}V_{l_1 l_2 l}^{(3)}(\mathbf{p},\mathbf{q})(\sigma^x \otimes I_{N_s})_{l,l_3}$.

For $n=4$, the quartic term 
\begin{align}
\mathcal{H}_{4} =& \frac{1}{N_{c}}\sum_{\alpha,\alpha'}\sum_{\mathbf{k},\mathbf{p},\mathbf{q}}[\frac{-1}{4}(b_{\mathbf{k},\alpha}^{\dagger}b_{\mathbf{p},\alpha'}^{\dagger}b_{\mathbf{q},\alpha'}^{\dagger}b_{\mathbf{k}+\mathbf{p}-\mathbf{q},\alpha'}\widetilde{J}_{\mathbf{k},\alpha\alpha'}^{--}+b_{\mathbf{k},\alpha}^{\dagger}b_{\mathbf{p},\alpha}^{\dagger}b_{\mathbf{q},\alpha'}^{\dagger}b_{\mathbf{k}+\mathbf{p}-\mathbf{q},\alpha}\widetilde{J}_{-\mathbf{q},\alpha\alpha'}^{--}+\rm{H.c.})\\
&+\frac{1}{2}(b_{\mathbf{k},\alpha}^{\dagger}b_{\mathbf{p},\alpha'}^{\dagger}b_{\mathbf{q},\alpha}b_{\mathbf{k}+\mathbf{p}-\mathbf{q},\alpha'}\widetilde{J}_{\mathbf{k}-\mathbf{q},\alpha\alpha'}^{zz}-b_{\mathbf{k},\alpha}^{\dagger}b_{\mathbf{p},\alpha'}^{\dagger}b_{\mathbf{q},\alpha'}b_{\mathbf{k}+\mathbf{p}-\mathbf{q},\alpha'}\widetilde{J}_{\mathbf{k},\alpha\alpha'}^{-+}-b_{\mathbf{k},\alpha}^{\dagger}b_{\mathbf{p},\alpha}^{\dagger}b_{\mathbf{q},\alpha}b_{\mathbf{k}+\mathbf{p}-\mathbf{q},\alpha'}\widetilde{J}_{\mathbf{k}+\mathbf{p}-\mathbf{q},\alpha\alpha'}^{-+})]\nonumber
\end{align}
describes the collision of magnons. In a compact form,
\begin{equation}
\mathcal{H}_{4}=\frac{1}{N_{c}}\frac{1}{4!}\sum_{\mathbf{k},\mathbf{p},\mathbf{q}}\sum_{l_1,l_2,l_3,l_4}V_{l_1 l_2 l_3l_4}^{(4)}(\mathbf{k},\mathbf{p},\mathbf{q}):B_{\mathbf{k},l_1}^{\dagger}B_{\mathbf{p},l_2}^{\dagger}B_{\mathbf{k}+\mathbf{q},l_3}B_{\mathbf{p}-\mathbf{q},l_4}:,
\end{equation}
with the interaction tensor $V_{l_1 l_2 l_3l_4}^{(4)}(\mathbf{k},\mathbf{p},\mathbf{q})$. Here, $:...:$ denotes the normal ordering with respect to the spin coherent state $|\{\phi_{\mathbf{r}}^{c},\theta_{\mathbf{r}}^{c}\}\rangle$, i.e., the vacuum state that satisfies $b_{\mathbf{k}}|\{\phi_{\mathbf{r}}^{c},\theta_{\mathbf{r}}^{c}\}\rangle=0$.

\section{The proof of curvature formula}
\label{Appproof}
In this Appendix, we provide the proof of the curvature formula. The detailed computations in the three key steps of the proof listed in the main text are shown in the first three subsections. In the fourth subsection, we generalize the theorem to the case of pseudo-Goldstone (PG) modes with finite momentum. 

In magnetic systems, some approximate symmetries are preserved at the classical level, but violated by quantum and thermal corrections. Specifically, the classical
energies $E_c$ of the states $|\{\phi_{\mathbf{r}}^{c},\theta_{\mathbf{r}}^{c}\}\rangle$ connected by symmetry transformations are degenerate, the degeneracy is lifted when the quantum and
thermal fluctuations are considered. 

To select the true equilibrium state at finite temperature, we consider the Hamiltonian Eq.~\eqref{eq:HSa} to first order in $O(S)$ and obtain the free energy 
\begin{equation}
 F=E_{g}+T\sum_{\mathbf{k},m}\ln(1-e^{-Sd_{\mathbf{k},m}/T}),\label{eq:F2}
\end{equation}
where $E_{g}=S(S+1)\mathcal{H}_{0}+S\sum_{\mathbf{k}}\text{tr}D_{\mathbf{k}}/2$. Minimizing the free energy $F$, we obtain the equilibrium configuration $(\phi_{\mathbf{r}}^{0},\theta_{\mathbf{r}}^{0})$. In the rotated
frame, the Hamiltonian is $H(\phi_{\mathbf{r}}^{0},\theta_{\mathbf{r}}^{0})=\hat{U}_{R}^{\dagger}(\phi_{\mathbf{r}}^{0},\theta_{\mathbf{r}}^{0})H\hat{U}_{R}(\phi_{\mathbf{r}}^{0},\theta_{\mathbf{r}}^{0})=\sum_{n=0}^{\infty}S^{2-n/2}\mathcal{H}_{n}$. 

The spontaneous breaking of approximate symmetries, i.e., accidental degeneracies at the classical level $E_c$, gives rise to zero modes of $M_{\mathbf{k}_{0}}$, i.e., PG modes. We denote the elements $\widetilde{d}_{\mathbf{k}_{0},m'}$
for type-I PG modes as $(\widetilde{d}_{\mathbf{k}_{0}})_{\mathrm{ps}}$, where the corresponding mode functions $(u_{\mathbf{k}_{0}})_{\mathrm{ps}}$
and $(v_{\mathbf{k}_{0}})_{\mathrm{ps}}$ are given by columns in the matrices $u_{\mathbf{k}_{0}}$ and $v_{\mathbf{k}_{0}}$, respectively. The single-magnon operators
describing the pseudo-Goldstone mode are $\gamma_{-\mathbf{k}_{0}}=-B_{\mathbf{k}_{0}}^{\dagger}(\sigma^{z}\otimes I_{N_{s}})(v_{\mathbf{k}_{0}})_{\mathrm{ps}}$
and $\gamma_{\mathbf{k}_{0}}^{\dagger}=B_{\mathbf{k}_{0}}^{\dagger}(\sigma^{z}\otimes I_{N_{s}})(u_{\mathbf{k}_{0}})_{\mathrm{ps}}$.

For the type-I PG mode, the Hamiltonian contains the quadratic term $S(\widetilde{d}_{\mathbf{k}_{0}})_{\mathrm{ps}}O_{\mathbf{k}_{0},x}^{\dagger}O_{\mathbf{k}_{0},x}$,
where the canonical position operator
\begin{equation}
O_{\mathbf{k}_{0},x}=\frac{1}{\sqrt{2}}(\gamma_{-\mathbf{k}_{0}}+\gamma_{\mathbf{k}_{0}}^{\dagger})=B_{\mathbf{k}_{0}}^{\dagger}\sigma^{z}\chi_{\mathbf{k}_{0},\phi},
\end{equation}
and $\chi_{\mathbf{k}_{0},\phi}\equiv[(u_{\mathbf{k}_{0}})_{\mathrm{ps}}-(v_{\mathbf{k}_{0}})_{\mathrm{ps}}]/\sqrt{2}$.
We can also define the canonical momentum operator
\begin{equation}
O_{\mathbf{k}_{0},p}=\frac{-i}{\sqrt{2}}(\gamma_{\mathbf{k}_{0}}-\gamma_{-\mathbf{k}_{0}}^{\dagger})=\chi_{\mathbf{k}_{0},\theta}^{\dagger}\sigma^{z}B_{\mathbf{k}_{0}}
\end{equation}
by $\chi_{\mathbf{k}_{0},\theta}\equiv i[(u_{\mathbf{k}_{0}})_{\mathrm{ps}}+(v_{\mathbf{k}_{0}})_{\mathrm{ps}}]/\sqrt{2}$, which satisfies the commutation relation $[O_{\mathbf{k}_{0},x},O_{\mathbf{k}_{0},p}]=i$. For the type-II PG mode, the Hamiltonian has the form $0\times(\gamma_{\mathbf{k}_{0}}^{\dagger}\gamma_{\mathbf{k}_{0}}+\gamma_{-\mathbf{k}_{0}}^{\dagger}\gamma_{-\mathbf{k}_{0}})$, which does not contain the pseudo-Goldstone mode at the quadratic order.

If the PG mode has zero momentum $\mathbf{k}_{0}=\mathbf{0}$, the rotation operator $\hat{R}=\hat{U}_{R}^{\dagger}(\phi_{\alpha}^{0},\theta_{\alpha}^{0})\hat{U}_{R}(\phi_{\alpha}^{c},\theta_{\alpha}^{c})$, which
connects the equilibrium configuration $(\phi_{\alpha}^{0},\theta_{\alpha}^{0})$
to arbitrary classically degenerate configurations $(\phi_{\alpha}^{c},\theta_{\alpha}^{c})$,
can always be parameterized as 
\begin{equation}
\hat{R}(\phi,\theta)=\exp[-i\sum_{j,\alpha}\omega_{\alpha}\mathbf{n}_{\alpha}(\phi,\theta)\cdot\hat{\mathbf{S}}_{j,\alpha})]\label{eqs:Rex}
\end{equation}
by $\phi$ and $\theta$, and the rotation is generated by the superposition of $S_{j,\alpha}^{x,y}$. For convenience, we define $\chi_{\phi(\theta)}\equiv\chi_{\mathbf{k}_{0}=\mathbf{0},\phi(\theta)}$. Without loss of generality, we can choose
\begin{equation}
\omega_{1}\mathbf{n}_{1}\cdot\hat{\mathbf{S}}_{j,1}=\sqrt{\frac{2}{N_c}}(\phi|\chi_{\phi,1}| \widetilde{S}_{j,1}^{x}+\theta|\chi_{\theta,1}|\widetilde{S}_{j,1}^{y})
\end{equation}
for the sublattice $\alpha=1$, where $\widetilde{S}_{j,1}^{x}=[{\text{Re}(\chi_{\phi,1})}\hat{S}_{j,1}^{x}+{\text{Im}(\chi_{\phi,1})}\hat{S}_{j,1}^{y}]/|\chi_{\phi,1}|$ and $\widetilde{S}_{j,1}^{y}=[{\text{Re}(\chi_{\theta,1})}\hat{S}_{j,1}^{x}+{\text{Im}(\chi_{\theta,1})}\hat{S}_{j,1}^{y}]/|\chi_{\theta,1}|$.
The relation between $(\omega_{\alpha},\mathbf{n}_{\alpha})$
and $(\phi,\theta)$ may be nonlinear for $\alpha\neq1$.

Using the HP transformation, we can connect the Hermitian operators $O_{\mathbf{0},x}$ and $O_{\mathbf{0},p}$
with spin operators $S_{j,\alpha}^{x}$ and $S_{j,\alpha}^{y}$ at
leading order $O(S^{1/2})$ as
\begin{align}
\sqrt{\frac{2}{N_{c}}}|\chi_{\phi,1}|\sum_{j,\alpha}\left[\frac{\text{Re}(\chi_{\phi,\alpha})}{|\chi_{\phi,1}|}S_{j,\alpha}^{x}+\frac{\text{Im}(\chi_{\phi,\alpha})}{|\chi_{\phi,1}|}S_{j,\alpha}^{y}\right]& \simeq\sqrt{S}O_{\mathbf{0},x} +O(S^{-1/2}),\\
  \sqrt{\frac{2}{N_{c}}}|\chi_{\theta,1}|\sum_{j,\alpha}\left[\frac{\text{Re}(\chi_{\theta,\alpha})}{|\chi_{\theta,1}|}S_{j,\alpha}^{x}+\frac{\text{Im}(\chi_{\theta,\alpha})}{|\chi_{\theta,1}|}S_{j,\alpha}^{y}\right] &\simeq\sqrt{S}O_{\mathbf{0},p}+O(S^{-1/2}).\nonumber 
\end{align}

For the type-I PG mode, the rotation along the direction $\text{Re}(\chi_{\phi,1})\hat{\mathbf{x}} + \text{Im}(\chi_{\phi,1})\hat{\mathbf{y}}$ in the
sublattice $\alpha=1$ does not cost the classical energy, while the
rotation along the direction $\text{Re}(\chi_{\theta,1})\hat{\mathbf{x}} + \text{Im}(\chi_{\theta,1})\hat{\mathbf{y}}$ changes the classical energy. As a result, only the rotation 
\begin{equation}
    \hat{R}(\phi,\theta=0)=\exp{[-i\phi\sqrt{S}O_{\mathbf{0},x}+O(\phi,S^{-1/2})]}
\end{equation}
preserves the classical energy. For the type-II PG mode, the rotation along any direction within the
$x$-$y$ plane does not change the classical energy, allowing for
arbitrary finite rotations $\phi$ and $\theta$. The rotation operator thus becomes
\begin{equation}
\hat{R}(\phi,\theta)=\exp{[-i\phi\sqrt{S}O_{\mathbf{0},x}-i\theta\sqrt{S}O_{\mathbf{0},p}+O(\phi,\theta,S^{-1/2})]}.
\end{equation}

For the PG mode with a finite momentum $\mathbf{k}_{0}\neq\mathbf{0}$, the treatment is similar to the zero momentum case. Here, the degenerate classical configurations with energy $E_c$ depend on the unit cell index $j$. We introduce a unit-cell dependent rotation operator $\hat{R}(\phi,\theta)=\exp[-i\sum_{j,\alpha}\omega_{j,\alpha}\mathbf{n}_{j,\alpha}(\phi,\theta)\cdot\hat{\mathbf{S}}_{j,\alpha})]$, which connects the equilibrium configuration $(\phi_{\mathbf{r}}^{0},\theta_{\mathbf{r}}^{0})$
to an arbitrary classically degenerate configuration $(\phi_{\mathbf{r}}^{c},\theta_{\mathbf{r}}^{c})$. Using the HP transformation, the rotation operator can be expressed as 
\begin{equation}
    \hat{R}(\phi,\theta)=\exp{\{-i\sqrt{\frac{S}{2}}[\phi(O_{\mathbf{k}_{0},x}+O_{\mathbf{k}_{0},x}^{\dagger})+\theta(O_{\mathbf{k}_{0},p}+O_{\mathbf{k}_{0},p}^{\dagger})]+O(\phi,\theta,S^{-1/2})\}}.
\end{equation}
For the type-I PG mode, only rotations along the $\phi$-direction in the sublattice $\alpha=1$ preserve $E_c$, leading to $\theta=0$. For the type-II PG mode, $E_c$ remains unchanged under $\hat{R}(\phi,\theta)$ with arbitrary $\phi$ and $\theta$. 

In the remainder of this section, we prove the \textit{theorem: At finite temperature $T$, the pseudo-Goldstone gap at zero momentum is 
\begin{equation}
\Delta=\frac{1}{S}\sqrt{\left(\frac{\partial^{2}F}{\partial\theta^{2}}%
\right)_{0}\left(\frac{\partial^{2}F}{\partial\phi^{2}}\right)_{0}-\left(%
\frac{\partial^{2}F}{\partial\theta\partial\phi}\right)_{0}^{2}},
\label{eqs:cur}
\end{equation}
determined by the curvature of the free energy $F(\phi,\theta)$.} Here, the subindex `0' indicates that the derivatives are evaluated at $\theta=\phi=0$, and $F(\phi ,\theta )=-T\ln Z(\phi,\theta)$ is obtained from the partition function $Z(\phi,\theta)=\text{Tr}\exp [-H_{R}(\phi ,\theta )/T]$. The quadratic Hamiltonian $H_R(\phi ,\theta)$ is obtained by truncating the expansion
\begin{equation}
\hat{R}^{\dagger }(\phi ,\theta )H(\phi _{\alpha }^{0},\theta _{\alpha }^{0})%
\hat{R}(\phi ,\theta )=\sum_{n=0}^{\infty }S^{2-n/2}\mathcal{H}_{n}(\phi
,\theta)
\end{equation}%
at $n=2$, where $\mathcal{H}_{n}(0,0)=\mathcal{H}_{n}$. For the type-I PG mode, although only $\hat{R}(\phi ,0)$ preserves the classical
energy, a finite $\theta $ in $\hat{R}(\phi ,\theta )$ is still necessary to compute the gap. 

The curvature formula is proved in three steps. First, we express the rotation operator $\hat{R}(\phi ,\theta )$ and the Hamiltonian $H(\phi,\theta)$ in terms of $\mathcal{H}_{n}$, i.e., the coefficients of $H(\phi _{\alpha }^{0},\theta _{\alpha }^{0})$ in the large-$S$ expansion. In the second step, we reformulate the curvature formula via $\mathcal{H}_{n}$. Finally, we prove the curvature formula by applying a perturbative expansion of $\mathcal{H}_{n>2}$.

It is worth noting that the proof of the curvature formula at finite momentum follows the same steps as in the zero-momentum case. This will be discussed in the final part of this section.

\subsection{Step1: Expansion of rotation operator and Hamiltonian}
We begin by expanding the generator of the rotation operator $\hat{R}(\phi ,\theta)$ Eq.~\eqref{eqs:Rex} in powers of $\phi$, $\theta$ and $S$ as
\begin{equation}
\hat{R}(\phi ,\theta)=\exp \left( -i\sum_{\zeta ,\eta ,\nu =0}^{\infty
}O_{\zeta ,\eta }^{(\nu )}\phi ^{\zeta }\theta ^{\eta }S^{1/2-\nu }\right).
\label{R}
\end{equation}%
Since the generator is a superposition
of $S_{j,\alpha }^{x}$ and $S_{j,\alpha }^{y}$, in principle the Hermitian
operator $O_{\zeta ,\eta }^{(\nu )}$ in the exponent
can be obtained using the Taylor expansion of $\omega _{\alpha }\mathbf{n}%
_{\alpha }$ and the HP representation of $S_{j,\alpha }^{x,y}$. In practice,
for a generic spin model, $O_{\zeta ,\eta }^{(\nu )}$ can be systematically
derived order by order by imposing the invariance of the classical
ground-state energy $E_c$ under the rotation $\hat{R}(\phi,\theta )$: $\mathcal{H}_{0}(\phi ,0)=\mathcal{H}_{0}$ for type-I PG modes and $\mathcal{H}_{0}(\phi ,\theta)=\mathcal{H}_{0}$ for type-II PG modes. 
Since all values of $\phi$ minimize the
classical energy, the saddle-point condition satisfies, namely, the
single-magnon term $\mathcal{H}_{1}(\phi,0)=0$ ($\mathcal{H}_{1}(\phi,\theta)=0$) vanishes for type-I (II) PG modes. This condition determines $O_{\zeta,\eta }^{(\nu)}$ via the
Baker-Campbell-Hausdorff (BCH) formula.  

For instance, at the linear order in $\phi$, $[O_{1,0}^{(0)},\mathcal{H}_{2}]=0$ fixes the canonical position operator $O_{1,0}^{(0)}=O_{\mathbf{0},x}=B_{\mathbf{0
}}^{\dagger }\sigma ^{z}\chi _{\phi }$. Due to the rotation $\hat{R}(0,\theta)$ along the direction orthogonal to $\phi$, the canonical momentum operator reads $O_{0,1}^{(0)}=O_{\mathbf{0},p}=B_{\mathbf{0}}^{\dagger}\sigma ^{z}\chi _{\theta }$. For
the type-I PG mode, under the rotation 
\begin{equation}
R(0,\theta)=\exp{\{-i\theta\sqrt{\frac{2}{N_c}%
}\sum_{j,\alpha}[\text{Re}(\chi_{\theta,\alpha} )S_{j,\alpha}^x+\text{Im}(
\chi_{\theta,\alpha}) S_{j,\alpha}^y]\}}    
\end{equation}
generated by $O_{0,1}^{(0)}$, the
zero mode Hamiltonian $S(\widetilde{d}_{\mathbf{0}})_{\mathrm{ps}%
}(O_{1,0}^{(0)})^{2}/2$ is not invariant. As a result, the classical
ground-state energy changes as $S^{2}\mathcal{H}_{0}(0,\theta )=S^{2}[%
\mathcal{H}_{0}+(\widetilde{d}_{\mathbf{0}})_{\mathrm{ps}}\theta ^{2}/2]$.
It can be straightforwardly proven that the change of $H$ generated by terms 
$O_{\zeta >0,\eta }^{(0)}$ and $O_{0,\eta >1}^{(0)}$ are either subleading $O(S^{\nu <2})$ or of higher order form $\theta ^{\zeta }\phi ^{\eta }$ with $\zeta +\eta >2$, which does not contribute to the second order derivative of 
$F$ at $\theta =\phi =0$. Therefore, only $O_{0,1}^{(0)}$ contributes to the gap of the type-I PG mode. At the quadratic order $\phi ^{2}$, $[\mathcal{H}
_{2},O_{2,0}^{(0)}]=i[[\mathcal{H}_{3},O_{1,0}^{(0)}],O_{1,0}^{(0)}]/2$
fixes $O_{2,0}^{(0)}=\chi _{\phi }^{\dagger }\Omega \chi _{\phi }/2$, where
\begin{equation}
\Omega _{l_{1},l_{2}}=\frac{i}{\sqrt{N_{c}}}\sum_{l_{3}=1}^{2N_{s}}V_{
l_{1}l_{2}l_{3}}^{(3)}(\mathbf{0},\mathbf{0})\left( M_{\mathbf{0}}^{-1}(\sigma^{z}\otimes I_{N_s})B_{\mathbf{0}}\right)_{l_{3}}
\end{equation}
with $M_{\mathbf{0}}^{-1}$ being the pseudo-inverse.   

For the type-II PG mode, the classical
ground-state energy remains invariant under the rotation $\hat{R}(\phi
,\theta )$, i.e., $\mathcal{H}_{0}(\phi ,\theta )=\mathcal{H}_{0}$.
Following the same procedure, we can obtain $O_{\zeta ,0}^{(0)}$, $O_{0,2}^{(0)}=\chi _{\theta }^{\dagger }\Omega \chi _{\theta }/2$,
and 
\begin{equation}
O_{1,1}^{(0)}=\frac{1}{2}(\chi _{\theta }^{\dagger }\Omega \chi _{\phi }+\chi _{\phi }^{\dagger }\Omega \chi _{\theta })=\chi _{\theta }^{\dagger }\Omega \chi _{\phi }
\end{equation}
by
imposing $\mathcal{H}_{1}(\phi ,\theta )=0$.
Now we have determined all the operators $O_{\zeta ,\eta }^{(0)}$ for $\nu =0$ in the
large-$S$ expansion. The higher order terms $O_{\zeta ,\eta }^{(\nu >0)}$
can be directly derived from the HP transformation. It is worth noting that the rotation operator in the work~\cite{Rau2018} (specifically, Eq.~(S48) therein)  only contains $O_{1,0}^{(\nu)}$ and $O_{0,1}^{(\nu)}$, while higher-order terms such as $O_{0,2}^{(\nu)}$, $O_{2,0}^{(\nu)}$ and $O_{1,1}^{(\nu)}$ are missing, which are crucial when the system has non-collinear order.

The Hamiltonian $\mathcal{H}_{2}(\phi ,\theta)=\sum_{\zeta ,\eta =0}^{2}h_{\zeta ,\eta }\phi
^{\zeta }\theta ^{\eta}$ up to the second order (with $\zeta +\eta =2$) contributes
to the PG gap as described by Eq.~\eqref{eqs:cur}. Here, $h_{0,0}=\mathcal{H}_{2}$ is the quadratic Hamiltonian from the spin-wave theory, and the
higher-order terms can be obtained using the BCH formula. For instance,
\begin{align}
h_{1,0}=&-i[\mathcal{H}_{3},O_{1,0}^{(0)}],\label{eq:h22}\\\nonumber
h_{0,1}=&-i[\mathcal{H}_{3},O_{0,1}^{(0)}],\\\nonumber
h_{2,0}=&-i[\mathcal{H}_{3},O_{2,0}^{(0)}]-\frac{1}{2}[[\mathcal{H}
_{4},O_{1,0}^{(0)}],O_{1,0}^{(0)}]-\frac{1}{2}[[\mathcal{H}_{2},O_{1,0}^{(1)}],O_{1,0}^{(0)}],\\\nonumber
h_{0,2}=&-i[\mathcal{H}_{3},O_{0,2}^{(0)}]-\frac{1}{2}[[\mathcal{H}
_{4},O_{0,1}^{(0)}],O_{0,1}^{(0)}]-\frac{1}{2}[[\mathcal{H}_{2},O_{0,1}^{(1)}],O_{0,1}^{(0)}],\\\nonumber
h_{1,1}=&-i[\mathcal{H}_{3},O_{1,1}^{(0)}]-\frac{1}{2}([[\mathcal{H}
_{4},O_{0,1}^{(0)}],O_{1,0}^{(0)}]+[[\mathcal{H}
_{4},O_{1,0}^{(0)}],O_{0,1}^{(0)}]+[[\mathcal{H}_{2},O_{0,1}^{(1)}],O_{1,0}^{(0)}]+[[\mathcal{H}_{2},O_{1,0}^{(1)}],O_{0,1}^{(0)}]).\nonumber
\end{align}

\subsection{Step2: Second-order derivatives of $F$}
We now compute the second-order derivatives of the free energy $F(\phi,\theta)$ with
respect to $\phi $ and $\theta$. The partition function $Z$ can be
expressed in the standard path integral form
\begin{equation}
Z(\phi,\theta)=\int\prod_{\mathbf{k},\tau,\alpha}d\psi_{\mathbf{k},\alpha}(\tau)d\psi_{\mathbf{k},\alpha}^{*}(\tau)\exp\left(-\int_{0}^{\frac{1}{T}}d\tau[\sum_{}\sum_{\mathbf{k},\alpha}\psi_{\mathbf{k},\alpha}^{*}(\tau)\partial_{\tau}\psi_{\mathbf{k},\alpha}(\tau)+H_R(\phi,\theta)]\right).
\end{equation}
where $H_R(\phi,\theta)$ is a $c$-number function obtained by replacing the field operators $b_{\mathbf{k},\alpha}$ and $b^{\dagger}_{\mathbf{k},\alpha}$ with the complex c-numbers $\psi_{\mathbf{k},\alpha}$ and $\psi_{\mathbf{k},\alpha}^{*}$.
By taking the
derivative of $F(\phi ,\theta )=-T \ln{Z(\phi,\theta)}$ with respect to $\phi $, we obtain $\mathcal{F}_{\phi
}\equiv (\partial ^{2}F/\partial \phi ^{2})_{0}=\sum_{L=1}^{3}\mathcal{F}%
_{\phi ,L}$ with 
\begin{align}
\mathcal{F}_{\phi ,1}&=-S\langle [[\mathcal{H}_{4},O_{1,0}^{(0)}],O_{1,0}^{(0)}]\rangle _{0},\label{eq:Fphi} 
\\\nonumber
\mathcal{F}_{\phi,2}&=-2iS\langle [\mathcal{H}_{3},O_{2,0}^{(0)}]\rangle _{0},\\\nonumber
\mathcal{F}_{\phi ,3}&=-TS^{2}(\int_{0}^{\frac{1}{T}}d\tau d\tau ^{\prime
}\langle \mathcal{T}_{\tau }h_{1,0}(\tau )h_{1,0}(\tau ^{\prime })\rangle
_{0}-\frac{\langle h_{1,0}\rangle _{0}^{2}}{T^{2}}),\nonumber
\end{align}
where $\langle X\rangle _{0}=\text{Tr}(\rho _{0}X)$ denotes the expectation
value of an operator $X$ in the thermal equilibrium state $\rho _{0}=e^{-S%
\mathcal{H}_{2}/T}/Z_{0}$ governed by the Hamiltonian $h_{0,0}$ without the
rotation, and $Z_{0}=$Tr$(e^{-S\mathcal{H}_{2}/T})$. Here, $h_{1,0}(\tau )=e^{S%
\mathcal{H}_{2}\tau }h_{1,0}e^{-S\mathcal{H}_{2}\tau }$, and $\mathcal{T}%
_{\tau }$ is the imaginary-time ordering operator. 
It is important to note that the last term in $h_{2,0}$ (cf. Eq.~\eqref{eq:h22})
does not contribute to $(\partial ^{2}F/\partial \phi ^{2})_{0}$,
because $\text{tr}(\rho _{0}[[\mathcal{H}_{2},O_{1,0}^{(1)}],O_{1,0}^{(0)}])=0$
follows from the Jacobi identity and the conditions $[\mathcal{H}_{2},O_{1,0}^{(0)}]=[\mathcal{H}_{2},\rho _{0}]=0$.

The second order derivative $\mathcal{F}_{\phi }=S\chi _{\phi }^{\dagger }\bar{\Sigma}_{\mathbf{0}}\chi _{\phi }$ is determined by $\bar{\Sigma}_{\mathbf{0}}=\sum_{L=1}^{3}\bar{\Sigma}
_{\mathbf{0}}^{(L)}$, where the three components $\bar{\Sigma}
_{\mathbf{0}}^{(L)}$ derived from $\mathcal{F}_{\phi,L}$ are
\begin{align}
    \bar{\Sigma}_{\mathbf{0},ll'}^{(1)}=&\frac{1}{2N_{c}}\sum_{\mathbf{p}}\sum_{l_1,l_2}V_{ll_1l_2l'}^{(4)}(\mathbf{0},\mathbf{p},\mathbf{p})[\langle B_{\mathbf{p}}^{\dagger}B_{\mathbf{p}}\rangle_{0}-\frac{1}{2}{(I-\sigma^z)}\otimes I_{N_s}]_{l_1l_2},\label{eq:selfenergy}\\\nonumber
    \bar{\Sigma}_{\mathbf{0},ll'}^{(2)}=&\frac{-1}{2N_{c}}\sum_{\mathbf{p}}\sum_{l_1,l_2,l_3,l_4}V_{ll'l_1}^{(3)}(\mathbf{0},\mathbf{0})\left(M_{\mathbf{0}}^{-1}\right)_{l_1l_2}\langle B_{\mathbf{p},l_3}^{\dagger}B_{\mathbf{p},l_4}\rangle_{0}\overline{V}_{l_3l_4l_2}^{(3)}(\mathbf{p},\mathbf{p}),\\\nonumber
     \bar{\Sigma}_{\mathbf{0},ll'}^{(3)}=&TS\frac{-1}{2N_c}\sum_{l_1,l_2,l_3,l_4}\int_{0}^{\frac{1}{T}}d\tau d\tau'\langle \mathcal{T}_{\tau}B_{\mathbf{p},l_2}(\tau)B_{\mathbf{p},l_3}^{\dagger}(\tau')\rangle_0 \langle \mathcal{T}_{\tau}B_{\mathbf{p},l_4}(\tau')B_{\mathbf{p},l_1}^{\dagger}(\tau)\rangle_0V_{l_3l_4l'}^{(3)}(\mathbf{p},\mathbf{p})\overline{V}_{l_1l_2l}^{(3)}(\mathbf{p},\mathbf{p}),
     \\\nonumber
     =&\frac{-1}{2N_{c}}\sum_{\mathbf{p}}\sum_{l_1,l_2,l_3,l_4,l_5,l_6}\frac{-n_{b}((\sigma^{z}\otimes SD_{\mathbf{p}})_{l_1l_1})+n_{b}((\sigma^{z}\otimes SD_{\mathbf{p}})_{l_2l_2})}{\left(\sigma^{z}\otimes D_{\mathbf{p}}\right)_{l_1l_1}-\left(\sigma^{z}\otimes D_{\mathbf{p}}\right)_{l_2l_2}}\\     
     &\times[(\sigma^{z}\otimes I_{N_s})U_{\mathbf{p}}^{\dagger}]_{l_2 l_3}V_{l_3l_4l'}^{(3)}(\mathbf{p},\mathbf{p})(U_{\mathbf{p}})_{l_4 l_1}[(\sigma^{z}\otimes I_{N_s})U_{\mathbf{p}}^{\dagger}]_{l_1l_5}\overline{V}_{l_5l_6l}^{(3)}(\mathbf{p},\mathbf{p})(U_{\mathbf{p}})_{l_6 l_2}. \nonumber
\end{align}
Here, $U_{\mathbf{p}}$ is the Bogoliubov transformation and $D_{\mathbf{p}}$ is the
positive spin-wave spectrum obtained by diagonalizing $\sigma^{z} M_{\mathbf{p}}$, as shown in Eq.~\eqref{B}. The Bose-Einstein distribution is given by
$n_{b}(\omega)=1/\left(e^{\omega/T}-1\right)$. 

In paper~\cite{Rau2018}, $\mathcal{F}_{\phi,2}$ is always zero because the operators $O_{2,0}^{(\nu)}$ are missed.  Additionally, the proof of curvature formula at zero temperature, i.e., Eq.~(10) in \cite{Rau2018}, does not account for the derivatives of ground states for the quadratic Hamiltonian $\mathcal{H}(\phi,\theta)$, leading to the omission of $\mathcal{F}_{\phi,3}$ in the zero-temperature limit.

For the type-I PG mode, only $O_{0,1}^{(0)}$ plays the role to the leading
order, thus, $\left( \partial ^{2}F/\partial \theta ^{2}\right) _{0}=S^{2}(%
\widetilde{d}_{\mathbf{0}})_{\mathrm{ps}}$ and $(\partial ^{2}F/\partial
\theta \partial \phi )_{0}=0$. For the type-II PG mode, we have $(\partial
^{2}F/\partial \theta ^{2})_{0}=S\chi _{\theta }^{\dagger }\bar{\Sigma}_{%
\mathbf{0}}\chi _{\theta }$ and $(\partial ^{2}F/\partial \theta \partial
\phi )_{0}=S\chi _{\theta }^{\dagger }\bar{\Sigma}_{\mathbf{0}}\chi _{\phi }$%
, both determined by the same matrix $\bar{\Sigma}_{\mathbf{0}}$. It follows from
Eq.~\eqref{eqs:cur} that
\begin{equation}
\Delta =\left\{ 
\begin{array}{c}
S^{1/2}\sqrt{(\widetilde{d}_{\mathbf{0}})_{\mathrm{ps}}\chi _{\phi
}^{\dagger }\bar{\Sigma}_{\mathbf{0}}\chi _{\phi }} \text{ \ \ \ \ \ \ \ \ \ \ \ \ \
\ \ \ \ \ \ \ type-I} \\ 
\sqrt{\chi _{\theta }^{\dagger }\bar{\Sigma}_{\mathbf{0}}\chi _{\theta }\chi
_{\phi }^{\dagger }\bar{\Sigma}_{\mathbf{0}}\chi _{\phi }-(\chi _{\theta
}^{\dagger }\bar{\Sigma}_{\mathbf{0}}\chi _{\phi })^{2}} \text{ \ \ type-II}%
\end{array}%
\right. ,  \label{eqs:gap}
\end{equation}%
which can be unified into a compact expression
\begin{equation}
\Delta =\sqrt{\chi _{\theta }^{\dagger }\bar{\Gamma} \chi _{\theta }\chi _{\phi
}^{\dagger }\bar{\Gamma} \chi _{\phi }-(\chi _{\theta }^{\dagger }\bar{\Gamma} \chi
_{\phi })^{2}},\label{eq:gapB}
\end{equation}%
where $\bar{\Gamma} \equiv SM_{\mathbf{0}}+\bar{\Sigma}_{\mathbf{0}}$. For the
type-I PG mode, $\chi _{\theta }^{\dagger }M_{\mathbf{0}}\chi _{\theta }=(\widetilde{d}_{\mathbf{0}})_{\mathrm{ps}}$ and $\chi _{\phi }^{\dagger }M_{\mathbf{0}}\chi _{\phi }=\chi _{\theta }^{\dagger }M_{\mathbf{0}}\chi _{\phi
}=0$ result in $\Delta =S^{1/2}\sqrt{(\widetilde{d}_{\mathbf{0}})_{\mathrm{ps%
}}\chi _{\phi }^{\dagger }\bar{\Sigma}_{\mathbf{0}}\chi _{\phi }}$ to the
leading order $O(S^{1/2})$. For the type-II PG mode, the projection $(\chi
_{\phi },\chi _{\theta })^{\dagger }M_{\mathbf{0}}(\chi _{\phi },\chi
_{\theta })$ of $M_{\mathbf{0}}$ onto the type-II PG subspace is zero,
resulting in $\Delta =\sqrt{\chi _{\theta }^{\dagger }\bar{\Sigma}_{\mathbf{0}%
}\chi _{\theta }\chi _{\phi }^{\dagger }\bar{\Sigma}_{\mathbf{0}}\chi _{\phi
}-(\chi _{\theta }^{\dagger }\bar{\Sigma}_{\mathbf{0}}\chi _{\phi })^{2}}$
to the leading order $O(S^{0})$.

\subsection{Step3: The standard field theoretical approach}\label{sec:FT}
Finally, we prove Eq.~\eqref{eqs:gap} using the standard field
theoretical approach. Expanding the Hamiltonian as $H(\phi_{\alpha }^{0},\theta
_{\alpha }^{0})\sim \sum_{n=0}^{4}S^{2-n/2}\mathcal{H}_{n}$ up to the quartic
order, we consider $\mathcal{H}_{2}$ as the unperturbed part, and treat $\mathcal{H}_{3}+\mathcal{H}_{4}$ perturbatively. By computing the Fourier
transform $G_{\mathbf{k}}(i\omega _{n})=\int_{0}^{1/T}d\tau G_{\mathbf{k}%
}(\tau )e^{i\omega _{n}\tau }$ of the Green function $G_{\mathbf{k}}(\tau
)=-\left\langle \mathcal{T}_{\tau }B_{\mathbf{k}}(\tau )B_{\mathbf{k}
}^{\dagger }(0)\right\rangle $, we can achieve the magnon spectrum from the
poles of $G_{\mathbf{k}}(i\omega_{n})$. Here, $\omega _{n}=2\pi nT$ ($n\in
Z $) is the Matsubara frequency, $\langle X\rangle =\text{Tr}(\rho X)$
denotes the expectation value of an operator $X$ in the thermal equilibrium
state $\rho =e^{-H(\phi _{\alpha }^{0},\theta _{\alpha }^{0})/T}/Z$ with $Z=$Tr$\exp [-H(\phi _{\alpha }^{0},\theta _{\alpha }^{0})/T]$, and $B_{\mathbf{k%
}}(\tau )=e^{H(\phi _{\alpha }^{0},\theta _{\alpha }^{0})\tau }B_{\mathbf{k}%
}e^{-H(\phi _{\alpha }^{0},\theta _{\alpha }^{0})\tau }$.

\begin{figure}	
		\centering		
		\includegraphics[width=0.6\textwidth]{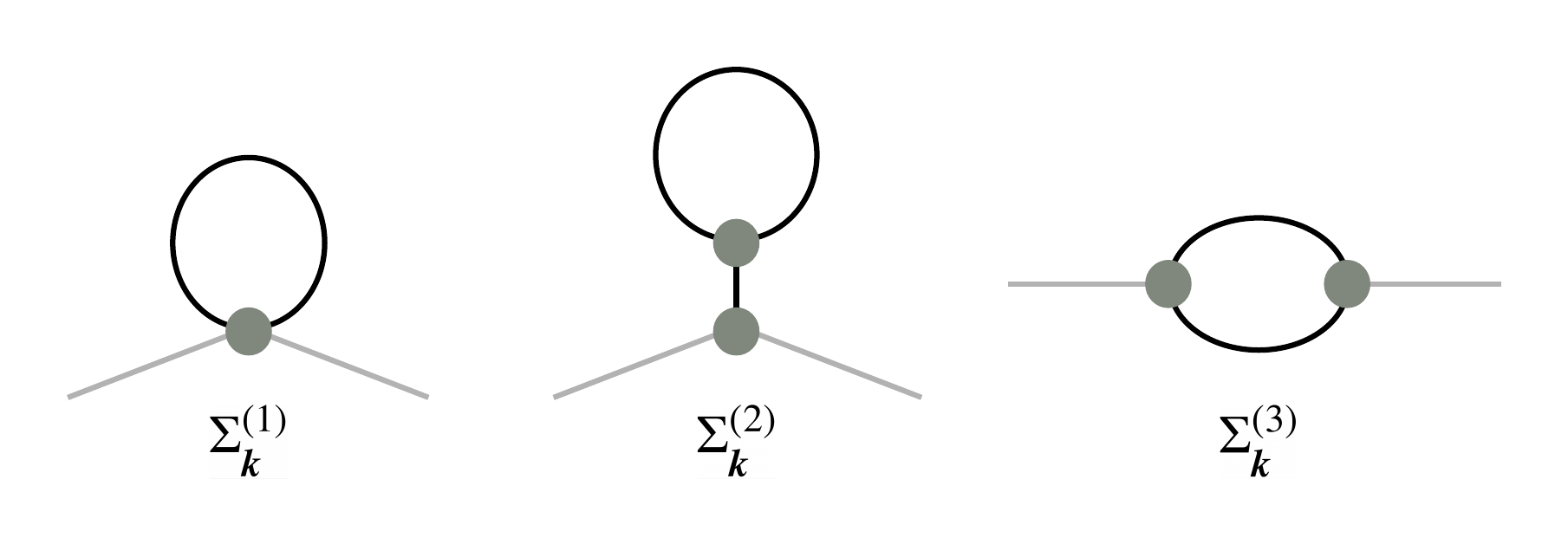}	
		\caption{Three Feynman diagrams contributed to the magnon self-energy.}
		\label{fig:selfenergy}
	\end{figure}

The Dyson expansion gives rise to $G_{\mathbf{k}}^{-1}(i\omega
_{n})=i\omega _{n}\sigma ^{z}\otimes I_{N_{s}}-SM_{\mathbf{k}}-\Sigma _{\mathbf{k}}(i\omega _{n})$, where the self-energy $\Sigma _{\mathbf{k}}(i\omega _{n})$ is obtained from one-particle irreducible diagrams. The leading contribution, $\Sigma _{
\mathbf{k}}(i\omega _{n})=\sum_{L=1}^{3}\Sigma _{\mathbf{k}}^{(L)}(i\omega_{n})$, contains three terms corresponding to the Feynman diagrams in Fig.~\ref{fig:selfenergy}:
the frequency-independent Hartree-Fock term
\begin{equation}
\Sigma_{\mathbf{k},ll'}^{(1)}=\frac{1}{2N_{c}}\sum_{\mathbf{p}}\sum_{l_1,l_2}V_{ll_1l_2l'}^{(4)}(\mathbf{k},\mathbf{p},\mathbf{p}-\mathbf{k})[\langle B_{\mathbf{p}}^{\dagger}B_{\mathbf{p}}\rangle_{0}-\frac{1}{2}{(I-\sigma^z)}\otimes I_{N_s}]_{l_1l_2}
\end{equation}
is contributed from the quartic-magnon interaction $\mathcal{H}_{4}$. The frequency-independent tadpole self-energy
\begin{equation}
\Sigma_{\mathbf{k},ll'}^{(2)}=\frac{-1}{2N_{c}}\sum_{\mathbf{p}}\sum_{l_1,l_2,l_3,l_4}V_{ll'l_1}^{(3)}(\mathbf{k},\mathbf{k})\left(M_{\mathbf{0}}^{-1}\right)_{l_1,l_2}\langle B_{\mathbf{p},l_3}^{\dagger}B_{\mathbf{p},l_4}\rangle\overline{V}_{l_3l_4l_2}^{(3)}(\mathbf{p},\mathbf{p})\label{eq:self2}
\end{equation}
may contribute significantly when the three-body
interaction vertex $\mathcal{H}_{3}\neq0$, where $M_{\mathbf{0}}^{-1}$ is the pseudo-inverse. The frequency-dependent
self-energy
\begin{align}
\Sigma _{\mathbf{k},ll'}^{(3)}(i\omega _{n})=&-\frac{1}{2N_{c}}\sum_{\mathbf{p}}\sum_{l_1,l_2,l_3,l_4,l_5,l_6}\frac{-n_{b}((\sigma^{z}\otimes SD_{\mathbf{p}})_{l_1l_1})+n_{b}((\sigma^{z}\otimes SD_{\mathbf{k}+\mathbf{p}})_{l_2l_2})}{i\omega_n/S+\left(\sigma^{z}\otimes D_{\mathbf{p}}\right)_{l_1l_1}-\left(\sigma^{z}\otimes D_{\mathbf{k}+\mathbf{p}}\right)_{l_2l_2}}\label{eq:self3} \\\nonumber
&\times[(\sigma^{z}\otimes I_{N_s})U_{\mathbf{k}+\mathbf{p}}^{\dagger}]_{l_2l_3}V_{l_3l_4l'}^{(3)}(\mathbf{k}+\mathbf{p},\mathbf{p})(U_{\mathbf{p}})_{l_4 l_1}[(\sigma^{z}\otimes I_{N_s})U_{\mathbf{p}}^{\dagger}]_{l_1l_5}\overline{V}_{l_5l_6l}^{(3)}(\mathbf{p},\mathbf{k}+\mathbf{p})(U_{\mathbf{k}+\mathbf{p}})_{l_6 l_2}
\end{align}
is obtained from the second-order perturbation of $\mathcal{H}_{3}$. Solving det$G_{\mathbf{k}}^{-1}(\omega)=0$, we obtain the magnon spectrum.

The correction of order $1/S$
to the spin-wave spectrum $d_{\mathbf{k},m}$ is obtained within the on-shell approximation \cite{Starykh2006s}. In this approximation, the self-energies are evaluated at the bare magnon
energy $d_{\mathbf{k},m}$, leading to the renormalized spectrum:
\begin{equation}
\omega_{\mathbf{k},m}=d_{\mathbf{k},m}+u_{\mathbf{k},m}^{\dagger}\Sigma_{\mathbf{k}}(d_{\mathbf{k},m})u_{\mathbf{k},m}=d_{\mathbf{k},m}+u_{\mathbf{k},m}^{\dagger}\left[\Sigma_{\mathbf{k}}^{(1)}+\Sigma_{\mathbf{k}}^{(2)}+\Sigma_{\mathbf{k}}^{(3)}(d_{\mathbf{k},m})\right]u_{\mathbf{k},m}.
\end{equation}
Here, $\Sigma_{\mathbf{k}}^{(3)}(d_{\mathbf{k},m})$ contributes to an imaginary part of $\omega_{\mathbf{k},m}$ describing the magnon decay rate.

To obtain the PG gap, we can expand $\Sigma_{\mathbf{k}_0}^{(3)}(i\omega)$
around $\omega=0$ as $\Sigma_{\mathbf{k}_0}^{(3)}(i\omega)=\Sigma_{\mathbf{k}_0}^{(3)}(0)+\omega\partial_{\omega}\Sigma_{\mathbf{k}_0}^{(3)}(i\omega)+O(\omega^{2})$. Since $i\omega _{n}$ always
appears in the denominator of $\Sigma _{\mathbf{k}_0}^{(3)}(i\omega _{n})$, to the leading order $O(S^{0})$, the self-energy $\Sigma _{\mathbf{k}_0}(i\omega
_{n})$ can be approximated by $\Sigma _{\mathbf{k}_0}(0)=\sum_{L=1}^{3}\Sigma
_{\mathbf{k}_0}^{(L)}(0)$.
Because the correction to the PG gap from other magnon bands is of higher order $O(S^{-1})$, we project the dispersion matrix $M_{\mathbf{k}_0}$ and the self-energy $\Sigma_{\mathbf{k}_0}(0)$ into the subspace of the PG mode spanned by $u_{{\mathbf{k}_0},{\mathrm{ps}}}$ and $v_{{\mathbf{k}_0},{\mathrm{ps}}}$. The PG gap is obtained by diagonalizing the matrix
\begin{equation}
\omega\sigma^{z}-(u_{{\mathbf{k}_0},{\mathrm{ps}}},v_{{\mathbf{k}_0},{\mathrm{ps}}})^{\dagger}[S M_{\mathbf{k}_0}+\Sigma_{\mathbf{k}_0}(0)](u_{{\mathbf{k}_0},{\mathrm{ps}}},v_{{\mathbf{k}_0},{\mathrm{ps}}}).    
\end{equation}
The positive eigenvalue
\begin{equation}
    \Delta=\mathrm{Im}(\chi_{\mathbf{k}_0,\phi}^{\dagger}\Gamma_{\mathbf{k}_0}\chi_{\mathbf{k}_0,\theta})+\sqrt{\chi_{\mathbf{k}_0,\theta}^{\dagger}\Gamma_{\mathbf{k}_0} \chi_{\mathbf{k}_0,\theta}\chi_{\mathbf{k}_0,\phi}^{\dagger}\Gamma_{\mathbf{k}_0}\chi_{\mathbf{k}_0,\phi}-\mathrm{Re}(\chi_{\mathbf{k}_0,\theta}^{\dagger}\Gamma_{\mathbf{k}_0}\chi_{\mathbf{k}_0,\phi})^{2}}\label{eq:delta}
\end{equation}
is the PG gap, where $\Gamma_{\mathbf{k}_0}\equiv S M_{\mathbf{k}_0}+\Sigma_{\mathbf{k}_0}(0)$. For the Hamiltonian Eq.~\eqref{eqs:H} with the inversion symmetry, we can prove $\mathrm{Im}(\chi_{\mathbf{k}_0,\phi}^{\dagger}\Gamma_{\mathbf{k}_0}\chi_{\mathbf{k}_0,\theta})=0$ using $\Gamma_{\mathbf{k}_0}=\Gamma_{-\mathbf{k}_0}$ and $(u_{\mathbf{k}_0})_{\mathrm{ps}}=(u_{-\mathbf{k}_0})_{\mathrm{ps}}$.

For the type-I PG mode, the dispersion matrix $M_{\mathbf{k}_0}$ in the PG-mode subspace is given by Eq.~\eqref{ND}, i.e.,
\begin{equation}
(\chi_{\mathbf{k}_0,\phi},\chi_{\mathbf{k}_0,\theta})^{\dagger}M_{\mathbf{k}_0}(\chi_{\mathbf{k}_0,\phi},\chi_{\mathbf{k}_0,\theta})=\frac{1}{2}(\widetilde{d}_{\mathbf{k}_0})_{\mathrm{ps}}(I_2-\sigma^z).
\end{equation}
 It follows from Eq.~\eqref{eq:delta} that
\begin{equation}
\Delta=\sqrt{S\chi_{\mathbf{k}_0,\theta}^{\dagger}M_{\mathbf{k}_0} \chi_{\mathbf{k}_0,\theta}\chi_{\mathbf{k}_0,\phi}^{\dagger}\Sigma_{\mathbf{k}_0}(0)\chi_{\mathbf{k}_0,\phi}}=\sqrt{S(\widetilde{d}_{\mathbf{k}_0})_{\mathrm{ps}}\chi_{\mathbf{k}_0,\phi}^{\dagger}\Sigma_{\mathbf{k}_0}(0)\chi_{\mathbf{k}_0,\phi}}.
\end{equation}
For the type II PG mode, the dispersion matrix becomes zero in the PG-mode subspace, as shown in Eq.~\eqref{D}. Thus, according to Eq.~\eqref{eq:delta},
the type-II PG gap is
\begin{equation}
    \Delta=\mathrm{Im}(\chi_{\mathbf{k}_0,\phi}^{\dagger}\Sigma_{\mathbf{k}_0}(0)\chi_{\mathbf{k}_0,\theta})+\sqrt{\chi_{\mathbf{k}_0,\theta}^{\dagger}\Sigma_{\mathbf{k}_0}(0)\chi_{\mathbf{k}_0,\theta}\chi_{\mathbf{k}_0,\phi}^{\dagger}\Sigma_{\mathbf{k}_0}(0)\chi_{\mathbf{k}_0,\phi}-\mathrm{Re}(\chi_{\mathbf{k}_0,\theta}^{\dagger}\Sigma_{\mathbf{k}_0}(0)\chi_{\mathbf{k}_0,\phi})^{2}}.\label{eq:type2gap}
\end{equation}

For PG\ modes with $\mathbf{k}_0=\mathbf{0}$, $\mathrm{Im}(\chi_{\phi}^{\dagger}\Gamma_{\mathbf{0}}\chi_{\theta})=0$ follows from $\Gamma_{\mathbf{0}}=(\sigma^{x}\otimes I_{N_s})\Gamma_{\mathbf{0}}^T(\sigma^{x}\otimes I_{N_s})$ and 
$(v_{\mathbf{0}})_{\mathrm{ps}}=(\sigma^{x}\otimes I_{N_s})(u_{\mathbf{0}})_{\mathrm{ps}}^{*}$. Thus, the PG gap $\Delta$ in Eq.~\eqref{eq:delta} is simplified as 
\begin{equation} \Delta=\sqrt{\chi_{\theta}^{\dagger}\Gamma_{\mathbf{0}} \chi_{\theta}\chi_{\phi}^{\dagger}\Gamma_{\mathbf{0}}\chi_{\phi}-(\chi_{\theta}^{\dagger}\Gamma_{\mathbf{0}}\chi_{\phi})^{2}}.\label{eq:gapC}
\end{equation}
A direct comparison shows $\Sigma _{\mathbf{0}}(0)=\bar{\Sigma}_{\mathbf{0}}$. As a result, the PG gap in Eq.~\eqref{eq:gapC} agrees with the PG gap obtained using the curvature formula, i.e., Eq.~\eqref{eq:gapB}. The theorem, i.e. Eq.~\eqref{eqs:cur}, is proved.

\subsection{The curvature formula at finite momenta}
For a spin system with inversion symmetry, following the same procedure, we can prove that the PG gap at finite momentum is also determined by Eq.~\eqref{eqs:cur}.
The Hermitian operators $O_{\zeta,\eta }^{(\nu)}$ in $\hat{R}(\phi,\theta)$, obtained by imposing the invariance of $E_c$, is different from those in the zero-momentum case. In the following, we show the relevant terms that determine the PG gap.

The canonical position and momentum operators are 
\begin{align}    
O_{1,0}^{(0)}=\frac{1}{\sqrt{2}}(B_{\mathbf{k}_0}^{\dagger}\sigma^{z}\chi _{\mathbf{k}_0,\phi}+\rm{H.c.}), \\
O_{0,1}^{(0)}=\frac{1}{\sqrt{2}}(B_{\mathbf{k}_0}^{\dagger}\sigma^{z}\chi _{\mathbf{k}_0,\theta}+\rm{H.c.}),
\end{align} 
which satisfy the commutation relation $[O_{1,0}^{(0)},O_{0,1}^{(0)}]=i$.
The bilinear operator 
\begin{equation}
O_{2,0}^{(0)}=\frac{1}{2}\chi_{\mathbf{k}_0,\phi}^{\dagger}\Omega_{\mathbf{k}_0}\chi_{\mathbf{k}_0,\phi}+\frac{1}{4}(\chi_{\mathbf{k}_0,\phi}^{\dagger}\bar{\Omega}_{\mathbf{k}_0}\chi_{-\mathbf{k}_0,\phi}+\rm{H.c.})
\end{equation}
generates the nonlinear dependence on $\phi$ in $
\omega_{j,\alpha }\mathbf{n}_{j,\alpha }$, where
\begin{align}
    \Omega _{\mathbf{k}_0,l_{1}l_{2}}&=i \frac{1}{\sqrt{N_{c}}} \sum_{l_{3}}V_{l_{1}l_{2}l_{3}}^{(3)}(\mathbf{k}_0,\mathbf{k}_0)\left(M_{\mathbf{0}}^{-1}(\sigma^{z}\otimes I_{N_s})B_{\mathbf{0}}\right)_{l_{3}},\\
    \bar{\Omega} _{\mathbf{k}_0,l_{1}l_{2}}&=i\frac{1}{\sqrt{N_{c}}}\sum_{l_{3}}V_{l_{1}l_{2}l_{3}}^{(3)}(\mathbf{k}_0,-\mathbf{k}_0)\left(M_{2\mathbf{k}_0}^{-1}(\sigma^{z}\otimes I_{N_s})B_{2\mathbf{k}_0}\right)
    _{l_{3}},    
\end{align}
and $M_{\mathbf{0}(2\mathbf{k}_0)}^{-1}$ is the pseudo-inverse. The bilinear operators 
\begin{align}
    O_{0,2}^{(0)}&=\frac{1}{2}\chi_{\mathbf{k}_0,\theta }^{\dagger }\Omega_{\mathbf{k}_0} \chi_{\mathbf{k}_0,\theta }+\frac{1}{4}(\chi_{\mathbf{k}_0,\theta}^{\dagger}\bar{\Omega}_{\mathbf{k}_0}\chi_{-\mathbf{k}_0,\theta}+\rm{H.c.}),\\
    O_{1,1}^{(0)}&=\frac{1}{2}(\chi _{\mathbf{k}_0,\theta }^{\dagger}\Omega_{\mathbf{k}_0} \chi_{\mathbf{k}_0,\phi }+\rm{H.c.})+\frac{1}{4}(\chi_{\mathbf{k}_0,\theta}^{\dagger}\bar{\Omega}_{\mathbf{k}_0}\chi_{-\mathbf{k}_0,\phi}+\chi_{\mathbf{k}_0,\phi}^{\dagger}\bar{\Omega}_{\mathbf{k}_0}\chi_{-\mathbf{k}_0,\theta}+\rm{H.c.})
\end{align}
are relevant only for type-II PG modes.

Next, we represent the Hamiltonian $H(\phi,\theta)$ in terms of $\mathcal{H}_{n}$ using Eq.~\eqref{eq:h22}. Following Eq.~\eqref{eq:Fphi}, we find $\mathcal{F}_{\phi }=S\chi _{\mathbf{k}_0,\phi }^{\dagger }\bar{\Sigma}_{\mathbf{k}_0}\chi _{\mathbf{k}_0,\phi }$ with $\bar{\Sigma}_{\mathbf{k}_0}={\Sigma}_{\mathbf{k}_0}(0)$. For the type-I PG mode, only $O_{0,1}^{(0)}$ contributes to the leading
order, and we have $\left( \partial ^{2}F/\partial \theta ^{2}\right) _{0}=S^{2}(\widetilde{d}_{\mathbf{k}_0})_{\mathrm{ps}}$ and $(\partial ^{2}F/\partial
\theta \partial \phi )_{0}=0$. For the type-II PG mode, we obtain $(\partial
^{2}F/\partial \theta ^{2})_{0}=S\chi _{\mathbf{k}_0,\theta }^{\dagger }\bar{\Sigma}_{\mathbf{k}_0}\chi _{\mathbf{k}_0,\theta }$ and $(\partial ^{2}F/\partial \theta \partial
\phi )_{0}=S\chi _{\mathbf{k}_0,\theta }^{\dagger }\bar{\Sigma}_{\mathbf{k}_0}\chi _{\mathbf{k}_0,\phi }$%
, completely determined by the same matrix $\bar{\Sigma}_{\mathbf{k}_0}$.
Finally, Eq.~\eqref{eq:delta} reduces to the curvature formula Eq.~\eqref{eqs:cur}, thus, the theorem for PG modes with non-zero momentum is proved.

For spin systems without inversion symmetry, the type-I PG gap at leading order remains governed by Eq.~\eqref{eqs:cur}. The type-II PG gap at finite momentum is given by
\begin{equation}
\Delta=\frac{1}{S}\left(%
\frac{\partial^{2}F'}{\partial\theta\partial\phi}\right)_{0}+\frac{1}{S}\sqrt{\left(\frac{\partial^{2}F}{\partial\theta^{2}}%
\right)_{0}\left(\frac{\partial^{2}F}{\partial\phi^{2}}\right)_{0}-\left(%
\frac{\partial^{2}F}{\partial\theta\partial\phi}\right)_{0}^{2}}\label{eq:Ck},
\end{equation}
where the second term follows the same procedure discussed above. We therefore focus on the first term, $(\partial ^{2}F'/\partial \theta \partial\phi )_{0}/S$, which originates from inversion symmetry breaking.
Here, $F'(\phi ,\theta)=-T\ln Z'(\phi,\theta)$ is obtained from the partition function $Z'(\phi,\theta)=\text{Tr}\exp (-H_{R}'(\phi ,\theta)/T)$. The quadratic Hamiltonian $H_R'(\phi ,\theta)$ is derived by truncating the expansion
\begin{equation}
 \hat{R'}^{\dagger}(\phi ,\theta)H(\phi _{\alpha }^{0},\theta _{\alpha }^{0})%
\hat{R'}(\phi ,\theta)=\sum_{n=0}^{\infty }S^{2-n/2}\mathcal{H}_{n}'(\phi,\theta)   
\end{equation}
at $n=2$. Under the rotation $\hat{R'}(\phi ,\theta)$, the classical ground-state energy remains invariant. We expand the generators of $\hat{R'}(\phi ,\theta)$ in powers of $\phi,\theta$ and $S$ as 
\begin{equation}
 \hat{R}'(\phi ,\theta)=\exp \left( -i\sum_{\zeta ,\eta ,\nu =0}^{\infty
}\bar{O}_{\zeta ,\eta }^{(\nu )}\phi ^{\zeta }\theta ^{\eta }S^{1/2-\nu }\right),   
\end{equation}
where $\bar{O}_{\zeta ,\eta }^{(\nu )}$ can be directly obtained by replacing $\chi_{\pm\mathbf{k}_0,\theta}$ with $\pm i \chi_{\pm\mathbf{k}_0,\theta}$ in ${O}_{\zeta ,\eta }^{(\nu )}$. For instance, 
\begin{align}
  \bar{O}_{1,0}^{(0)}&={O}_{1,0}^{(0)}=\frac{1}{\sqrt{2}}(B_{\mathbf{k}_0}^{\dagger}\sigma^{z}\chi _{\mathbf{k}_0,\phi}+\rm{H.c.}),\\
  \bar{O}_{0,1}^{(0)}&=\frac{1}{\sqrt{2}}(i B_{\mathbf{k}_0}^{\dagger}\sigma^{z}\chi _{\mathbf{k}_0,\theta}+\rm{H.c.}),\nonumber\\
  \bar{O}_{2,0}^{(0)}&={O}_{2,0}^{(0)}=\frac{1}{2}\chi_{\mathbf{k}_0,\phi}^{\dagger}\Omega_{\mathbf{k}_0}\chi_{\mathbf{k}_0,\phi}+\frac{1}{4}(\chi_{\mathbf{k}_0,\phi}^{\dagger}\bar{\Omega}_{\mathbf{k}_0}\chi_{-\mathbf{k}_0,\phi}+\rm{H.c.}),\nonumber\\
  \bar{O}_{0,2}^{(0)}&={O}_{0,2}^{(0)}=\frac{1}{2}\chi_{\mathbf{k}_0,\theta }^{\dagger }\Omega_{\mathbf{k}_0} \chi_{\mathbf{k}_0,\theta }+\frac{1}{4}(\chi_{\mathbf{k}_0,\theta}^{\dagger}\bar{\Omega}_{\mathbf{k}_0}\chi_{-\mathbf{k}_0,\theta}+\rm{H.c.}),\nonumber\\
  \bar{O}_{1,1}^{(0)}&=\frac{1}{2}(i\chi _{\mathbf{k}_0,\phi }^{\dagger}\Omega_{\mathbf{k}_0} \chi_{\mathbf{k}_0,\theta }+\rm{H.c.})+\frac{1}{4}(-i\chi_{\mathbf{k}_0,\theta}^{\dagger}\bar{\Omega}_{\mathbf{k}_0}\chi_{-\mathbf{k}_0,\phi}-i\chi_{\mathbf{k}_0,\phi}^{\dagger}\bar{\Omega}_{\mathbf{k}_0}\chi_{-\mathbf{k}_0,\theta}+\rm{H.c.}).\nonumber
\end{align}

By reformulating the second order derivative  $(\partial ^{2}F'/\partial \theta \partial\phi )_{0}$ via $\mathcal{H}_{n}$, we rigorously establish that 
\begin{equation}
  \frac{1}{S}\left(%
\frac{\partial^{2}F'}{\partial\theta\partial\phi}\right)_{0}=\mathrm{Im}(\chi_{\mathbf{k}_0,\phi}^{\dagger}\Sigma_{\mathbf{k}_0}(0)\chi_{\mathbf{k}_0,\theta}), 
\end{equation}
thereby completing the proof of the theorem for type-II PG modes with non-zero momentum.

\section{The triangular lattice XXZ model}
\label{AppXXZ}
In this Appendix, we apply the theorem to study the PG mode in the antiferromagnetic XXZ model on the triangular lattice at finite temperature, described by the Hamiltonian
\begin{equation}
H_{XXZ}=\sum_{\langle \mathbf{r},\mathbf{r}'\rangle}\hat{S}_{\mathbf{r}}^{x}\hat{S}_{\mathbf{r}'}^{x}+\hat{S}_{\mathbf{r}}^{y}\hat{S}_{\mathbf{r}'}^{y}+\kappa\hat{S}_{\mathbf{r}}^{z}\hat{S}_{\mathbf{r}'}^{z},
\end{equation}
where $\kappa>0$ and the sum $\sum_{\langle \mathbf{r},\mathbf{r}'\rangle}$ runs over all nearest-neighbor pairs. The basis vectors $\delta r=(1,0,0)$, $(-1/2,-\sqrt{3}/2,0)$ and $(-1/2,\sqrt{3}/2,0)$ connect the nearest neighbor sites.
The total $z$ component of spin, $\sum_{\mathbf{r}}\hat{S}_{\mathbf{r
}}^{z}$, commutes with $H_{XXZ}$, generating a $U(1)$ symmetry. For $\kappa
<1$, the classical spins arrange in a three-sublattice $120^{\circ }$
configuration in the $x$-$y$ plane, and the spin system has a single
Goldstone mode due to spontaneous breaking of the $U(1)$ symmetry. When $\kappa =1$, the
model corresponds to the well-known Heisenberg antiferromagnet, exhibiting the
$120^{\circ }$ order. In this case, the system has three Goldstone modes arising
from the spontaneous breaking of three global spin-rotation symmetries $\sum_{\mathbf{r}}\hat{S}%
_{\mathbf{r}}^{\mu=x,y,x }$. 
In this paper, we focus only on the regime where $\kappa\geq1$.

\subsection{Classical ground-state configuration}
\begin{figure}	
		\centering		
		\includegraphics[width=0.5\textwidth]{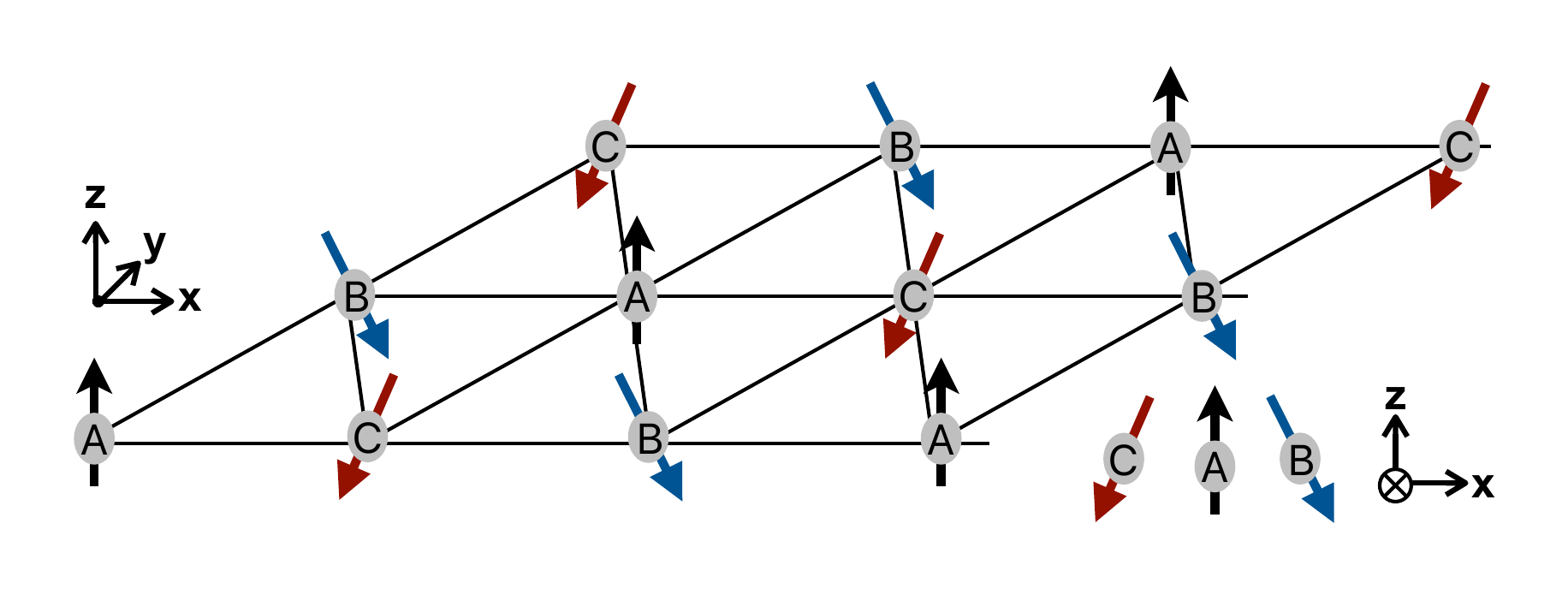}	
		\caption{Three sublattice structure of the triangular lattice XXZ antiferromagnet .}
		\label{fig:XXZ}
	\end{figure}

For a three-sublattice structure [cf. Fig.~\ref{fig:XXZ}], the classical
ground-state configuration is characterized by three polar angles $\phi_{\alpha}^c$ and three azimuthal angles $\theta_{\alpha}^c$, with $\alpha=A,B,C$.
The classical energy
\begin{align}
E_c=S^{2}\mathcal{H}_{0} & =NS^{2}[\sin(\theta_{A}^c)\sin(\theta_{B}^c)\cos(\phi_{A}^c-\phi_{B}^c)+\sin(\theta_{A}^c)\sin(\theta_{C}^c)\cos(\phi_{A}^c-\phi_{C}^c)+\sin(\theta_{B}^c)\sin(\theta_{C}^c)\cos(\phi_{B}^c-\phi_{C}^c)\nonumber \\
 & +\kappa(\cos(\theta_{A}^c)\cos(\theta_{B}^c)+\cos(\theta_{A}^c)\cos(\theta_{C}^c)+\cos(\theta_{B}^c)\cos(\theta_{C}^c))]
\end{align}
is invariant under global rotations along the $z$-direction, thus exhibiting $U(1)$
symmetry. Due to the
antiferromagnetic coupling, the mean-field solution is always coplanar. Without loss of generality, the spin configuration in the ground state can be set in the
$x$-$z$ plane, i.e., $\phi_{A}^c=\phi_{B}^c=\phi_{C}^c=0$.
The classical energy depends on the three angles $(\theta_{A}^c,\theta_{B}^c,\theta_{C}^c)$.

The polar angles can be obtained by minimizing the ground state energy, which leads to the saddle-point condition
\begin{equation}
\frac{\partial E_c}{\partial\theta_{A}^c}=\frac{\partial E_c}{\partial\theta_{B}^c}=\frac{\partial E_c}{\partial\theta_{C}^c}=0.\label{SP}
\end{equation}
Solving Eq.~\eqref{SP}, we obtain
\begin{align}
\cos(\theta_{B}^c) & =\kappa g_{1}(-\cos(\theta_{A}^c)-g_{2}\sin(\theta_{A}^c)),\nonumber \\
\sin(\theta_{B}^c) & =g_{1}(-\kappa^{2}\sin(\theta_{A}^c)+g_{2}\cos(\theta_{A}^c)),\nonumber \\
\cos(\theta_{C}^c) & =\kappa g_{1}(-\cos(\theta_{A}^c)+g_{2}\sin(\theta_{A}^c)),\nonumber \\
\sin(\theta_{C}^c) & =g_{1}(-\kappa^{2}\sin(\theta_{A}^c)-g_{2}\cos(\theta_{A}^c)),\label{eq:thetaBC}
\end{align}
where 
\begin{align}
g_{1}(\kappa,\theta_{A}^c)& =\frac{1}{(1+\kappa)(\kappa^{2}+(1-\kappa^{2})\cos^{2}(\theta_{A}^c))},\\
g_{2}(\kappa,\theta_{A}^c)& =\sqrt{(1+\kappa)^{2}(\kappa^{2}+(1-\kappa^{2})\cos^{2}(\theta_{A}^c))-\kappa^{2}}.
\end{align}
Under the condition Eq.~\eqref{eq:thetaBC}, the classical ground-state energy $E_c=-NS^{2}(\kappa^{2}+\kappa+1)/(1+\kappa)$ is degenerate for arbitrary $\theta_{A}^c$. The accident degeneracy indicates the presence of an approximate symmetry at the
classical level. Different classical configurations are connected by the rotation $\delta U_R=\exp[i\int_{0}^{\delta\theta_{A}}d\theta_{A}^c(\hat{S}_{\mathbf{r}_{j,A}}^{y}+i\partial_{\theta_{A}^c}\theta_{B}^c\hat{S}_{\mathbf{r}_{j,B}}^{y}+i\partial_{\theta_{A}^c}\theta_{C}^c\hat{S}_{\mathbf{r}_{j,C}}^{y})].$ in the $x-z$ plane.

\subsection{Equilibrium state configurations}
Since $\delta U_R$ is not a true symmetry of the system, the accidental degeneracy at the classical level is lifted when quantum and thermal fluctuations are included. In this subsection, we study equilibrium state configurations, incorporating the effects of these fluctuations.

The transformed triangular XXZ Hamiltonian under $\hat{U}_R(\phi_{\alpha}^c,\theta_{\alpha}^c)$ reads
\begin{align}
H_{XXZ}(\phi_{\alpha}^c,\theta_{\alpha}^c) & =\sum_{j,\delta r}[H_{\mathbf{r}_{j,A},\mathbf{r}_{j,A}+\delta r}(\theta_{A}^c,\theta_{B}^c)+H_{\mathbf{r}_{j,A},\mathbf{r}_{j,A}-\delta r}(\theta_{A}^c,\theta_{C}^c)+H_{\mathbf{r}_{j,B},\mathbf{r}_{j,B}+\delta r}(\theta_{B}^c,\theta_{C}^c)],
\end{align}
where
\begin{equation}
H_{\mathbf{r},\mathbf{r}'}(\theta_{A}^c,\theta_{B}^c)=\left(\begin{array}{ccc}
\hat{S}_{\mathbf{r}}^{x} & \hat{S}_{\mathbf{r}}^{y} & \hat{S}_{\mathbf{r}}^{z}\end{array}\right)\left(\begin{array}{ccc}
\cos(\theta_{A}^c)\cos(\theta_{B}^c)+\kappa\sin(\theta_{A}^c)\sin(\theta_{B}^c) & 0 & \cos(\theta_{A}^c)\sin(\theta_{B}^c)-\kappa\sin(\theta_{A}^c)\cos(\theta_{B}^c)\\
0 & 1 & 0\\
\sin(\theta_{A}^c)\cos(\theta_{B}^c)-\kappa\cos(\theta_{A}^c)\sin(\theta_{B}^c) & 0 & \sin(\theta_{A}^c)\sin(\theta_{B}^c)+\kappa\cos(\theta_{A}^c)\cos(\theta_{B}^c)
\end{array}\right)\left(\begin{array}{c}
\hat{S}_{\mathbf{r}'}^{x}\\
\hat{S}_{\mathbf{r}'}^{y}\\
\hat{S}_{\mathbf{r}}^{z}
\end{array}\right).
\end{equation}

Applying the HP transformation to $H_{XXZ}(\phi_{\alpha}^c,\theta_{\alpha}^c)$, we obtain the quadratic Hamiltonian
\begin{equation}
\mathcal{H}_{2}(\phi_{\alpha}^c,\theta_{\alpha}^c)=S\mathcal{H}_{0}+\frac{S}{2}\sum_{k}B_{k}^{\dagger}M_{k}B_{k}
\end{equation}
at order $O(S)$, where in the Nambu representation $B_{k}=(b_{k,A}, b_{k,B}, b_{k,C}, b_{-k,A}^{\dagger}, b_{-k,B}^{\dagger}, b_{-k,C}^{\dagger})^T$, and the dispersion matrix is $M_{k}=\left(\begin{array}{cc}
\varepsilon_{k} & \Delta_{k}\\
\Delta_{k}^{\dagger} & \varepsilon_{-k}^{\mathrm{T}}
\end{array}\right)$, and
\begin{align}
\varepsilon_{k} & =\frac{1}{2}\left(\begin{array}{ccc}
\varepsilon_{k}^{A,B,C} & \varepsilon_{k}^{A,B} & \left(\varepsilon_{k}^{A,C}\right)^{*}\\
\left(\varepsilon_{k}^{A,B}\right)^{*} & \varepsilon_{k}^{B,C,A} & \varepsilon_{k}^{B,C}\\
\varepsilon_{k}^{A,C} & \left(\varepsilon_{k}^{B,C}\right)^{*} & \varepsilon_{k}^{C,A,B}
\end{array}\right),
\end{align}
\begin{align}
\Delta_{k} & =\frac{1}{2}\left(\begin{array}{ccc}
0 & \Delta_{k}^{A,B} & \left(\Delta_{k}^{A,C}\right)^{*}\\
\left(\Delta_{k}^{A,B}\right)^{*} & 0 & \Delta_{k}^{B,C}\\
\Delta_{k}^{A,C} & \left(\Delta_{k}^{B,C}\right)^{*} & 0
\end{array}\right).
\end{align}
The matrix elements are
\begin{align}
\varepsilon_{k}^{A,B,C} & =-6\left(\kappa\cos(\theta_{A}^c)\left(\cos(\theta_{B}^c)+\cos(\theta_{C}^c)\right)+\sin(\theta_{A}^c)\left(\sin(\theta_{B}^c)+\sin(\theta_{C}^c)\right)\right),\\
\varepsilon_{k}^{A,B} & =\left(\kappa\sin(\theta_{A}^c)\sin(\theta_{B}^c)+\cos(\theta_{A}^c)\cos(\theta_{B}^c)+1\right)f_{k},\\
\Delta_{k}^{A,B} & =\left(\kappa\sin(\theta_{A}^c)\sin(\theta_{B}^c)+\cos(\theta_{A}^c)\cos(\theta_{B}^c)-1\right)f_{k},
\end{align}
where $f_{k}=\sum_{\delta r}e^{ik\cdot \delta r}=e^{ik_{x}}+2e^{-ik_{x}/2}\cos(\sqrt{3}k_{y}/2)$.

For $k\neq0$, $M_{k}$ is diagonalized as $U_{k}^{\dagger}M_{k}U_{k}=\mathrm{diag}(D_{k},D_{-k})$ by a Bogoliubov transformation
$U_{k}$ satisfying $U_{k}(\sigma^{z}\otimes I_3)U_{k}^{\dagger}=(\sigma^{z}\otimes I_3)$, where the corresponding eigenvectors to eigenvalues $d_{k,m}$
and $d_{-k,m}$ are $u_{k,m}$
and $v_{-k,m}=\sigma^{x}u_{-k,m}^{*}$, i.e., the columns of $U_{k}$, and $m$ labels the three magnon bands. The inversion symmetry of the model implies $d_{-k}=d_{k}$.

We introduce the Bogoliubov operators $\gamma_{k,m}=u_{k,m}^{\dagger}\sigma^{z}B_{k}$ 
to describe the annihilation of magnons with wave vector $k$. The Hamiltonian at order $O(S)$ becomes
\begin{equation}
H_{XXZ}(\phi_{\alpha}^c,\theta_{\alpha}^c)=S(S+1)\mathcal{H}_{0}+E_{\mathrm{zp}}(\theta_{A}^c)+S\sum_{k\neq0,m}d_{k,m}(\theta_{A}^c)\gamma_{k,m}^{\dagger}\gamma_{k,m}+\mathcal{H}_{2,k=0}(\theta_{A}^c)+O(\frac{1}{\sqrt{S}}),
\label{eq:Hxxz2}
\end{equation} 
where $E_{\mathrm{zp}}(\theta_{A}^c)$ is the quantum zero point energy, and $d_{k,m}(\theta_{A}^c)$ is the dispersion relation of the magnon.
The term $\mathcal{H}_{2,k=0}(\theta_{A}^c)$ describes magnon excitations with zero momentum. There are two possibilities: (a) for $\theta_{A}^c \neq\theta_{A}^{0}\in\left\{ 0,\pi,\arccos\left(\pm\kappa/(1+\kappa)\right)\right\}$
\begin{equation}
\mathcal{H}_{2,k=0}(\theta_{A}^c)=\frac{1}{4}{S}\widetilde{d}_{t}(\theta_{A}^c)(\gamma_{t}^{\dagger}+\gamma_{t})^{2}+S\sum_{j=1,2}d_{gj}(\theta_{A}^c)\gamma_{gj}^{\dagger}\gamma_{gj}, \label{t0}
\end{equation}
where one zero mode and two gapped modes with energies $S d_{gj=1,2}$ appear;
(b) For $\theta_{A}^c=\theta_{A}^{0}$,
\begin{equation}
    \mathcal{H}_{2,k=0}(\theta_{A}^{0})=Sd_{g1}\gamma_{g1}^{\dagger}\gamma_{g1}+\frac{1}{4}S\widetilde{d}_{t}(\gamma_{t}^{\dagger}+\gamma_{t})^{2}+\frac{1}{4}S(\widetilde{d}_{0})_{\mathrm{ps}}(\gamma_{\mathrm{ps}}^{\dagger}+\gamma_{\mathrm{ps}})^{2}, \label{tc}
\end{equation}
where only one mode is gapped with the energy $S d_{g1}$. From Eqs.~\eqref{t0} and \eqref{tc}, we find that the true Goldstone mode (denoted by $\gamma_{t}$ and $\gamma_{t}^{\dagger}$) exists for any classical contribution, while the PG mode (denoted by $\gamma_{\mathrm{ps}}$ and $\gamma_{\mathrm{ps}}^{\dagger}$) only appears at some specific angles $\theta_{A}^{0}$. This implies that the PG mode is unstable under the rotation along the $y$-axis.

By incorporating the zero-point energy $E_{\mathrm{zp}}(\theta_{A}^c)$, we obtain the ground-state energy
\begin{equation}
E_{g}(\theta_{A}^c)=S(S+1)\mathcal{H}_{0}+\frac{S}{2}\sum_{k\neq0,m}d_{k,m}(\theta_{A}^c)+\sum_{j=1,2}\frac{S}{2}d_{gj}(\theta_{A}^c)
\end{equation}
for $\theta_{A}^c \neq\theta_{A}^{0}$, and
\begin{equation}
E_{g}(\theta_{A}^{0})=S(S+1)\mathcal{H}_{0}+\frac{S}{2}\sum_{k\neq0,m}d_{k,m}(\theta_A^{0})+\frac{S}{2}d_{g1}
\end{equation}
for $\theta_{A}^c=\theta_{A}^{0}$. Minimizing the ground-state energy with respect to $\theta_{A}^c$, we find that the minimum is achieved at $\theta_{A}^c=\theta_{A}^{0}$. 

\begin{figure}	
		\centering		
		\includegraphics[width=0.8\textwidth]{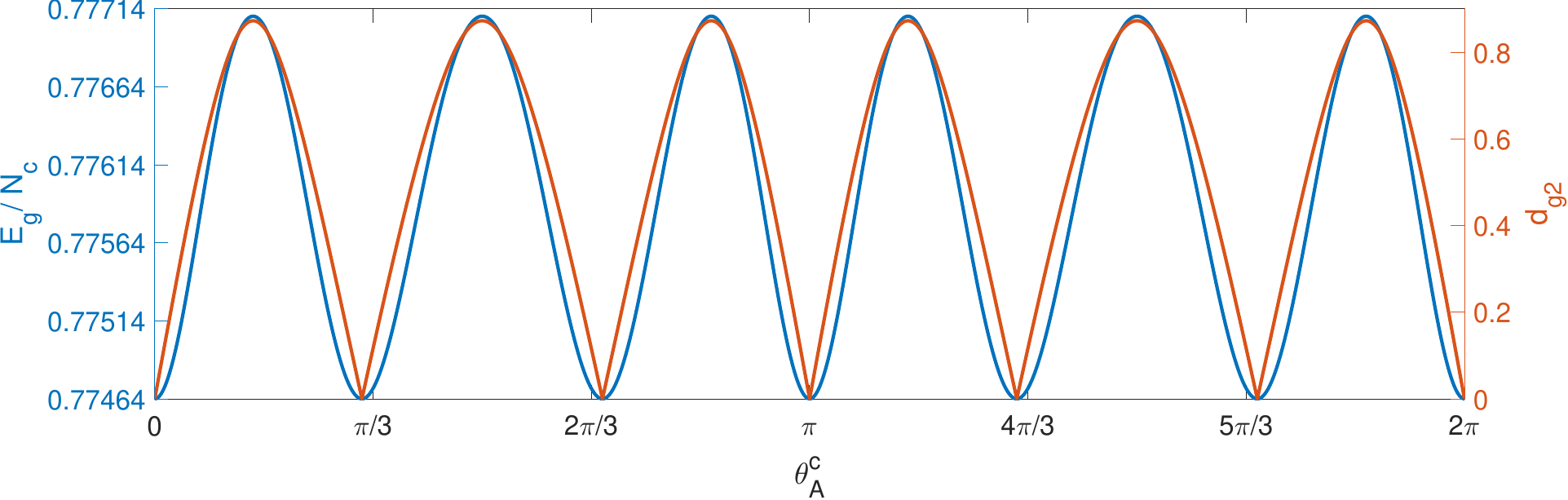}
		\caption{The ground-state energy density $E_{\mathrm{g}}/N_c$ and the spin-wave spectrum $d_{g2}$ vs $\theta_{A}$ ($\kappa=1.2$ and $S=0.5$).}
		\label{fig:Equds}
	\end{figure}

To get an explicit picture of the energy landscape, in Fig.~\ref{fig:Equds}, we show $E_g(\theta_A^{c})$ and $d_{g2}(\theta_A^{c})$. Here, the gap closes when the minimum of $E_g(\theta_A^{c})$ is reached. In the vicinity of $\theta_{A}=\theta_{A}^{0}$, the ground state energy can be expanded to the second order of $\delta \theta_A=\theta_A-\theta_{A}^{0}$ as: $E_{g}(\theta_{A})\simeq E_{g}(\theta_{A}^{0})+{\partial_{\theta_{A}}^{2}E_{g}(\theta_{A}^{0})}\delta\theta_{A}^{2}/2$. Without loss of generality, we break the $12$-fold symmetry and choose $\theta _{A}^{0}=0$, $\theta
_{B}^{0}=\arccos(-\kappa /(1+\kappa ))$, $\theta _{C}^{0}=2\pi -\theta
_{B}^{0}$, and $\phi_{\alpha }^{0}=0$ as the ground-state configuration, representing a coplanar order as shown in Fig.~\ref{fig:XXZ}. The remaining $11$
configurations can be obtained by permuting the three sublattices and applying reflection symmetry with respect to the $x$-$y$ plane. At finite temperature, minimizing the free energy $F(\theta_A^c)$ in Eq.~\eqref{eq:F2} gives rise to the same equilibrium state configurations $(\mathbf{\phi}_{\alpha}^{0}=0,\mathbf{\theta}_{\alpha}^{0})$.

For further analysis, we show the explicit form of $d_{g1}={3\sqrt{2-2\kappa-4\kappa^{2}+4\kappa^{4}}}/{(1+\kappa)}$, $\widetilde{d}_{t}={(3+6\kappa)}/{(1+\kappa)}$ and $(\widetilde{d}_{0})_{\mathrm{ps}}={3}/{(2+2\kappa)}$ in the equilibrium state parameterized by $\theta _{A}^{0}=0$, $\theta
_{B}^{0}=\arccos(-\kappa /(1+\kappa ))$, $\theta _{C}^{0}=2\pi -\theta
_{B}^{0}$, and $\phi_{\alpha }^{0}=0$. 
The magnon annihilation and creation  operators are 
\begin{equation}
(\gamma_t^{\dagger},\gamma_\mathrm{ps}^{\dagger},\gamma_{g1}^{\dagger},\gamma_t,\gamma_\mathrm{ps},\gamma_{g1})=B_{0}^{\dagger}(\sigma^{z}\otimes I_3)U_{0}(\sigma^{z}\otimes I_3),  
\end{equation}
where 
\begin{equation}
    U_{0}=(
u_{t},(u_{0})_{\mathrm{ps}} ,u_{g} ,(\sigma^{x}\otimes I_3)u_{t}^{*} , (\sigma^{x}\otimes I_3)(u_{0})_{\mathrm{ps}}^*, (\sigma^{x}\otimes I_3)u_{g}^*)
\end{equation}
 satisfies the condition $U_{0}(\sigma^{z}\otimes I_3)U_{0}^{\dagger}=(\sigma^{z}\otimes I_3)$. The explicit form of vectors are
\begin{equation}
\left(\begin{array}{ccc}
u_{t} & (u_{0})_{\mathrm{ps}} & u_{g}\end{array}\right)=\left(\begin{array}{ccc}
0 & (u_{0})_{\mathrm{ps},1} & u_{g,1}\\
\frac{1}{\sqrt{2}} & \frac{1+6\kappa+14\kappa^{2}+8\kappa^{3}}{2\left(1+\kappa\right)\left(1+4\kappa\right)}(u_{0})_{\mathrm{ps},1} & u_{g,2}\\
\frac{-1}{\sqrt{2}} & \frac{1+6\kappa+14\kappa^{2}+8\kappa^{3}}{2\left(1+\kappa\right)\left(1+4\kappa\right)}(u_{0})_{\mathrm{ps},1} & u_{g,2}\\
0 & \frac{3+4\kappa}{1+4\kappa}(u_{0})_{\mathrm{ps},1} & u_{g,4}\\
0 & \frac{-1+2\kappa+10\kappa^{2}+8\kappa^{3}}{2\left(1+\kappa\right)\left(1+4\kappa\right)}(u_{0})_{\mathrm{ps},1} & u_{g,5}\\
0 & \frac{-1+2\kappa+10\kappa^{2}+8\kappa^{3}}{2\left(1+\kappa\right)\left(1+4\kappa\right)}(u_{0})_{\mathrm{ps},1} & u_{g,5}
\end{array}\right),
\end{equation}
where
\begin{align}
  (u_{0})_{\mathrm{ps},1}&=\frac{-i(1+4\kappa)}{2\sqrt{\frac{2}{1+\kappa}\left(1+2\kappa\right)\left(-1+2\kappa^{2}\left(1+\kappa\right)\right)}},  \\\nonumber
  u_{g,1}&=-i\frac{\sqrt{2}\left(-1-\kappa+2\kappa^{3}\right)+2\kappa\sqrt{1+\kappa\left(-1-2\kappa+2\kappa^{3}\right)}}{2\left(2-2\kappa-4\kappa^{2}+4\kappa^{4}\right)^{3/4}}, \\\nonumber
u_{g,2}&=i\frac{2^{3/4}-2^{3/4}\kappa^{2}+2^{1/4}\sqrt{1+\kappa\left(-1-2\kappa+2\kappa^{3}\right)}}{4\left(1+\kappa\left(-1-2\kappa+2\kappa^{3}\right)\right)^{3/4}}, \\\nonumber
u_{g,4}&=\frac{i\kappa}{2\left(\frac{1}{2}-\frac{\kappa}{2}-\kappa^{2}+\kappa^{4}\right)^{1/4}}-\frac{i\left(-1-\kappa+2\kappa^{3}\right)}{2*2^{1/4}\left(1+\kappa\left(-1-2\kappa+2\kappa^{3}\right)\right)^{3/4}}, \\\nonumber
u_{g,5}&=-i\frac{-\sqrt{2}+\sqrt{2}\kappa^{2}+\sqrt{1+\kappa\left(-1-2\kappa+2\kappa^{3}\right)}}{2\left(2-2\kappa-4\kappa^{2}+4\kappa^{4}\right)^{3/4}}.
\end{align}
In this basis, the dispersion
matrix becomes
\begin{equation}
U_{0}^{\dagger}M_{0}U_{0}=\left(\begin{array}{cccccc}
\frac{\widetilde{d}_{t}}{2} & 0 & 0 & \frac{\widetilde{d}_{t}}{2} & 0 & 0\\
0 & \frac{(\widetilde{d}_{0})_{\mathrm{ps}}}{2} & 0 & 0 & \frac{(\widetilde{d}_{0})_{\mathrm{ps}}}{2} & 0\\
0 & 0 & d_{g1} & 0 & 0 & 0\\
\frac{\widetilde{d}_{t}}{2} & 0 & 0 & \frac{\widetilde{d}_{t}}{2} & 0 & 0\\
0 & \frac{(\widetilde{d}_{0})_{\mathrm{ps}}}{2} & 0 & 0 & \frac{(\widetilde{d}_{0})_{\mathrm{ps}}}{2} & 0\\
0 & 0 & 0 & 0 & 0 & d_{g1}
\end{array}\right).\label{eq:xxzM}
\end{equation}
Here, the spontaneous breaking of the $U(1)$ symmetry generates the type-I true Goldstone mode, whose subspace is spanned by $u_{t}$ and $(\sigma^{x}\otimes I_3) u_{t}^*$; while the accident degeneracy at the classical level produces the type-I PG mode, whose subspace is
spanned by $(u_{0})_{\mathrm{ps}}$ and $(\sigma^{x}\otimes I_3)(u_{0})_{\mathrm{ps}}^*$, in agreement with Eq.~\eqref{tc}. 

 \subsection{Pseudo-Goldstone gap}
In this subsection, we derive the PG gap using the curvature formula. We first construct the rotation transformation that connects $H_{XXZ}(\phi_{\alpha}^{0},\theta_{\alpha}^{0})$ to $H_{XXZ}(\phi_{\alpha}^{c},\theta_{\alpha}^{c})$, i.e.,
\begin{equation}
H_{XXZ}(\phi_{\alpha}^{c},\theta_{\alpha}^{c})=H_{XXZ}(\phi,\theta=0)=\hat{R}^{\dagger}(\phi,\theta=0)H_{XXZ}(\phi_{\alpha}^{0},\theta_{\alpha}^{0})\hat{R}(\phi,\theta=0).
\end{equation}
Here, the rotation operator has the form
\begin{equation}
\hat{R}(\phi,\theta=0)=\exp[-i\sum_{j,\alpha}\omega_{\alpha}(\phi)\hat{S}_{j,\alpha}^{y}],
\end{equation}
where $\omega_{A}(\phi)=\theta_{A}^c-\theta_{A}^{0}=f_{A}\phi$, and
\begin{equation}
f_{A}=\sqrt{\frac{2}{N_{c}}}|\chi_{\phi,1}|=\sqrt{\frac{2(1+\kappa)(1+2\kappa)}{[-1+2\kappa^{2}(1+\kappa)]{N_{c}}}}. 
\end{equation}
The rotation angles in sublattices $B$ and $C$ are 
\begin{align}
    \omega_{B}(\phi)&=\theta_{B}^c-\theta_{B}^{0}=\frac{1}{n!}\sum_{n=1}^{\infty}\frac{\partial^{n}\theta_{B}^c}{\partial(\theta_{A}^{c})^n} f_{A}^{n}\phi^{n}, \\\nonumber
    \omega_{C}(\phi)&=\theta_{C}^c-\theta_{C}^{0}=\frac{1}{n!}\sum_{n=1}^{\infty}\frac{\partial^{n}\theta_{C}^c}{\partial(\theta_{A}^{c})^n}f_{A}^{n}\phi^{n}.
\end{align}
From the relation between ($\theta_{B}^c,\theta_{C}^c$) and $\theta_{A}^c$ in Eq.~\eqref{eq:thetaBC}, we find
\begin{align}
\frac{\partial\theta_{B}^c}{\partial\theta_{A}^c}\rvert_{\theta_{A}^c=0} & =\frac{\partial\theta_{C}^c}{\partial\theta_{A}^c}\rvert_{\theta_{A}^c=0}=\kappa,\\
\frac{\partial^{2}\theta_{B}^c}{(\partial\theta_{A}^c)^{2}}\rvert_{\theta_{A}^c=0} & =\frac{\kappa-\kappa^{3}}{\sqrt{1+2\kappa}}=-\frac{\partial^{2}\theta_{C}^c}{(\partial\theta_{A}^c)^{2}}\rvert_{\theta_{A}^c=0}.
\end{align}

Using the HP transformation, the rotation operator can be expressed as
\begin{equation}
\hat{R}(\phi,\theta=0)=e^{-i\sum_{\zeta=1}^{\infty}O_{\zeta,0}^{(\nu)}\phi^{\zeta}S^{1/2-\nu}}.\label{R0}
\end{equation}
The first-order term is
\begin{equation}
O_{1,0}^{(0)}=i\sqrt{\frac{N_c}{2}}f_{A}(b_{0,A}+\kappa b_{0,B}+\kappa b_{0,C})+{\rm{H.c.}}= B_{0}^{\dagger}\sigma^{z}\chi_{\phi}.
\end{equation}
Higher-order terms ($\zeta\geq2$) are given by
\begin{equation}
O_{\zeta\geq2,0}^{(0)}=i\sqrt{\frac{N_c}{2}}\frac{1}{\zeta!}f_{A}^{\zeta}\left[\left(\frac{\partial^{\zeta}\theta_{B}}{\partial\theta_{A}^{\zeta}}\right)_{0}b_{0,B}+\left(\frac{\partial^{\zeta}\theta_{C}}{\partial\theta_{A}^{\zeta}}\right)_{0}b_{0,C}\right]+\rm{H.c.}.\label{eq:O200}
\end{equation}
For example, the bilinear operator
\begin{align}
 O_{2,0}^{(0)}&=\frac{i}{\sqrt{2N_{c}}}\frac{(1+\kappa)(\kappa-\kappa^{3})\sqrt{1+2\kappa}}{-1+2\kappa^{2}(1+\kappa)}\left(b_{0,B}-b_{0,C}\right)+\rm{H.c.}
\end{align}


The spin-wave spectra $d_{k,m}(\theta_A^c)$ and the free energy $F(\theta_A^c)$ have been obtained numerically in the previous subsection. From the curvature formula, it follows that the type-I PG gap reads $\Delta=\sqrt{(\widetilde{d}_{0})_{\mathrm{ps}}(\partial_{\phi}^2F(\phi,\theta))_0 }=f_A\sqrt{(\widetilde{d}_{0})_{\mathrm{ps}}(\partial_{\theta_A^c}^2F(\theta_A^c))_0 }$. To compute the second derivatives of the free energy, we expand the Hamiltonian $H_{XXZ}(\mathbf{\phi}_{\alpha}^{0},\mathbf{\theta}_{\alpha}^{0})=\sum_{n=0}^{\infty}S^{2-n/2}\mathcal{H}_{n}$
using the HP transformation at the equilibrium state configuration. Here, the quadratic term is given in the previous subsection, and the cubic term reads
\begin{equation}
\mathcal{H}_{3}=\sum_{{p},{q},l_1,l_2,l_3}\frac{V_{l_1 l_2 l_3}^{(3)}(\mathbf{p},{q})}{3!\sqrt{N_{c}}}B_{{p},l_1}^{\dagger}B_{{q},l_2}B_{{p}-\mathbf{q},l_3},
\end{equation}
with the interaction tensors
\begin{equation}
V_{l_1l_21}^{(3)}(p,q)=V_{\kappa}\left(\begin{array}{cccccc}
0 & \kappa f_{q} & -\kappa\text{ }f_{-q} & 0 & \kappa f_{q} & -\kappa f_{-q}\\
0 & -f_{p-q}^{*} & 0 & \kappa f_{-p} & 0 & 0\\
0 & 0 & f_{p-q} & -\kappa\text{ }f_{p} & 0 & 0\\
0 & 0 & 0 & 0 & 0 & 0\\
0 & 0 & 0 & \kappa f_{-p} & -f_{p-q}^{*} & 0\\
0 & 0 & 0 & -\kappa f_{p} & 0 & f_{p-q}
\end{array}\right)_{l_1 l_2},
\end{equation}
\begin{equation}
V_{l_1l_22}^{(3)}(p,q)=V_{\kappa}\left(\begin{array}{cccccc}
\kappa f_{p-q} & 0 & 0 & 0 & -f_{p} & 0\\
-f_{q}^{*} & 0 & \kappa\text{ }f_{q} & -f_{-q} & 0 & \kappa f_{q}\\
0 & 0 & -\kappa f_{p-q}^{*} & 0 & \kappa f_{-p} & 0\\
0 & 0 & 0 & \kappa f_{p-q} & -f_{-p}^{*} & 0\\
0 & 0 & 0 & 0 & 0 & 0\\
0 & 0 & 0 & 0 & \kappa f_{-p} & -\kappa f_{p-q}^{*}
\end{array}\right)_{l_1 l_2},
\end{equation}
\begin{equation}
V_{l_1l_23}^{(3)}(p,q)=V_{\kappa}\left(\begin{array}{cccccc}
-\kappa\text{ }f_{p-q}^{*} & 0 & 0 & 0 & 0 & f_{-p}\\
0 & \kappa\text{ }f_{p-q} & 0 & 0 & 0 & -\kappa f_{p}\\
f_{q} & -\kappa f_{q}^{*} & 0 & f_{q} & -\kappa f_{-q} & 0\\
0 & 0 & 0 & -\kappa\text{ }f_{p-q}^{*} & 0 & f_{-p}\\
0 & 0 & 0 & 0 & \kappa\text{ }f_{p-q} & -\kappa f_{-p}^{*}\\
0 & 0 & 0 & 0 & 0 & 0
\end{array}\right)_{l_1 l_2},
\end{equation}
 $V_{l_1l_24}^{(3)}(p,q)=V_{l_1l_21}^{(3)\dagger}(q,p),V_{l_1l_25}^{(3)}(p,q)=V_{l_1l_22}^{(3)\dagger}(q,p),V_{l_1l_26}^{(3)}(p,q)=V_{l_1l_23}^{(3)\dagger}(q,p)$,
and $V_{\kappa}=-\sqrt{1+2\kappa}/[\sqrt{2}(1+\kappa)]$.
The explicit form of the quartic term 
\begin{equation}
    \mathcal{H}_{4}=\frac{1}{N_{c}}\frac{1}{4!}\sum_{{k},{p},{q}}\sum_{l_1,l_2,l_3,l_4}V_{l_1 l_2 l_3l_4}^{(4)}({k},{p},{q})B_{{k},l_1}^{\dagger}B_{{p},l_2}^{\dagger}B_{{k}+{q},l_3}B_{{p}-{q},l_4}
\end{equation}
is
\begin{align}
\mathcal{H}_{4} & =\frac{1}{4N_{c}}\sum_{k,p,q}[2\frac{-\kappa^{2}}{1+\kappa}b_{k,A}^{\dagger}b_{p,A}b_{q,B}^{\dagger}b_{k+q-p,B}f_{k-p}+2\frac{-\kappa^{2}}{1+\kappa}b_{k,A}^{\dagger}b_{p,A}b_{q,C}^{\dagger}b_{k+q-p,C}f_{k-p}^{*} \\
& +2\frac{-1-\kappa+\kappa^{2}}{1+\kappa}b_{k,B}^{\dagger}b_{p,B}b_{q,C}^{\dagger}b_{k+q-p,C}f_{k-p} -\frac{1}{2}\frac{-1-2\kappa}{1+\kappa}(b_{k,A}^{\dagger}b_{q,A}b_{p,A}b_{k-q-p,B}f_{k-q-p}+b_{k,B}^{\dagger}b_{q,B}b_{p,B}b_{k-q-p,A}f_{k-q-p}^{*}\nonumber \\
 & +b_{k,A}^{\dagger}b_{q,A}b_{p,A}b_{k-q-p,C}f_{k-q-p}^{*}+b_{k,C}^{\dagger}b_{q,C}b_{p,C}b_{k-q-p,A}f_{k-q-p}+b_{k,B}^{\dagger}b_{q,B}b_{p,B}b_{k-q-p,C}f_{k-q-p}\nonumber \\
 & +b_{k,C}^{\dagger}b_{q,C}b_{p,C}b_{k-q-p,B}f_{k-q-p}^{*})-\frac{1}{2}\frac{1}{1+\kappa}(b_{k,B}^{\dagger}b_{q,A}^{\dagger}b_{p,A}b_{k+q-p,A}f_{k}^{*}+b_{k,A}^{\dagger}b_{q,B}^{\dagger}b_{p,B}b_{k+q-p,B}f_{k}\nonumber \\
 & +b_{k,C}^{\dagger}b_{q,A}^{\dagger}b_{p,A}b_{k+q-p,A}f_{k}+b_{k,A}^{\dagger}b_{q,C}^{\dagger}b_{p,C}b_{k+q-p,C}f_{k}^{*}+b_{k,C}^{\dagger}b_{p,B}^{\dagger}b_{q,B}b_{k+p-q,B}f_{k}^{*}+b_{k,B}^{\dagger}b_{p,C}^{\dagger}b_{q,C}b_{k+p-q,C}f_{k})]+\rm{H.c.}.\nonumber
\end{align}

It can be verified that under the rotation Eq.~\eqref{R0} the classical energy $E_c$ is invariant. The derivatives of the free energy follow from Eq.~\eqref{eq:Fphi} as $(\partial_{\phi}^2F(\phi,\theta))_0=\mathcal{F}_{\phi }=S\chi _{\phi }^{\dagger }\bar{\Sigma}_{\mathbf{0}}\chi _{\phi }$ with the self-energy $\bar{\Sigma}_{\mathbf{0}}=\sum_{L=1}^{3}\bar{\Sigma}
_{\mathbf{0}}^{(L)}$. Here,
\begin{equation}
\bar{\Sigma}
_{\mathbf{0}}^{(1)}=\left(\begin{array}{cc}
\delta h_{k} & \delta\Delta_{k}\\
\delta\Delta_{(0,k)}^{\dagger} & \delta h_{-k}^{\mathrm{T}}
\end{array}\right),\delta h_{k}=\left(\begin{array}{ccc}
\tilde{h}^{\text{AA}} & \tilde{h}_{k}^{\text{AB}} & \tilde{h}_{k}^{\text{AC}}\\
\tilde{h}_{k}^{\text{AB}*} & \tilde{h}^{\text{BB}} & \tilde{h}_{k}^{\text{BC}}\\
\tilde{h}_{k}^{\text{AC}*} & \tilde{h}_{k}^{\text{BC}*} & \tilde{h}^{\text{CC}}
\end{array}\right),\delta\Delta_{k}=\left(\begin{array}{ccc}
\tilde{\Delta}^{\text{AA}} & \tilde{\Delta}_{k}^{\text{AB}} & \tilde{\Delta}_{k}^{\text{AC}}\\
\tilde{\Delta}_{-k}^{\text{AB}} & \tilde{\Delta}^{\text{BB}} & \tilde{\Delta}_{k}^{\text{BC}}\\
\tilde{\Delta}_{-k}^{\text{AC}} & \tilde{\Delta}_{-k}^{\text{BC}} & \tilde{\Delta}^{\text{CC}}
\end{array}\right),
\end{equation}
are determined by
\begin{align}
\tilde{h}^{\text{AA}} & =\frac{1}{4N_{c}}\sum_{p}[\frac{-6\kappa^{2}}{1+\kappa}\left(\left\langle b_{p,B}^{\dagger}b_{p,B}\right\rangle_0 +\left\langle b_{p,C}^{\dagger}b_{p,C}\right\rangle_0 \right)-\frac{1}{1+\kappa}\left(f_{p}^{*}\left\langle b_{p,B}^{\dagger}b_{p,A}\right\rangle_0 +f_{p}\left\langle b_{p,C}^{\dagger}b_{p,A}\right\rangle_0 \right)\\
& -\frac{-1-2\kappa}{1+\kappa}\left(f_{p}^{*}\left\langle b_{p,A}b_{-p,B}\right\rangle_0 +f_{p}\left\langle b_{p,A}b_{-p,C}\right\rangle_0 \right)]+\rm{H.c.},\nonumber\\
\tilde{h}_{k}^{\text{AB}} & =\frac{1}{4N_{c}}\sum_{p}\left[\frac{-4\kappa^{2}}{1+\kappa}\left\langle b_{p,B}^{\dagger}b_{p,A}\right\rangle_0 f_{k-p}-\frac{f_{k}}{1+\kappa}\left(\left\langle b_{p,B}^{\dagger}b_{p,B}\right\rangle_0 +\left\langle b_{p,A}^{\dagger}b_{p,A}\right\rangle_0 \right)+\frac{1}{2}\frac{1+2\kappa}{1+\kappa}\left(\left\langle b_{p,A}b_{-p,A}\right\rangle_0 +\left\langle b_{p,B}^{\dagger}b_{-p,B}^{\dagger}\right\rangle_0 \right)f_{k}\right],\nonumber\\
\tilde{h}_{k}^{\text{AC}} & =\frac{1}{4N_{c}}\sum_{p}\left[\frac{-4\kappa^{2}}{1+\kappa}\left\langle b_{p,C}^{\dagger}b_{p,A}\right\rangle_0 f_{k-p}^{*}-\frac{f_{k}^{*}}{1+\kappa}\left(\left\langle b_{p,C}^{\dagger}b_{p,C}\right\rangle_0 +\left\langle b_{p,A}^{\dagger}b_{p,A}\right\rangle_0 \right)+\frac{1}{2}\frac{1+2\kappa}{1+\kappa}\left(\left\langle b_{p,A}b_{-p,A}\right\rangle_0 +\left\langle b_{p,C}^{\dagger}b_{-p,C}^{\dagger}\right\rangle_0 \right)f_{k}^{*}\right],\nonumber\\
\tilde{h}_{k}^{\text{BC}} & =\frac{1}{4N_{c}}\sum_{p}[4\frac{-1-\kappa+\kappa^{2}}{1+\kappa}\left\langle b_{p,C}^{\dagger}b_{p,B}\right\rangle_0 f_{k-p}-\frac{1}{1+\kappa}\left(\left\langle b_{p,C}^{\dagger}b_{p,C}\right\rangle_0 +\left\langle b_{p,B}^{\dagger}b_{p,B}\right\rangle_0 \right)f_{k}\nonumber\\
&-\frac{1}{2}\frac{-1-2\kappa}{1+\kappa}\left(\left\langle b_{p,B}b_{-p,B}\right\rangle_0 +\left\langle b_{p,C}^{\dagger}b_{-p,C}^{\dagger}\right\rangle_0 \right)f_{k}],\nonumber
\end{align}
\begin{align}
\tilde{\Delta}^{\text{AA}} & =\frac{1}{2N_{c}}\sum_{p}\left[-\frac{1}{2}\frac{1}{1+\kappa}\left(\left\langle b_{p,B}b_{-p,A}\right\rangle_0 f_{p}+\left\langle b_{p,C}b_{-p,A}\right\rangle_0 f_{p}^{*}\right)+\frac{1}{2}\frac{1+2\kappa}{1+\kappa}\left(\left\langle b_{p,B}^{\dagger}b_{p,A}\right\rangle_0 f_{p}^{*}+\left\langle b_{p,C}^{\dagger}b_{p,A}\right\rangle_0 f_{p}\right)\right],\\
\tilde{\Delta}^{\text{BB}} & =\frac{1}{2N_{c}}\sum_{p}\left[-\frac{1}{2}\frac{1}{1+\kappa}\left(\left\langle b_{p,A}b_{-p,B}\right\rangle_0 f_{p}^{*}+\left\langle b_{p,C}b_{-p,B}\right\rangle_0 f_{p}\right)+\frac{1}{2}\frac{1+2\kappa}{1+\kappa}\left(\left\langle b_{p,A}^{\dagger}b_{p,B}\right\rangle_0 f_{p}+\left\langle b_{p,C}^{\dagger}b_{p,B}\right\rangle_0 f_{p}^{*}\right)\right],\nonumber\\
\tilde{\Delta}_{k}^{\text{AB}} & =\frac{1}{4N_{c}}\sum_{p}\left[\frac{-4\kappa^{2}}{1+\kappa}\left\langle b_{p,A}b_{-p,B}\right\rangle_0 f_{k-p}-\frac{1}{2}\frac{f_{k}}{1+\kappa}\left(\left\langle b_{p,A}b_{-p,A}\right\rangle_0 +\left\langle b_{p,B}b_{-p,B}\right\rangle \right)+\frac{1+2\kappa}{1+\kappa}\left(\left\langle b_{p,A}^{\dagger}b_{p,A}\right\rangle_0 +\left\langle b_{p,B}^{\dagger}b_{p,B}\right\rangle_0 \right)f_{k}\right],\nonumber\\
\tilde{\Delta}_{k}^{\text{AC}} & =\frac{1}{4N_{c}}\sum_{p}\left[\frac{-4\kappa^{2}}{1+\kappa}\left\langle b_{p,A}b_{-p,C}\right\rangle_0 f_{k-p}^{*}-\frac{1}{2}\frac{f_{k}^{*}}{1+\kappa}\left(\left\langle b_{p,A}b_{-p,A}\right\rangle_0 +\left\langle b_{p,C}b_{-p,C}\right\rangle_0 \right)+\frac{1+2\kappa}{1+\kappa}\left(\left\langle b_{p,A}^{\dagger}b_{p,A}\right\rangle_0 +\left\langle b_{p,C}^{\dagger}b_{p,C}\right\rangle_0 \right)f_{k}^{*}\right],\nonumber\\
\tilde{\Delta}_{k}^{\text{BC}} & =\frac{1}{4N_{c}}\sum_{p}[4\frac{-1-\kappa+\kappa^{2}}{1+\kappa}\left\langle b_{p,B}b_{-p,C}\right\rangle_0 f_{k-p}-\frac{1}{2}\frac{1}{1+\kappa}\left(\left\langle b_{p,B}b_{-p,B}\right\rangle_0 +\left\langle b_{p,C}b_{-p,C}\right\rangle_0 \right)f_{k}\nonumber\\
&+\frac{1+2\kappa}{1+\kappa}\left(\left\langle b_{p,B}^{\dagger}b_{p,B}\right\rangle_0 +\left\langle b_{p,C}^{\dagger}b_{p,C}\right\rangle_0 \right)f_{k}].\nonumber
\end{align}
Due to the permutation symmetries, $\tilde{h}^{\text{CC}}=\tilde{h}^{\text{BB}}$,
$\tilde{\Delta}^{\text{BB}}=\tilde{\Delta}^{\text{CC}}$, $\delta h_{-k}^{\mathrm{T}}=\delta h_{k}$, 
and $\delta\Delta_{-k}^{\dagger}=\delta\Delta_{k}$.

The second term $\bar{\Sigma}_{\mathbf{0}}^{(2)}$ can be simplified as
\begin{equation}
\bar{\Sigma}_{\mathbf{0}}^{(2)}=\sum_{l_3}{\frac{1}{N_{c}}}V_{l_1l_2l_3}^{(3)}(k,k)\left(\frac{1}{-M_{0}}v\right)_{l_3}\frac{\varGamma_{\kappa}}{\sqrt{2}},
\end{equation}
where $v=(0,1,-1,0,1,-1)$, $({-1}/{M_{0}})v=-v{(1+\kappa)}/{(3+6\kappa)}$, and
\begin{equation}
\varGamma_{\kappa}=\frac{-\sqrt{1+2\kappa}}{1+\kappa}\sum_{k}\left[f_{k}\left(\kappa\left\langle b_{k,C}b_{-k,B}\right\rangle_0 -\left\langle b_{k,B}b_{-k,A}\right\rangle_0 +\kappa\left\langle b_{k,B}^{\dagger}b_{k,C}\right\rangle_0 -\left\langle b_{k,A}^{\dagger}b_{k,B}\right\rangle_0 \right)+3\kappa\left(\left\langle b_{k,A}^{\dagger}b_{k,A}\right\rangle_0 -\left\langle b_{k,C}^{\dagger}b_{k,C}\right\rangle_0 \right)\right].
\end{equation}

The third term $\bar{\Sigma}_{\mathbf{0}}^{(3)}$ is
\begin{align}
\bar{\Sigma}_{\mathbf{0}}^{(3)}=&\frac{-1}{2N_{c}}\sum_{\mathbf{p}}\sum_{l_1,l_2,l_3,l_4,l_5,l_6}\frac{-n_{b}((\sigma^{z}\otimes SD_{\mathbf{p}})_{l_1l_1})+n_{b}((\sigma^{z}\otimes SD_{\mathbf{p}})_{l_2l_2})}{\left(\sigma^{z}\otimes D_{\mathbf{p}}\right)_{l_1l_1}-\left(\sigma^{z}\otimes D_{\mathbf{p}}\right)_{l_2l_2}} \\\nonumber   
    & \times[(\sigma^{z}\otimes I_{N_s})U_{\mathbf{p}}^{\dagger}]_{l_2 l_3}V_{l_3l_4l'}^{(3)}(\mathbf{p},\mathbf{p})(U_{\mathbf{p}})_{l_4 l_1}[(\sigma^{z}\otimes I_{N_s})U_{\mathbf{p}}^{\dagger}]_{l_1l_5}\overline{V}_{l_5l_6l}^{(3)}(\mathbf{p},\mathbf{p})(U_{\mathbf{p}})_{l_6 l_2},
\end{align}
where the interaction tensor $\overline{V}_{l_1 l_2 l_3}^{(3)}(\mathbf{p},\mathbf{q})$ is defined as $\overline{V}_{l_1 l_2 l_3}^{(3)}(\mathbf{p},\mathbf{q})=\sum_{l}V_{l_1 l_2 l}^{(3)}(\mathbf{p},\mathbf{q})(\sigma^x \otimes I_{N_s})_{l,l_3}$.

Finally, from Eqs.~\eqref{eqs:gap} and ~\eqref{eq:xxzM}, we obtain the type-I PG gap $\Delta=S^{1/2}\sqrt{(\widetilde{d}_{\mathbf{0}})_{\mathrm{ps}}\chi _{\phi}^{\dagger }\bar{\Sigma}_{\mathbf{0}}\chi _{\phi }}$ with $(\widetilde{d}_{\mathbf{0}})_{\mathrm{ps}}={3}/{(2+2\kappa)}$, which can also be obtained by computing three Feynman diagrams in Fig.~\ref{fig:selfenergy}, as shown in Sec.~\ref{sec:FT}.

\section{Order-by-disorder models}
\label{AppObD}
In this Appendix, we apply the theorem to study the PG mode in two order-by-disorder models at finite temperature. In the first subsection, we focus on the Heisenberg-compass model on a square lattice. In the second subsection, we study the Heisenberg Dzyaloshinskii-Moriya (DM) model. 

\subsection{Heisenberg-compass model on square lattice}
\begin{figure}[H]	
		\centering		
		\includegraphics[width=0.6\textwidth]{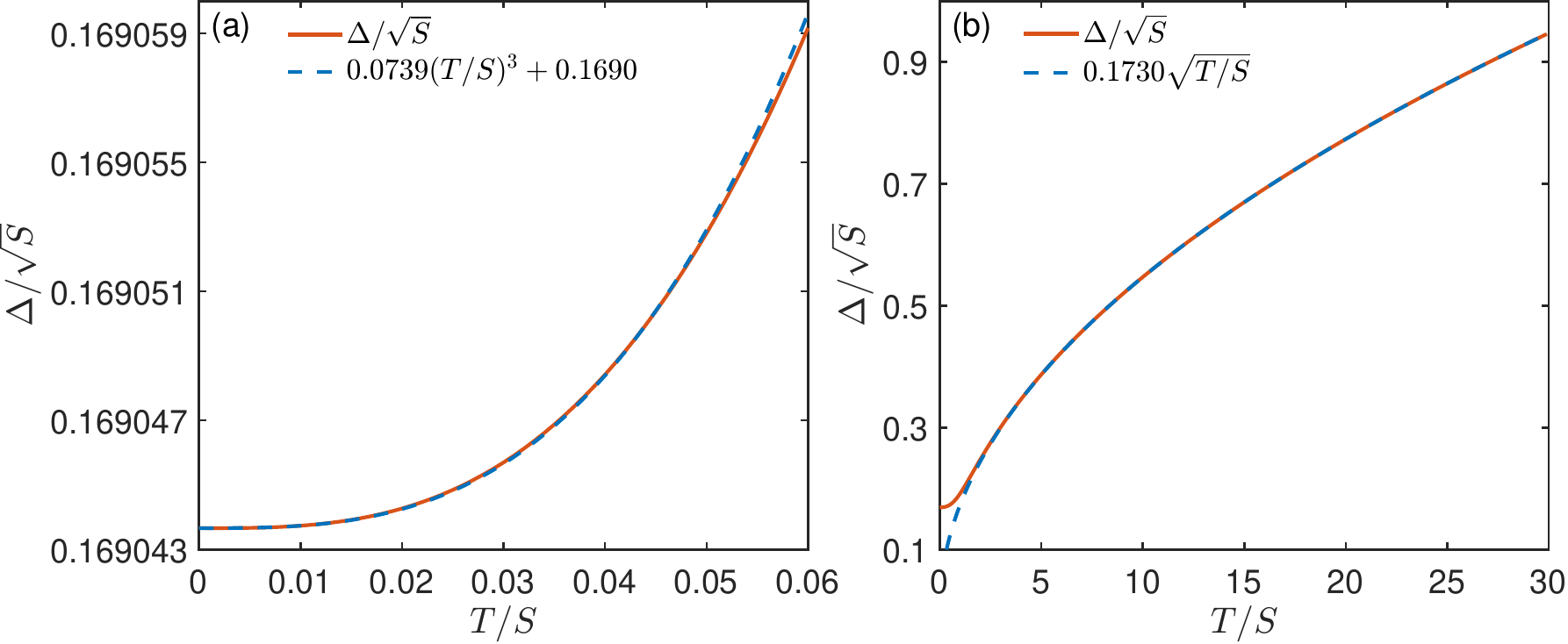}	
		\caption{Pseudo-Goldstone gap scaled by $\sqrt{S}$, $\Delta/\sqrt{S}$, as a function of temperature for $\kappa=-0.5$. The PG gap follows $0.0739(T/S)^3+0.1690$ in the low-T limit, and $0.1730\sqrt{T/S}$ in the classical limit.}
		\label{fig:k05}
\end{figure}

\begin{figure}[H]
		\centering		
		\includegraphics[width=0.6\textwidth]{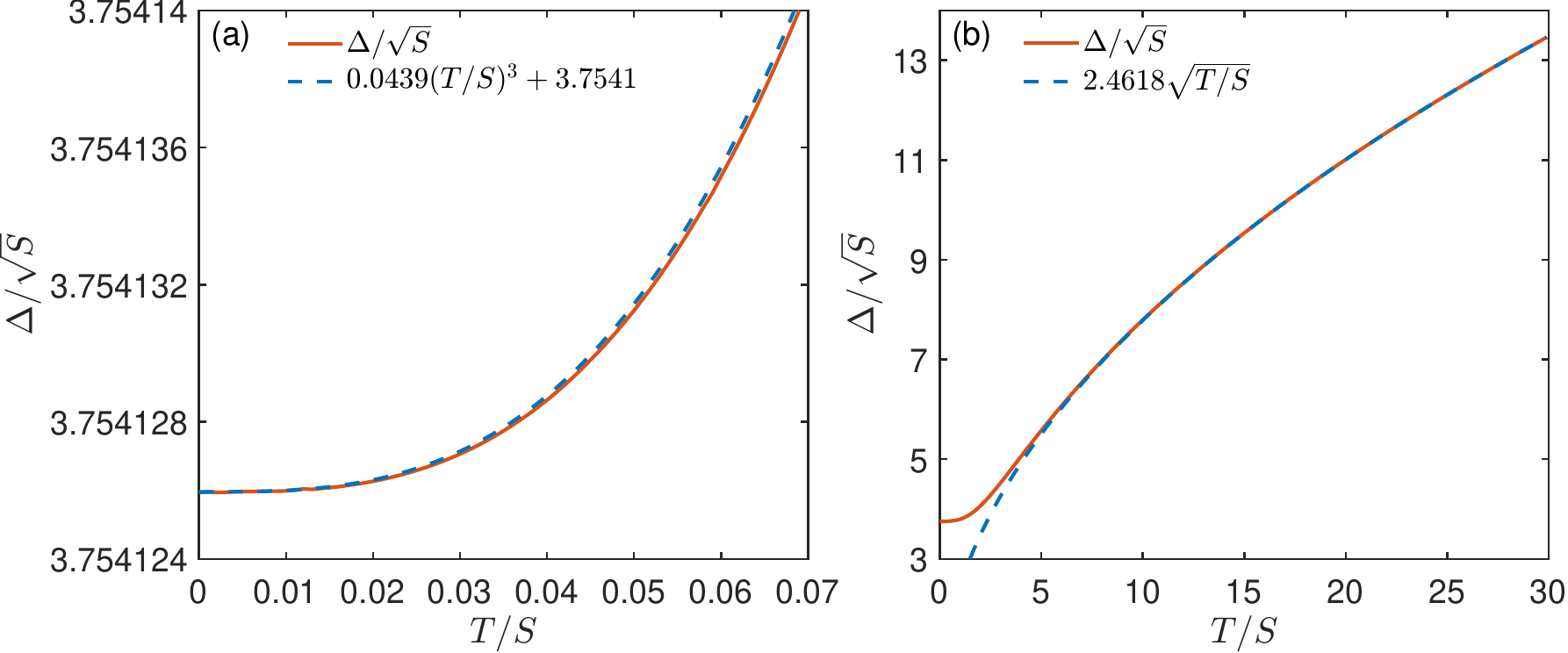}	
		\caption{ Pseudo-Goldstone gap scaled by $\sqrt{S}$, $\Delta/\sqrt{S}$, as a function of temperature for $\kappa=-5$. The PG gap follows $0.0439(T/S)^3+3.7541$ in the low-T limit, and $2.4618\sqrt{T/S}$ in the classical limit. 
        }
		\label{fig:k5}
\end{figure}

We consider the ferromagnetic Heisenberg-compass model on a square lattice, described by the Hamiltonian
\begin{equation}
    H=-\sum_{\langle \mathbf{r},\mathbf{r}'\rangle}\sum_{\mu}\hat{S}_{\mathbf{r}}^{\mu}\hat{S}_{\mathbf{r}'}^{\mu}+\kappa \sum_{\mathbf{r}}\left( \hat{S}_{\mathbf{r}}^{x}\hat{S}_{\mathbf{r}+\mathbf{x}}^{x}+\hat{S}_{\mathbf{r}}^{y}\hat{S}_{\mathbf{r}+\mathbf{y}}^{x}\right),
\end{equation}
where $\kappa<0$, the sum $\sum_{\langle \mathbf{r},\mathbf{r}'\rangle}$ runs over all nearest-neighbor pairs, and $(\mathbf{x},\mathbf{y})$ denote the unit lattice vectors along the $(x,y)$ directions. The total spin along the $z$-axis, i.e.,
$\sum_{\mathbf{r}}\hat{S}_{\mathbf{r
}}^{z}$, does not commute with $H$. At the classical level, the ground-state configuration is ferromagnetic with an arbitrary spin orientation in the $\hat{\mathbf{x}}$-$\hat{\mathbf{y}}$ plane, implying that $\sum_{\mathbf{r}}\hat{S}_{\mathbf{r
}}^{z}$ is an approximate symmetry.  This accidental degeneracy is lifted by quantum and thermal fluctuations. In the quantum ground state, the spins align with the $\hat{\mathbf{x}}$ and $\hat{\mathbf{y}}$ directions in the square lattice, and a type-I PG mode emerges. The type-I PG gaps for $\kappa=-0.5,-5$ are shown in Fig.~\ref{fig:k05} and~\ref{fig:k5}. At $T=0$, the PG gap for $\kappa=-0.5$ matches the result listed in the main text table of Ref.~\cite{Rau2018}. In the classical limit ($T\gg d_{\mathbf{k},m}$), the PG gap for $\kappa=-5$ agrees with the result obtained from the self-consistent mean-field theory in Ref.~\cite{Khatua2023}. Our approach is capable of covering a wide range of temperatures.


\subsection{Heisenberg DM model on pyrochlore lattice}
\begin{figure}[H]	
		\centering		
		\includegraphics[width=0.6\textwidth]{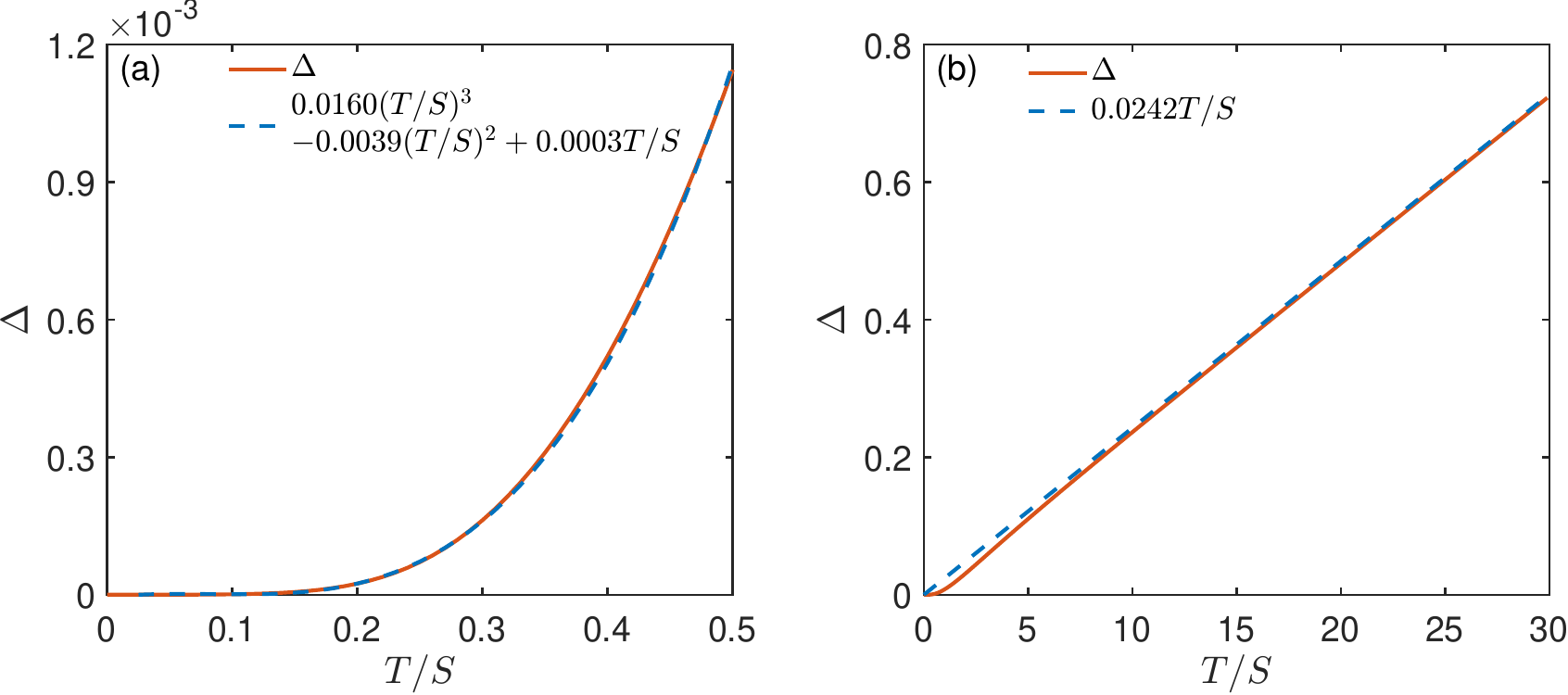}	
		\caption{Pseudo-Goldstone gap scaled by $S$, $\Delta$, as a function of temperature for $\kappa=1$. The PG gap follows $0.0160(T/S)^3-0.0039(T/S)^2+0.0003T/S$ in the low-T limit, and $0.0242{T/S}$ in the classical limit.}
		\label{fig:k1}
\end{figure}

We consider the Heisenberg DM model on a pyrochlore lattice, described by the Hamiltonian
\begin{equation}
    H_{\mathrm{DM}}=-\sum_{\langle \mathbf{r},\mathbf{r}'\rangle}\sum_{\mu}\hat{S}_{\mathbf{r}}^{\mu}\hat{S}_{\mathbf{r}'}^{\mu}+\kappa \sum_{\langle \mathbf{r},\mathbf{r}'\rangle} \epsilon_{\mu_1\mu_2\mu_3}\mathbf{d}_{ij}^{\mu_1} \hat{S}_{\mathbf{r}}^{\mu_2}\hat{S}_{\mathbf{r}'}^{\mu_2},
\end{equation}
where  $\epsilon_{\mu_1\mu_2\mu_3}$ is antisymmetric tensor, and $\mathbf{d}_{ij}$ are the nearest-neighbor DM vectors determined by Moriya's rules~\cite{hickey2024}. The Hamiltonian $H_{\mathrm{DM}}$ does not preserve the global $SU(2)$ symmetry generated by $\sum_{\mathbf{r}}\hat{S}_{\mathbf{r
}}^{\mu}$. For $-1<\kappa<2$ $(\kappa\neq0)$, ferromagnetic states with arbitrary spin orientation not only minimize $E_c$ but also remain exact ground states of $H_{\mathrm{DM}}$, implying that quantum fluctuations can not lift the degeneracy induced by the approximate symmetry. However, quantum corrections do affect excited states since the DM interaction explicitly breaks the $SU(2)$ symmetry. As a result, the accidental degeneracy is lifted by thermal fluctuations at finite temperature. For $-1<\kappa<1.2$ $(\kappa\neq0)$, thermal fluctuations select the equilibrium configuration $(\phi^0,\theta^0)=(\pi/4,\mathrm{acos}(1/\sqrt{3}))$ that minimizes the free energy. For $1.2<\kappa<2$, the twelve degenerate equilibrium configurations emerge with spins aligned along the six axes of a single tetrahedron in the pyrochlore lattice, e.g., $(\phi^0,\theta^0)=(0,\pi/4),(\pi/2,\pi/4), (\pi/4,\pi/2)$, etc. The type-II PG gap for $\kappa=1$ is shown in Fig.~\ref{fig:k1}.

\end{document}